\documentclass[twocolumn]{aastex631}

\pdfoutput=1
\usepackage{commath}
\usepackage{epsfig}
\newcommand{\Ms}{{\ensuremath{M_{\odot} }}}

\submitjournal{AAS journals}

\shorttitle{Radio Emission from High-$z$ AGNs}
\shortauthors{Latif et al.}

\begin{document}

\title{Radio Emission from High-Redshift Active Galactic Nuclei in the JADES and CEERS Surveys}

\correspondingauthor{Muhammad A. Latif}
\email{latifne@gmail.com}

\author[0000-0003-2480-0988]{Muhammad A. Latif}
\affiliation{Physics Department, College of Science, United Arab Emirates University, PO Box 15551, Al-Ain, UAE}
\author{Ammara Aftab}

\affiliation{Physics Department, College of Science, United Arab Emirates University, PO Box 15551, Al-Ain, UAE}

\author[0000-0001-6646-2337]{Daniel J. Whalen}

\affiliation{Institute of Cosmology and Gravitation, Portsmouth University, Dennis Sciama Building, Portsmouth PO1 3FX}

\begin{abstract}

Recent calculations indicate that radio emission from quasars at $z \sim$ 6 - 7 could be detected at much earlier stages of evolution, at $z \sim$ 14 - 15, by the Next-Generation Very Large Array (ngVLA) and the Square Kilometer Array (SKA).  However, the {\em James Webb Space Telescope} has now discovered less luminous active galactic nuclei (AGNs) at $z >$ 4 and a few massive black holes (BHs) at $z >$ 10, which may be the progenitors of supermassive black holes (SMBHs) but at different stages of growth.  Radio detections of these new AGNs would provide complementary measures of their properties and those of their host galaxies.  Here we estimate radio flux densities for 19 new AGNs found by the JADES, CEERS and UNCOVER surveys.  We find that ngVLA should be able to detect most of these sources in targeted surveys with integration times of 10 - 100 hr (and in just 1 hr for a few of them) but most would require at least 100 hr of SKA time in spite of its greater sensitivities at low frequencies.  In some cases, radio emission from the BH can be distinguished from that of H II regions and supernovae in their host galaxies, which could be used to estimate their star formation rates.  Such detections would be yet another example of the useful synergies between near infrared and radio telescopes in SMBH science in the coming decade.

\end{abstract}

\keywords{active galactic nuclei -- quasars: supermassive black holes --- early universe --- dark ages, reionization, first stars --- galaxies: formation --- galaxies: high-redshift}

\section{Introduction}

\label{sec:intro}

The detection of more than 300 quasars at $z >$ 6 has revealed the existence of $10^9~ \Ms$ black holes (BHs) as little as 700 Myr after the big bang (\citealt{yang20,wang21} -- see \citealt{fbs23} for a recent review).  Their initial discovery in 2003 posed challenges to cold dark matter cosmologies \citep{lf16,titans,ivh20} but recent cosmological simulations have now shown a number of scenarios in which they can form \citep{smidt18,may19,zhu22,latif22b}. However, the {\em James Webb Space Telescope} ({\em JWST}) has now discovered less luminous active galactic nuclei (AGNs) at $z >$ 4 along with a few massive BHs at $z >$ 10.  \citet{Maio23} found 12 new AGN identified by the presence of broad emission lines at $z >$ 4 in the {\em JWST} JADES survey. The BHs are a few times $10^5-10^7 ~ \Ms$, less massive than quasars at the same redshifts but, interestingly, similar in mass at the low end to those expected for direct collapse black holes at birth \citep[DCBHs;][]{hos13,tyr17,latif22b,herr23a,pat23}. They are overmassive relative to their host galaxies when compared to the local $\rm M_{BH} - M_{star}$ relation and their bolometric luminosities range from $\rm 10^{44}- 10^{45} ~erg/s$.  They reside in galaxies with sub-solar metallicities. 
% found 12 new AGNs through a broad line region

Quasars have also been detected in radio at $z >$ 6 \citep{bd20,mcg06,wil10,ban21,zng22}. These radio-loud sources have typical BH masses of $\gtrsim$ 10$^8$ \Ms\ and grow at nearly  the Eddington limit \citep{ban21}.  Their radio emission originates from compact cores rather than extended jets. Recent work indicates that these massive objects could be detected at earlier stages of evolution at $z \lesssim 13 - 14$ by the Next-Generation Very Large Array (ngVLA) and the Square Kilometer Array (SKA) in the coming decade \citep[][see also \citealt{wet21a}]{W23,L24}.  Detections of SMBH candidates in the near infrared (NIR) and radio at these higher redshifts will be crucial to determining their origin \citep{Vol23,Sch23} and probing their evolution at early times \citep[e.g.,][]{wf12,jet13}.
%at lower mass ($M_{\rm BH}\geq 10^6~\Ms$) 

Here, we examine the prospects for detection of less luminous AGNs at high redshifts by ngVLA and the SKA by estimating radio flux densities for the 12 new AGNs found by \citet{Maio23} in the {\em JWST} JADES survey at $z =$ 5.6 - 10.6 and 7 other AGNs found in CEERS, UNCOVER and Abell 2744 lensing surveys at $z =$ 5.42 - 8.7 \citep{Bod23,Kok23,Furt23,Lamb23,Lars23,Gould23,Maio23a}.  Besides emission due to BH accretion, our estimates include thermal bremsstrahlung from H II regions and synchrotron emission from supernovae (SNe) when star formation rates (SFRs) for their host galaxies are known. In Section 2 we discuss how we estimate radio emission from AGNs and their host galaxies.  We present radio flux densities and examine their prospects for detection in the coming decade in Section 3 and conclude in Section 4.

\begin{table*}
\caption{JADES and CEERS AGN Data \citep{Maio23}}
\begin{center}
\begin{tabular}{| c | c | c | c| c| c| c|c|}
\hline
\hline
Source ID & $z$  & log $M_{\rm BH}$  & $\rm L/L_{Edd}$ & log $L_{\rm bol}$ & SFR & lensing factor & Reference  \\
    &         &            (\Ms)            &       & (erg/s)           &  (\Ms/yr) & $\mu$ & \\
\hline
J-10013704 &5.919 &	5.65 &	1.06 &	43.8 &- &- & \cite{Maio23} \\
J-8083 &4.6482 &	7.25 &	0.16 &	44.6 &- &- & \cite{Maio23}\\
J-3608 &5.268 &	6.82 &	0.11 &	44.0 &- &- & \cite{Maio23}\\
J-11836 &4.409&	7.13&	0.2&	44.5 &- &- & \cite{Maio23}\\
J-20621 &4.681&	7.3	&  0.18&	44.7 &- &- & \cite{Maio23}\\
J-73488 &4.1332&	7.71&	0.16&	45.0 &- &- & \cite{Maio23} \\
J-77652 &5.229&	6.86&	0.38&	44.5 &- &- & \cite{Maio23} \\
J-61888 &5.874&	7.22&	0.32&	44.8 &- &- & \cite{Maio23}\\
J-62309 &5.172&	6.56&	0.39&	44.2 &- &- & \cite{Maio23} \\
J-53757 &4.4480&	7.69&	0.05&	44.4 &- &- &  \cite{Maio23}\\
J-954 &6.760&	7.9&	0.42&	45.6 &- &- & \cite{Maio23} \\
J-1093 &5.595 &7.36 & 0.2 & 44.8 &- &- & \cite{Maio23}\\
GN-z11 &	10.6 &	6.2	& 5.5	& 45.0	& 25   &- & \cite{Maio23a} \\
 &	 &		& & 	&  \cite{Nath23}  & &  \\
UHZ1 &	10.1 &	7.602 &	1	& 45.7 &	1.13  & 3.81 & \cite{Bod23} \\
 &	 &	  & &   & \cite{Gould23} &  &  \\
CEERS 1670 &	5.242 &	7.113 &	0.15 &	44.4 &	3.6 & - & \cite{Koc23} \\
CEERS 3210 &	5.624 &	7.6	 & 0.29 &	- &	-& 	- & \cite{Koc23} \\
CEERS 3210 Av=4 & 	5.624 &	6.95 &	3.5 &- &- &	- & \\
CEERS 1019&  8.679 & 6.95 & 1.3 & 45.1 & 30 & -  & \cite{Lars23}\\
Abell 2744-QSO1 &	7.045 &	7.477 &	0.3 &	45.0 &	40 & 3.52 & \cite{Furt23}\\
UNCOVER-20466 &	8.50 &	8.16 &	0.4 & 45.811 &	- &- & \cite{Kok23} \\
\hline
\end{tabular}
\label{tbl:tbl1}
\end{center}
\end{table*}

\section{Numerical Method}

\label{sec:method}

We calculate radio flux densities for the 19 AGNs summarized in Table~\ref{tbl:tbl1} at 0.1 - 10 GHz with fundamental planes (FPs) of BH accretion.  FPs are empirical relationships between BH mass, $M_\mathrm{BH}$, its nuclear radio luminosity at 5 GHz, $L_\mathrm{R}$, and its nuclear X-ray luminosity at 2 - 10 keV, $L_\mathrm{X}$ \citep{merl03}.  They cover six orders of magnitude in BH mass, down to 10$^5$ \Ms\ intermediate-mass black holes \citep{gul14}.  
%[IMBHs;][]
\subsection{BH Radio Flux Density}

To calculate the BH radio flux density in a given band in the observer frame we first use an FP to find $L_\mathrm{R}$ in the source frame, which requires $L_\mathrm{X}$ and $M_\mathrm{BH}$.  We determine $L_\mathrm{X}$ from $L_{\mathrm{bol}}$, the bolometric luminosity of the BH, from Equation 21 of \citet{marc04},
\begin{equation}
\mathrm{log}\left(\frac{L_\mathrm{bol}}{L_\mathrm{X}}\right) = 1.54 + 0.24 \mathcal{L} + 0.012 \mathcal{L}^2 - 0.0015 \mathcal{L}^3,
\end{equation}
where $L_\mathrm{bol}$ is in units of solar luminosity and $\mathcal{L} = \mathrm{log} \, L_\mathrm{bol} - 12$.  $M_\mathrm{BH}$ and $L_{\mathrm{bol}}$ for all 20 sources are listed in Table~\ref{tbl:FPs}.  $L_\mathrm{R}$ can then be found from $L_\mathrm{X}$ from the FP,
\begin{equation}
\mathrm{log} \, L_\mathrm{R} (\mathrm{erg \, s^{-1}})= A \, \mathrm{log} \, L_\mathrm{X} (\mathrm{erg \, s^{-1}}) + B \, \mathrm{log} \, M_\mathrm{BH} (\mathrm{M}_{\odot})+ C,
\end{equation}
where $A$, $B$, and $C$ are taken from \citet[][MER03]{merl03}, \citet[][KOR06]{kord06}, \citet[][GUL09]{gul09}, \citet[][PLT12]{plot12}, and \citet[][BON13]{bonchi13} and are shown in Table~\ref{tbl:FPs}. We also use the FP in Equation 19 in \citet[][GUL19]{gul19}, which has a different form:
\begin{equation}
R \, = \, -0.62 + 0.70 \, X + 0.74 \, \mu,
\end{equation}
where $R =$ log($L_\mathrm{R}/10^{38} \mathrm{erg/s}$), $X =$ log($L_\mathrm{X}/10^{40} \mathrm{erg/s}$) and $\mu =$ log($M_\mathrm{BH}/10^{8}$\Ms). 

\begin{table}
\caption{Fundamental Plane Coefficients }
\begin{center}
\begin{tabular}{cccc}
\hline
\\
FP & A & B & C \\
\hline
\\
MER03   &  0.60  & 0.78  &  7.33  \\
KOR06   &  0.71  &  0.62  &  3.55 \\
GUL09   &  0.67  &  0.78  &  4.80  \\
PLT12    &  0.69  &  0.61  &  4.19  \\
BON13   &  0.39  &  0.68  & 16.61 
\\
\hline
\end{tabular}
\end{center}
\label{tbl:FPs}
\end{table}

%Radio flux that is cosmologically redshifted into a given observer band today does not originate from 5 GHz in the source frame at high redshifts
Radio flux from high redshift sources in a given receiver band today is usually redshifted from a frequency other than 5 GHz in the source frame, so we determine the source frame flux from $L_\mathrm{R} =$ $\nu L_{\nu}$, assuming that the spectral luminosity $L_{\nu} \propto \nu^{-\alpha}$.  We consider $\alpha =$ 0.7 and 0.3 to enclose a reasonable range of spectral profiles \citep{ccb02,glou21}.  The spectral flux at $\nu$ in the Earth frame is then determined from the spectral luminosity at $\nu'$ in the rest frame from
\begin{equation}
F_\nu = \frac{L_{\nu'}(1 + z)}{4 \pi {d_\mathrm L}^2},
\end{equation}
where $\nu' = (1+z) \nu$ and $d_\mathrm L$ is the luminosity distance.  Here, we assume second-year \textit{Planck} cosmological parameters:  $\Omega_{\mathrm M} = 0.308$, $\Omega_\Lambda = 0.691$, $\Omega_{\mathrm b}h^2 = 0.0223$, $\sigma_8 =$ 0.816, $h = $ 0.677 and $n =$ 0.968 \citep{planck2}.  We scale this flux by the magnification factor, $\mu$, for the two lensed sources in Table~\ref{tbl:tbl1}.
 
\subsection{Supernova Radio Flux Density}

Synchrotron emission from young SN remnants in the host galaxy can mimic AGNs, particularly at lower frequencies.  We estimate the non-thermal radio flux density from SNe due to star formation from \citet{con92}:
\begin{equation}
\left(\frac{L_{\mathrm{N}}}{\mathrm{W \, Hz^{-1}}}\right) \, \sim \, 4.4 \times 10^{34} \left(\frac{\nu}{\mathrm{GHz}}\right)^{-\beta} \left[\frac{\mathrm{SFR}(M > 5 \, \mathrm{M_{\odot}})}{\mathrm{M_{\odot}} \, \mathrm{yr^{-1}}} \right],
\label{eq:SNe}
\end{equation}
where $\beta \sim$ 0.8 is the nonthermal spectral index and SFRs for sources for which they are known are listed in Table~\ref{tbl:tbl1}.  Equation~\ref{eq:SNe} may overstimate the SN flux density from the host galaxies because core-collapse SNe \citep{jet09b,latif22a} only produce nJy radio flux densities \citep{mw12} in the diffuse H II regions of high-redshift halos \citep{wan04,wet08a}. \cite{con92} also assume that all stars $>$ 5 \Ms\ contribute to SN flux. This assumption somewhat overestimates the total SN flux because the minimum mass for CC SNe is thought to be 8 \Ms.  Since we further assume that all stars in the host galaxy are above 5 \Ms\ for simplicity, our SN flux density estimates should be taken to be upper limits.

\subsection{H II Region Radio Emission}

Thermal bremsstrahlung in H II regions due to massive star formation also produces continuum radio emission whose spectral density can be calculated from the ionizing photon emission rate in the H II region, $Q_{\mathrm{Lyc}}$:
\begin{equation}
L_{\nu} \, \lesssim \, \left(\frac{Q_{\mathrm{Lyc}}}{6.3 \times 10^{52} \, \mathrm{s^{-1}}}\right) \left(\frac{T_{\mathrm{e}}}{10^4 \mathrm{K}}\right)^{0.45} \left(\frac{\nu}{\mathrm{GHz}}\right)^{-0.1}
\end{equation}
in units of 10$^{20}$ W Hz$^{-1}$ \citep{con92}, where $Q_{\mathrm{Lyc}} (s^{-1}) =$ SFR (\Ms \, yr$^{-1}$) $/$ $1.0 \times 10^{-53}$ \citep{ken98} and $T_{\mathrm{e}}$ is the electron temperature, taken to be 10$^4$ K.  As with the BH flux densities, we scale the SN and H II region flux densities by the lensing factors of their host galaxies where applicable.  For AGNs from \citet{Maio23} for which SFRs are not available we consider 1, 3 and 10 \Ms\ yr$^{-1}$, as found for similar galaxies in numerical simulations and observations \citep{L18,L20}.

\subsection{SKA / ngVLA Sensitivity Limits}

\begin{table}
\caption{Top three rows: planned sensitivity limits for SKA1 at 500 MHz, 1.5 GHz, 6.5 GHz, and 12.5 GHz.  Bottom three rows: same but for ngVLA. They are in units of nJy/beam for 1 hr, 10 hr, and 100 hr integration times (see Table 3 of \citealt{B19} and \citealt{pr18}).
}
\label{tab:sens}
\begin{center}
\begin{tabular}{lcccc}
\hline
\\
 & 500 MHz & 1.5 GHz & 6.5 GHz  & 12.5 GHz \\
\hline
\\
1 hr     & 4400  & 2000 & 1300 & 1200  \\
10 hr   & 1391  & 632   & 411   &  379   \\
100 hr & 440    & 200   & 130   & 120    \\
\\
\hline
\\
1 hr     & ---  & 382 & 220 & 220  \\
10 hr   & ---  & 121 & 70   &  70   \\
100 hr & ---  & 38   & 22   &  22   \\
\\
\hline
\vspace{-0.3in}
\end{tabular}
\end{center}
\end{table}

We show SKA1 and ngVLA sensitivity limits for integration times of 1 hr, 10 hrs and 100 hrs in Table~\ref{tab:sens}. The SKA rms limits are listed for each frequency band in Table 3 of \citet{B19} and the ngVLA 5$\sigma$ rms values are taken from \cite{pr18}. The SKA limits are based on SKA1-Low and SKA1-Mid. SKA1-Low will have 131,072 dipole antennae grouped into 512 stations, each with 256 antennae. They will cover 50-350 MHz with a baseline of 65 km.  SKA1-Mid will consist of 133 dishes 15-m in radius and, in combination with 64 existing MeerKAT dishes (13.5-m each), will cover 350 MHz to 15 GHz for baselines of up to 150 km.
%\footnote{https://www.skao.int/en/explore/telescopes}. \footnote{https://www.aspbooks.org/a/volumes/table_of_contents/?book_id=592}

The proposed ngVLA reference design (Long Base Array) will have 244 18-m diameter dishes with a baseline up to $\sim$ 8,000 km. The sensitivity of ngVLA will be 10 times that of VLA and band 3 of ALMA \citep{pr18}. Similarly, SKA sensitivities will be at least a factor of a few better than those of VLA and uGMRT \citep[Figure 6 of][]{B19}. The long baseline of ngVLA in particular is expected to be helpful in selecting contamination from SF.

\section{Results} 

\label{sec:res}
\begin{figure*} 
\begin{center}
\includegraphics[scale=0.45]{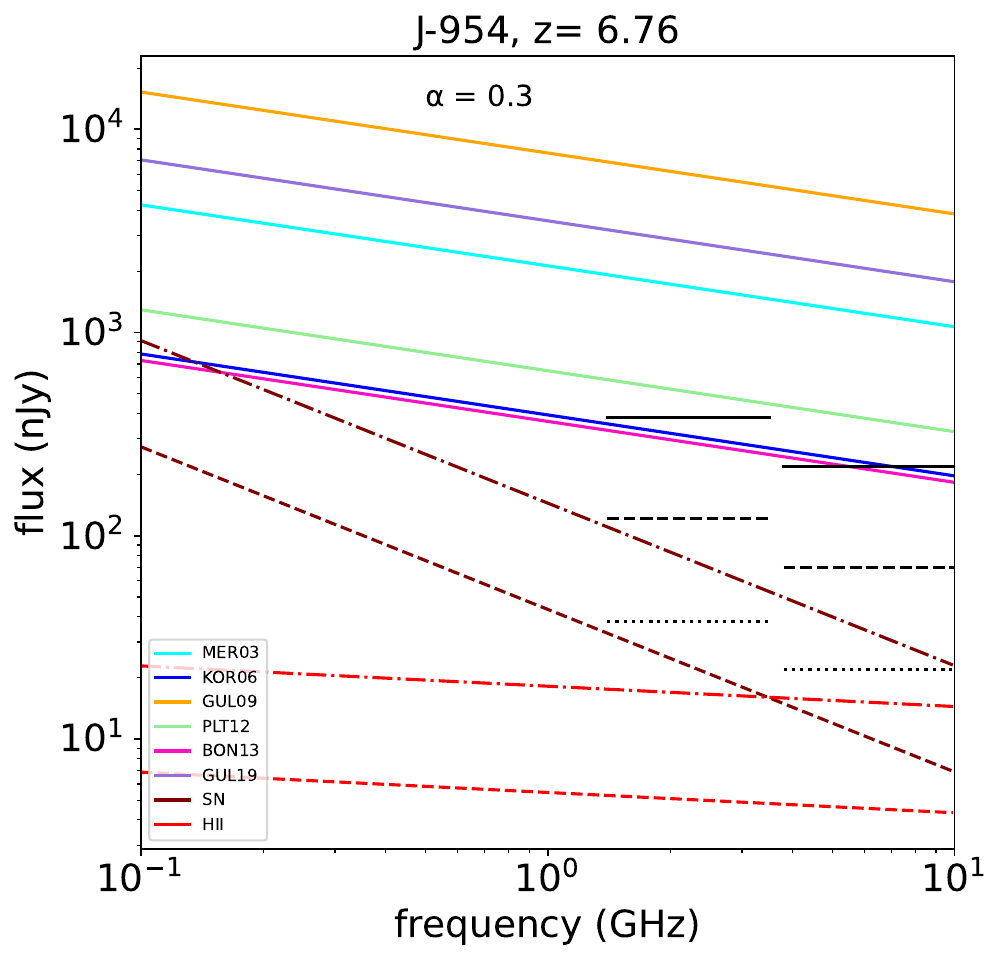}
\includegraphics[scale=0.45]{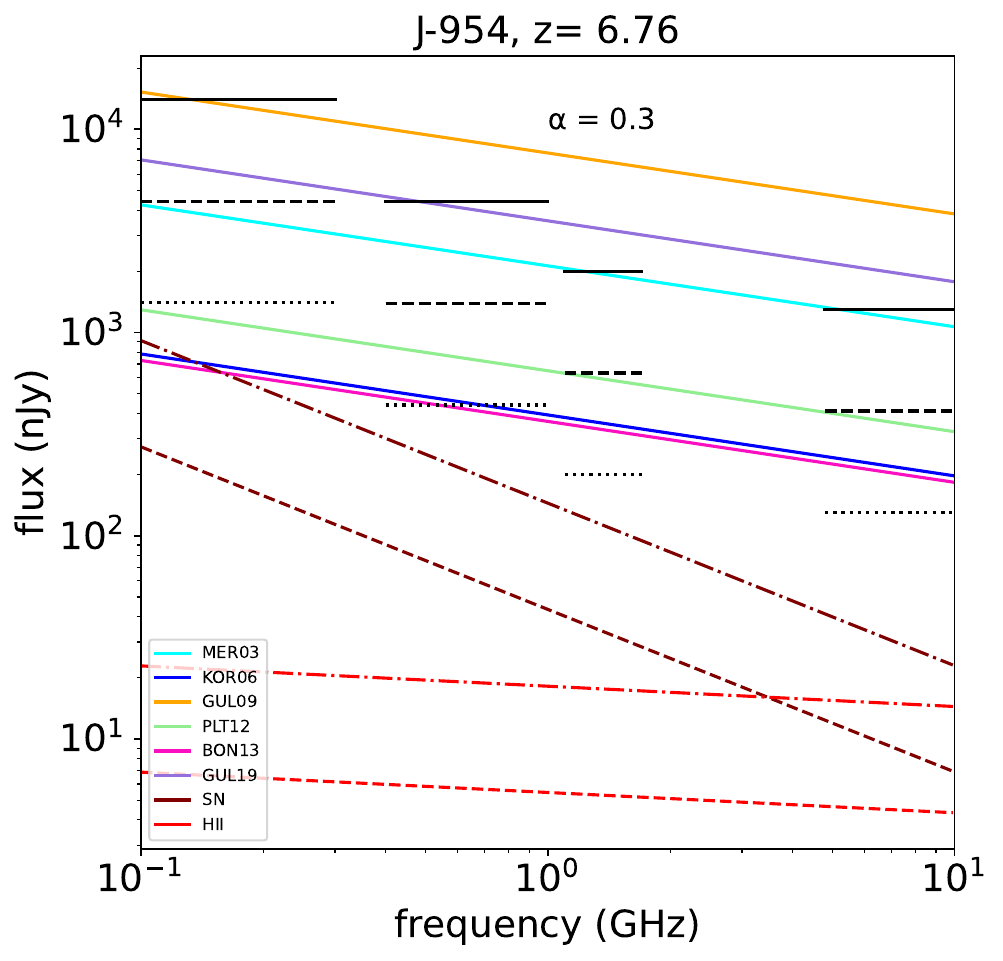}
\includegraphics[scale=0.45]{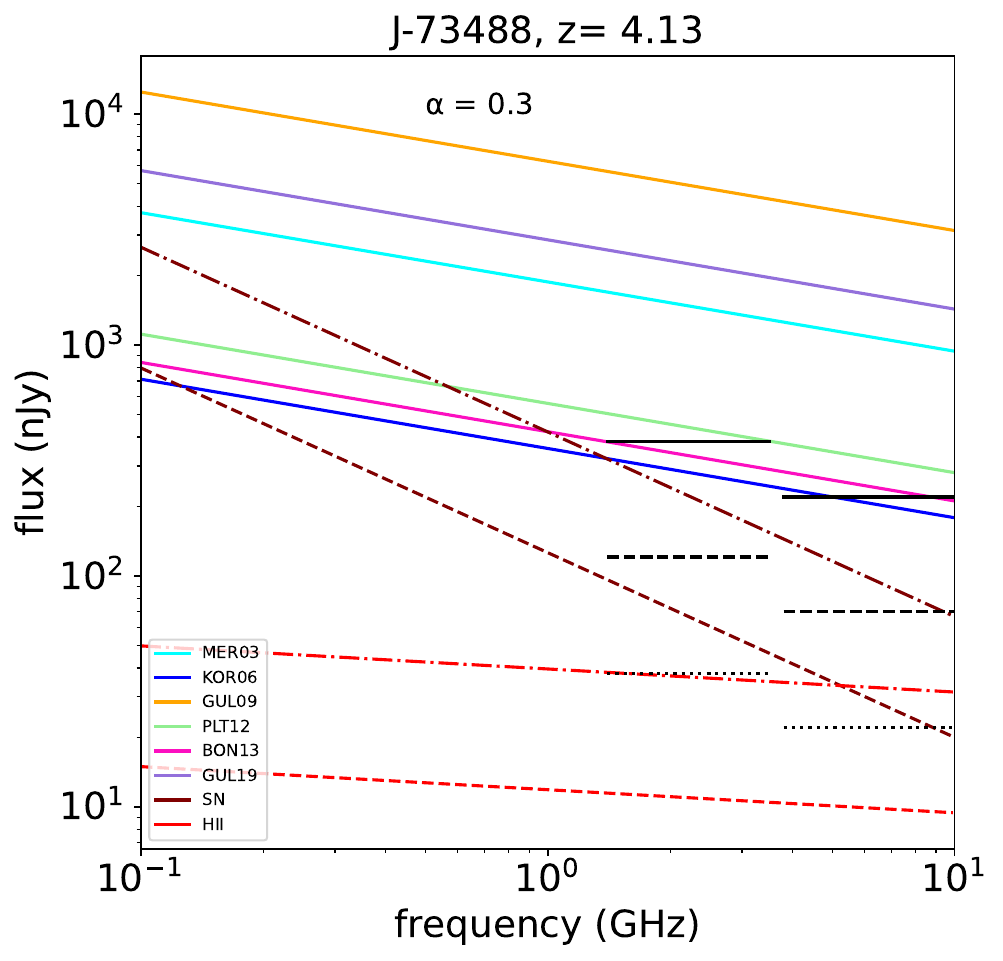}
\includegraphics[scale=0.45]{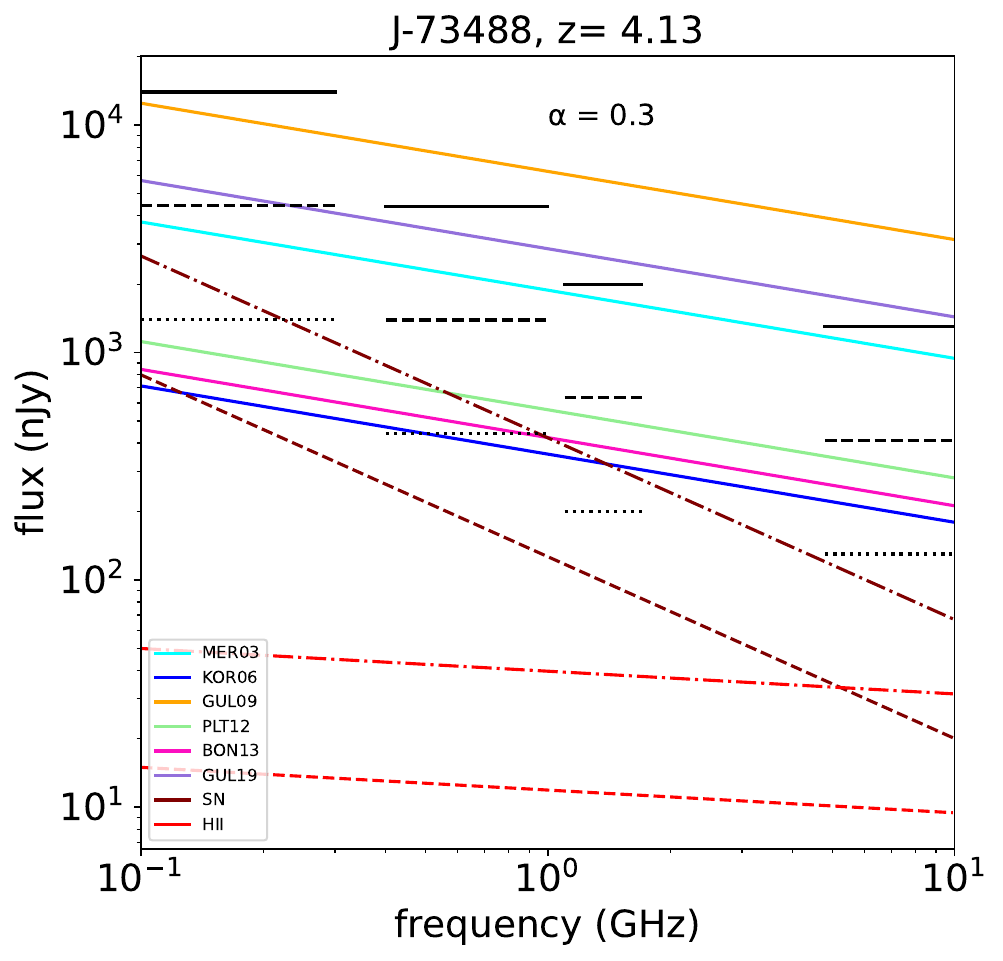}
\includegraphics[scale=0.45]{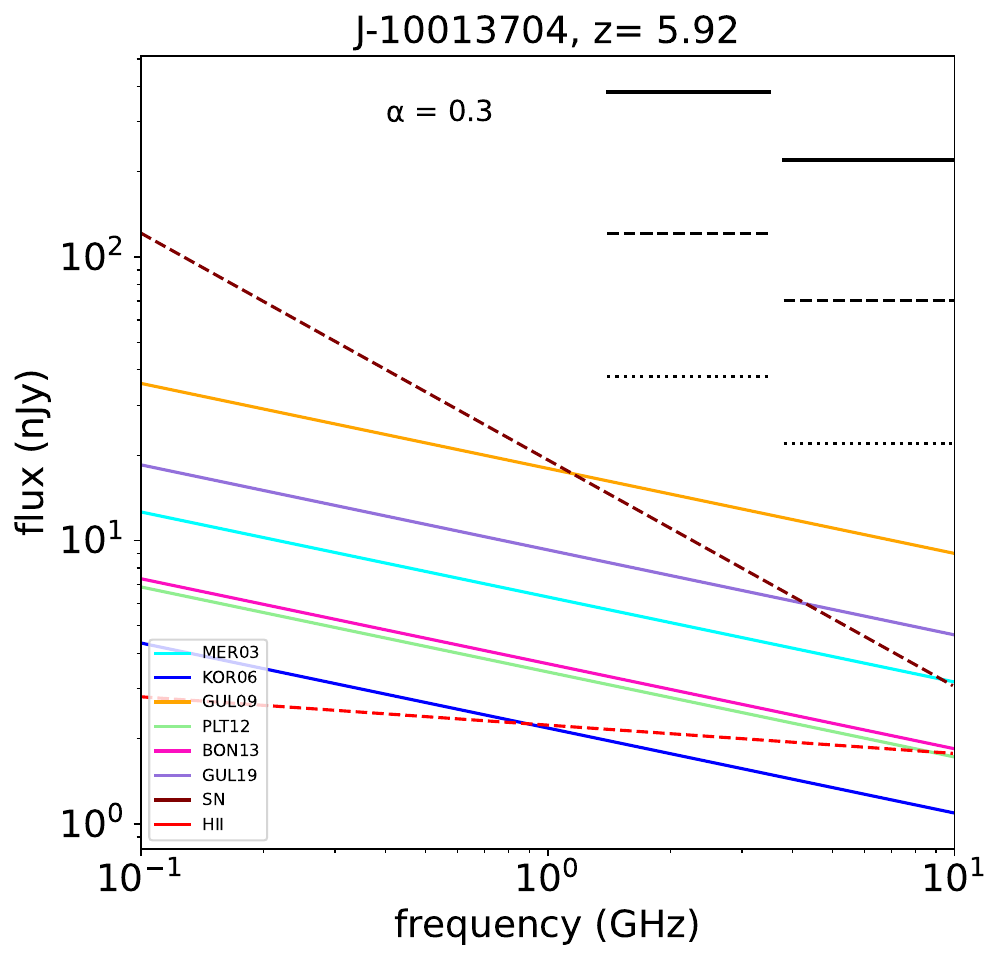}
\includegraphics[scale=0.45]{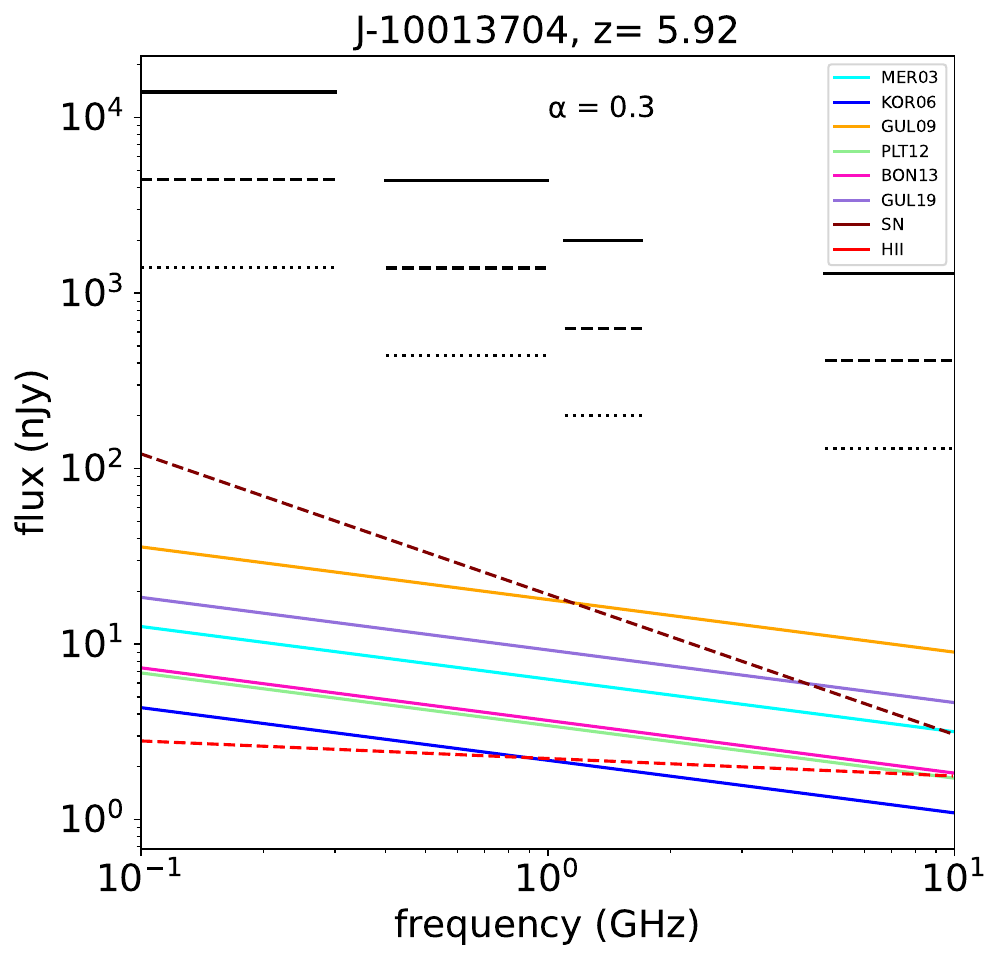} 
\end{center}
\vspace{-0.1cm}
\caption{Radio flux densities for AGNs observed in JADES for $\alpha$ = 0.3 with detection limits for ngVLA (left column) and SKA (right column). The dotted, dashed and dot-dashed red and brown lines are H II region and SN flux densities for SFRs of 1, 3 and 10 \Ms\ yr$^{-1}$, respectively. The black solid, dashed and dotted horizontal bars show ngVLA and SKA detection limits for integration times of 1, 10 and 100 hr, respectively.}
\label{fig:f1}
\end{figure*}

\begin{figure*} 
\begin{center}
\includegraphics[scale=0.45]{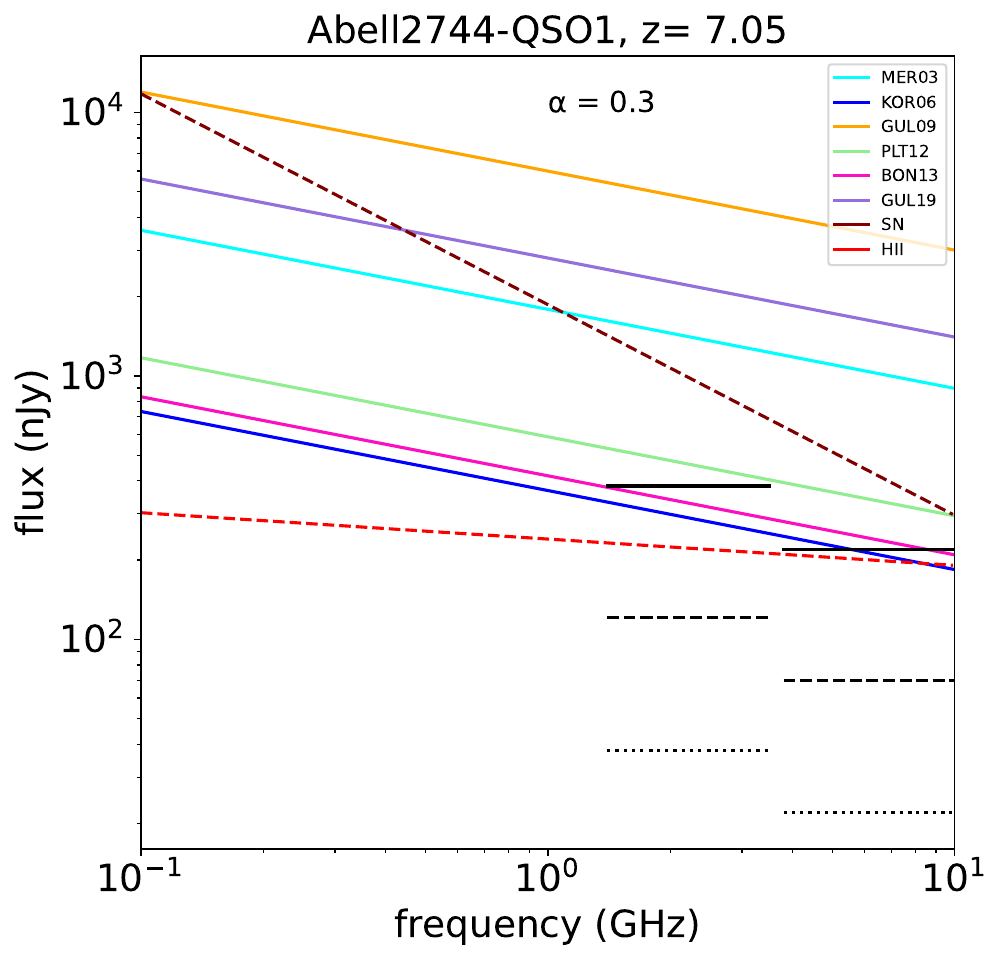}
\includegraphics[scale=0.45]{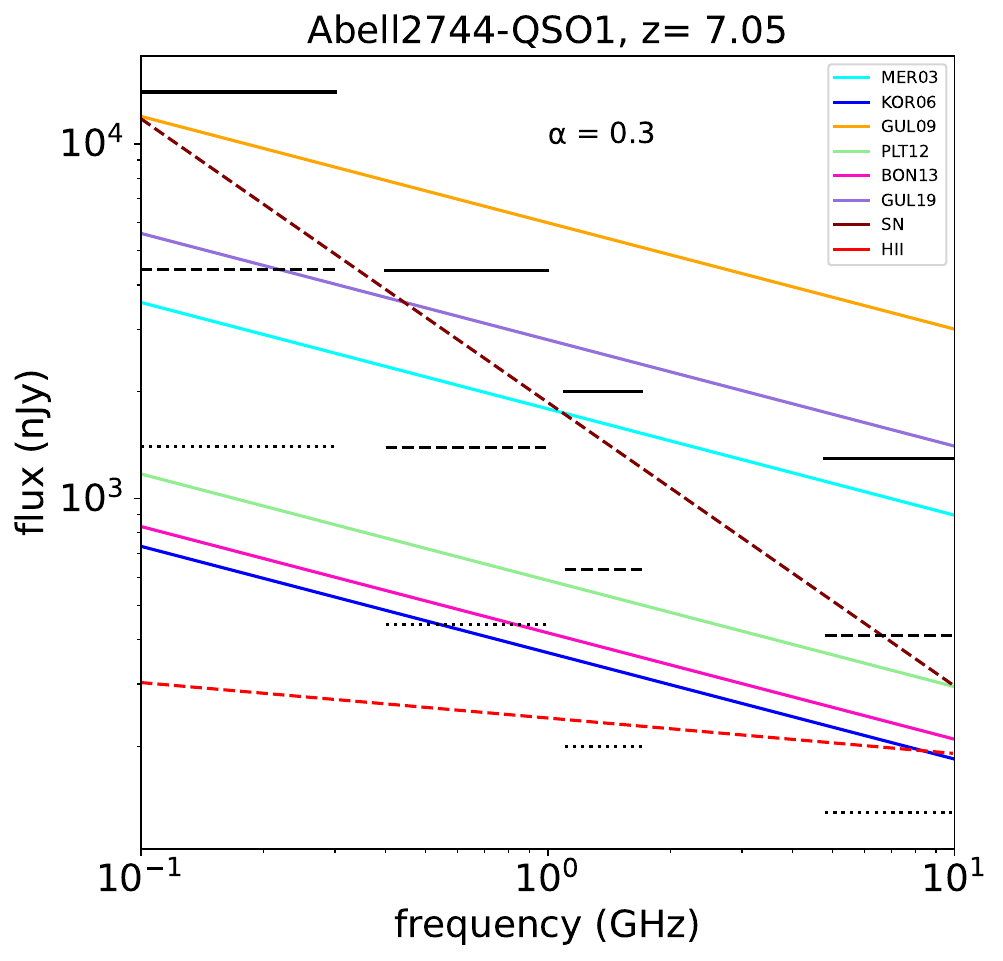}
\includegraphics[scale=0.45]{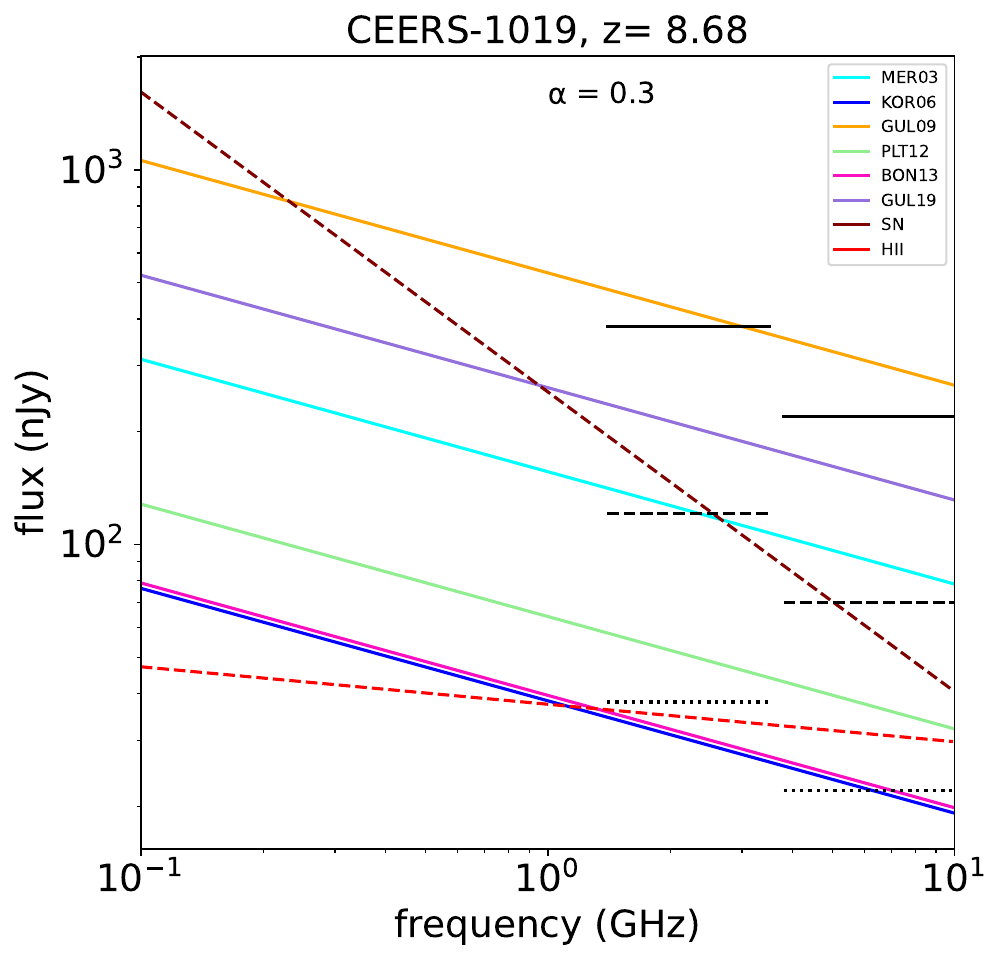}
\includegraphics[scale=0.45]{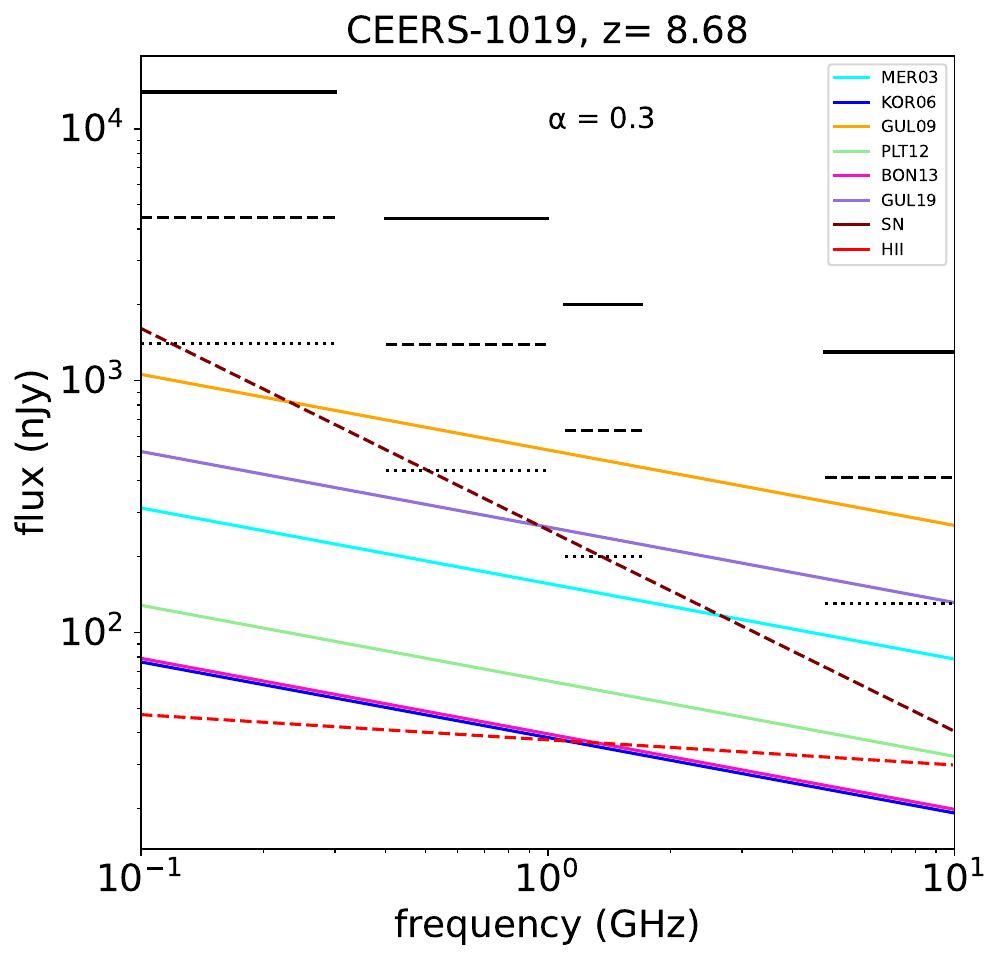}
\includegraphics[scale=0.45]{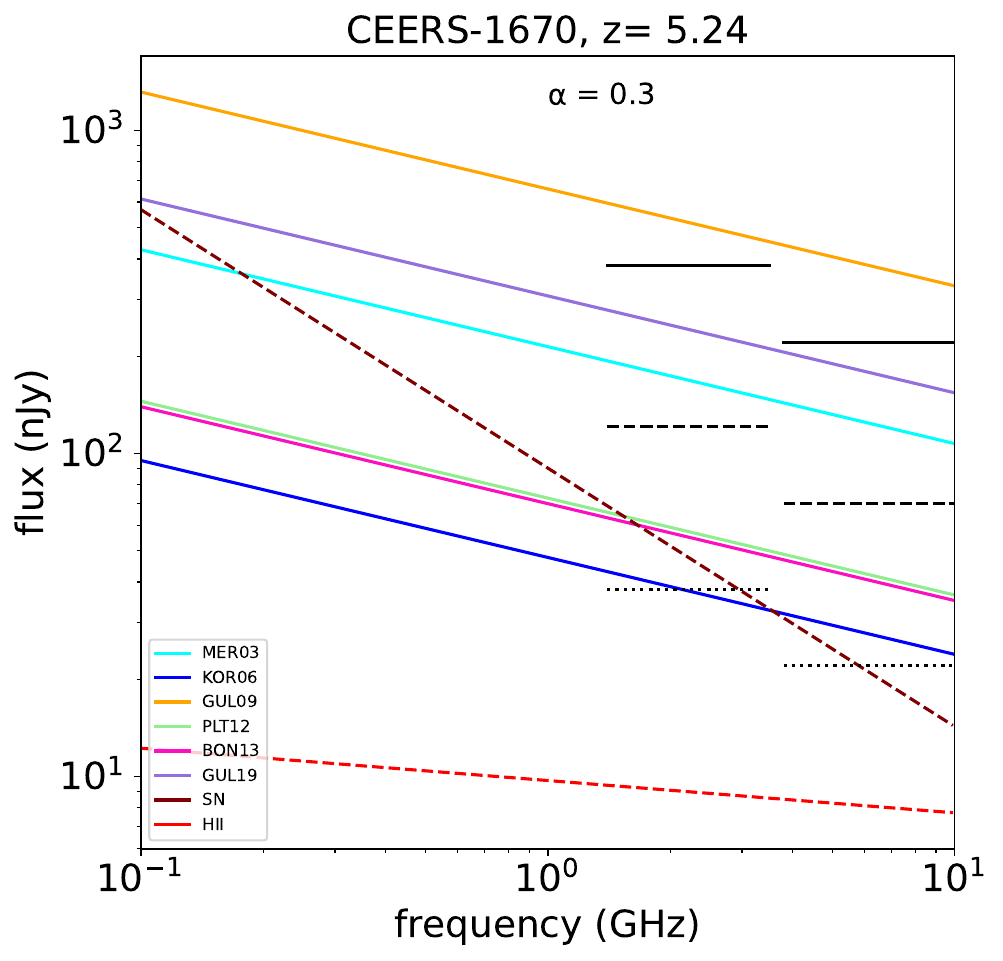}
\includegraphics[scale=0.45]{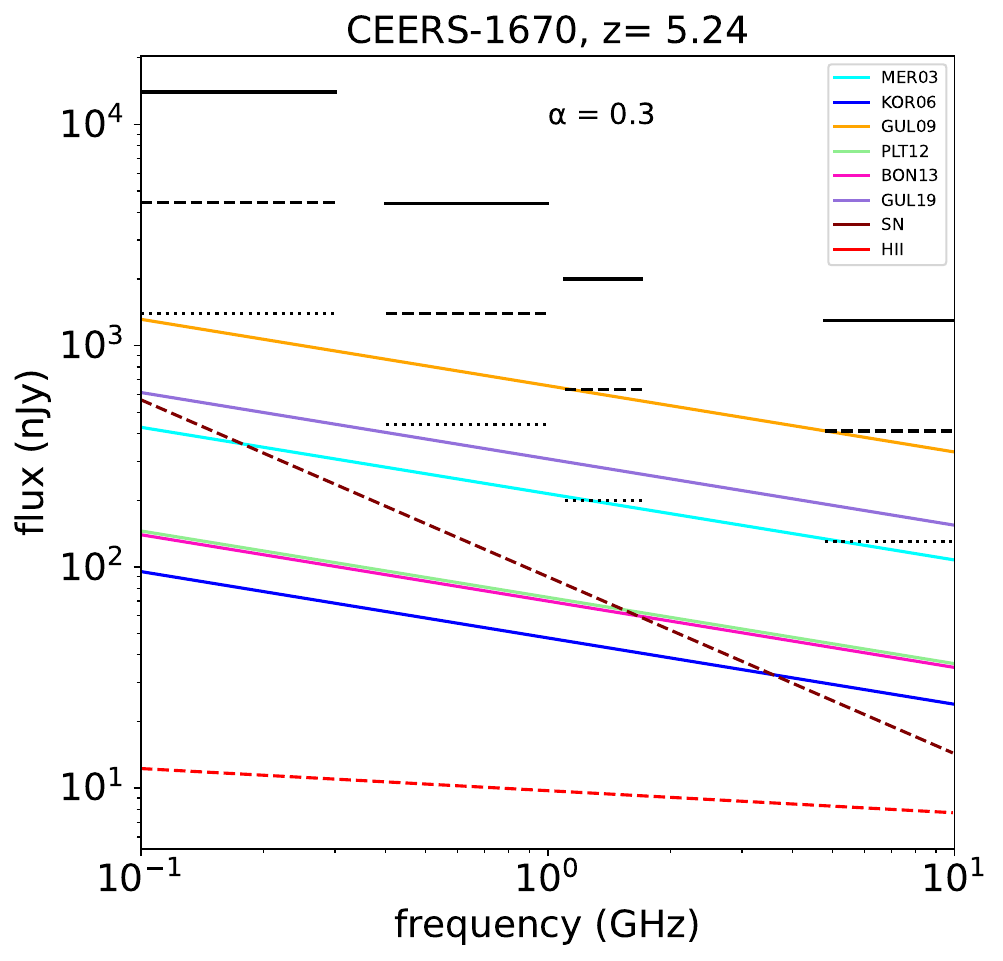} 
\end{center}
\vspace{-0.1cm}
\caption{Radio flux densities for AGNs observed in CEERS, JADES and UNCOVER for $\alpha$ = 0.3 with detection limits for ngVLA (left column) and SKA (right column). The dotted, dashed and dot-dashed red and brown lines are H II region and SN flux densities for SFRs of 1, 3 and 10 \Ms\ yr$^{-1}$, respectively. The black solid, dashed and dotted horizontal bars show ngVLA and SKA detection limits for integration times of 1, 10 and 100 hr, respectively.}
\label{fig:f2}
\end{figure*}

\begin{figure*} 
\begin{center}
\includegraphics[scale=0.45]{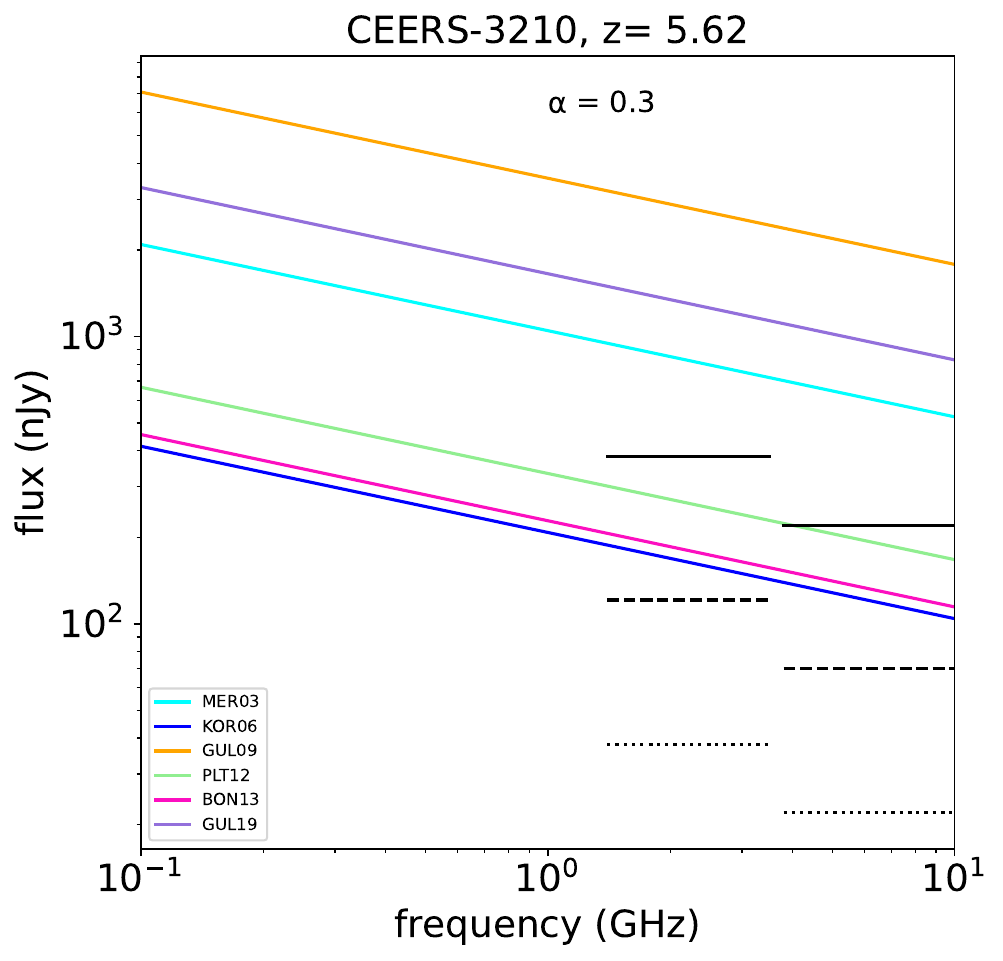}
\includegraphics[scale=0.45]{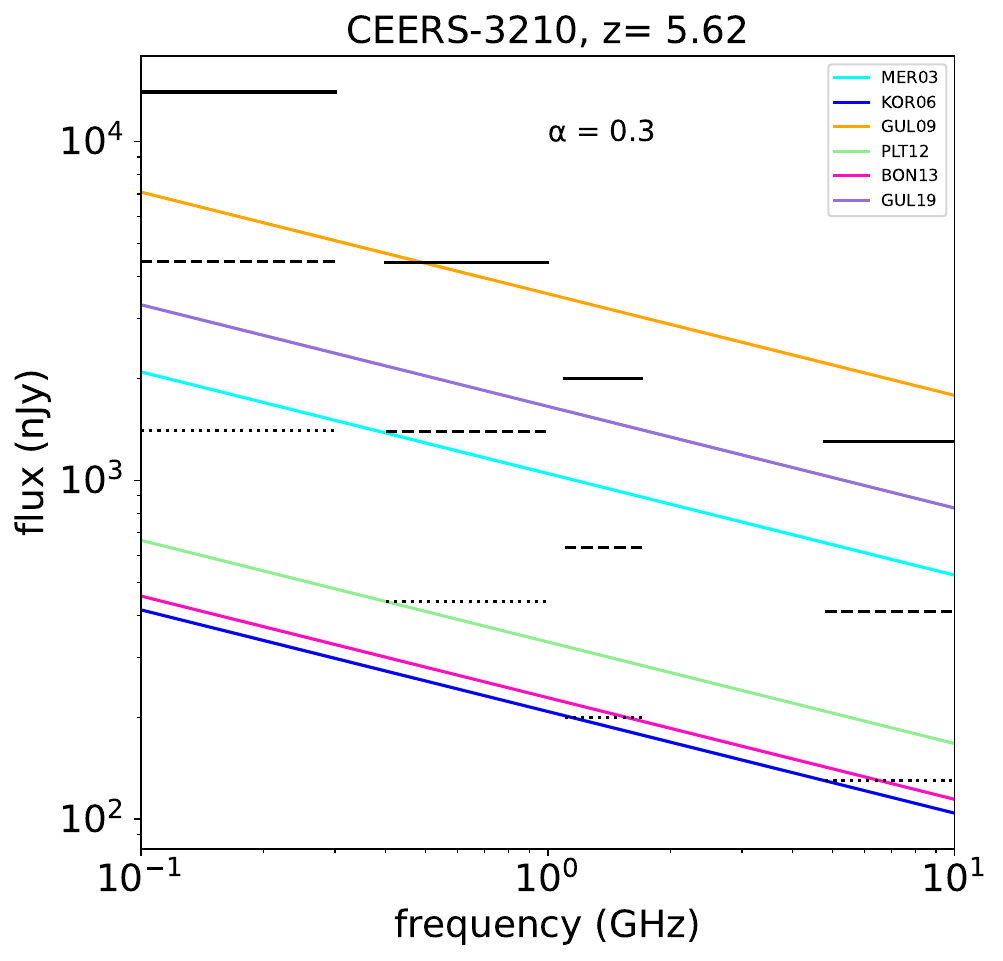} 
\includegraphics[scale=0.45]{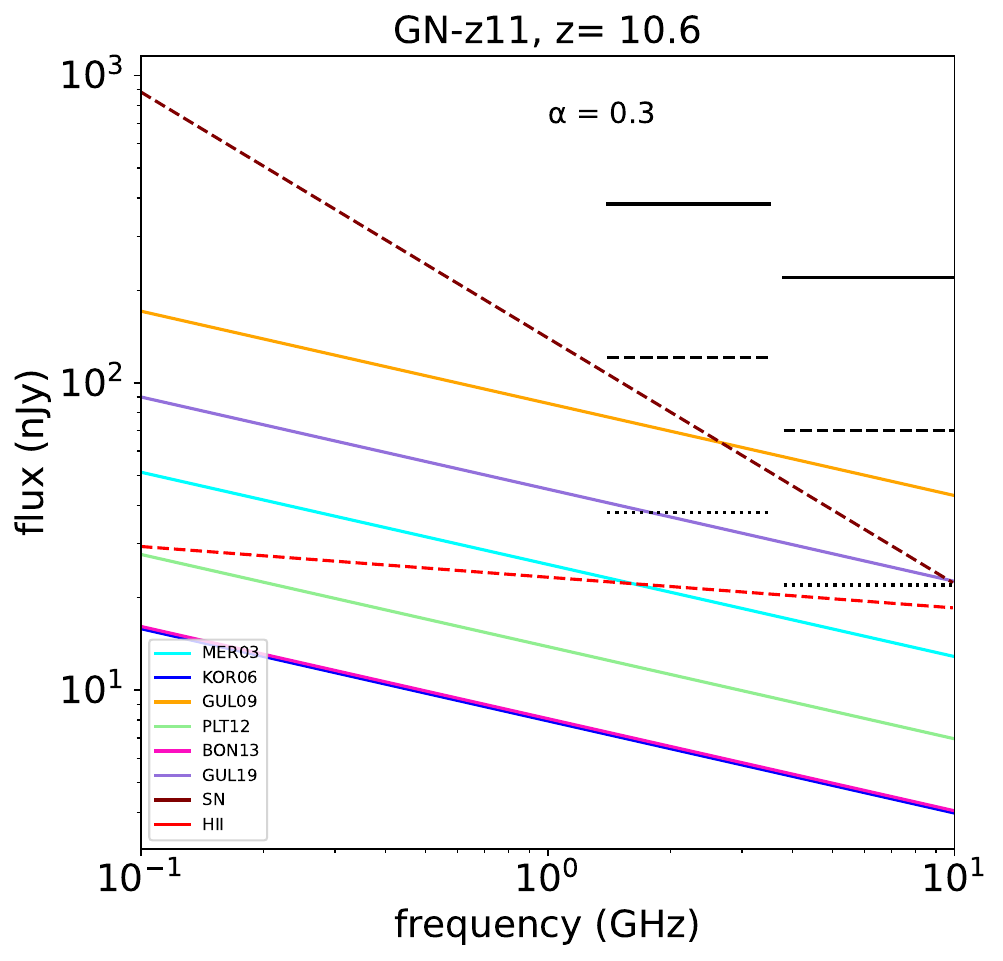}
\includegraphics[scale=0.45]{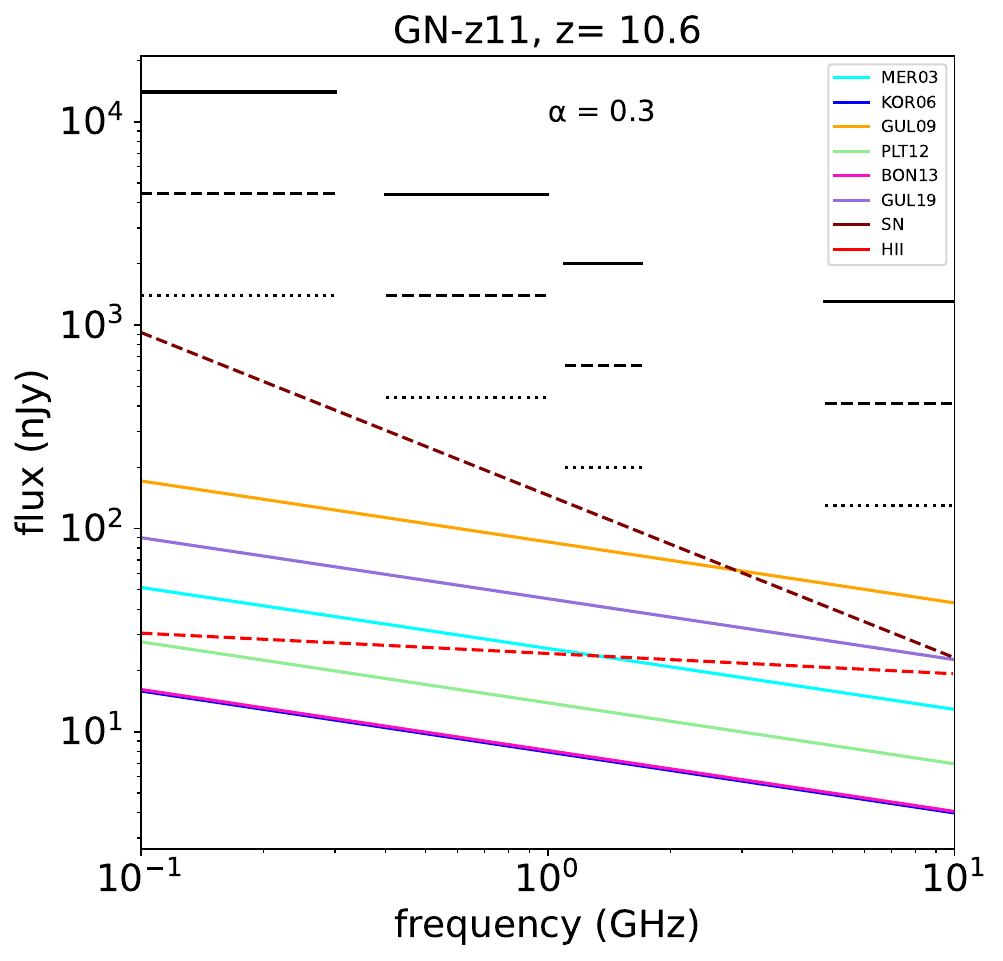}
\includegraphics[scale=0.45]{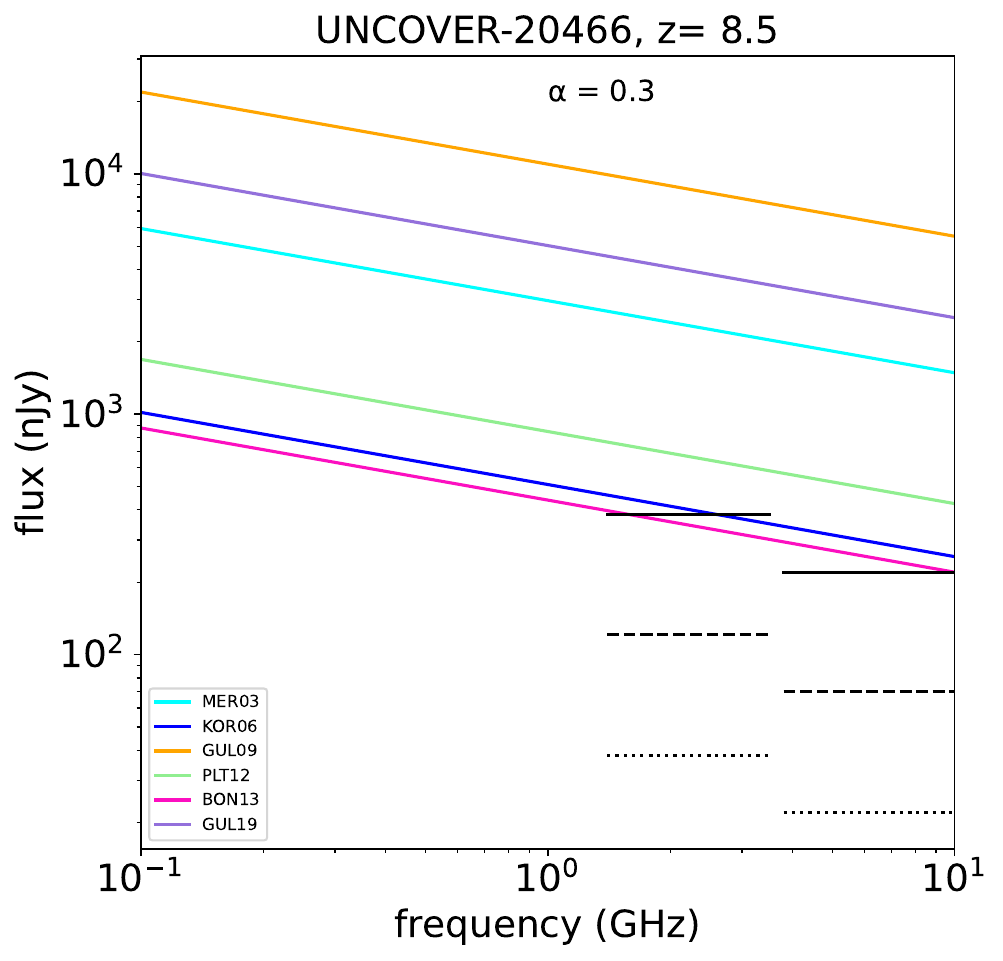}
\includegraphics[scale=0.45]{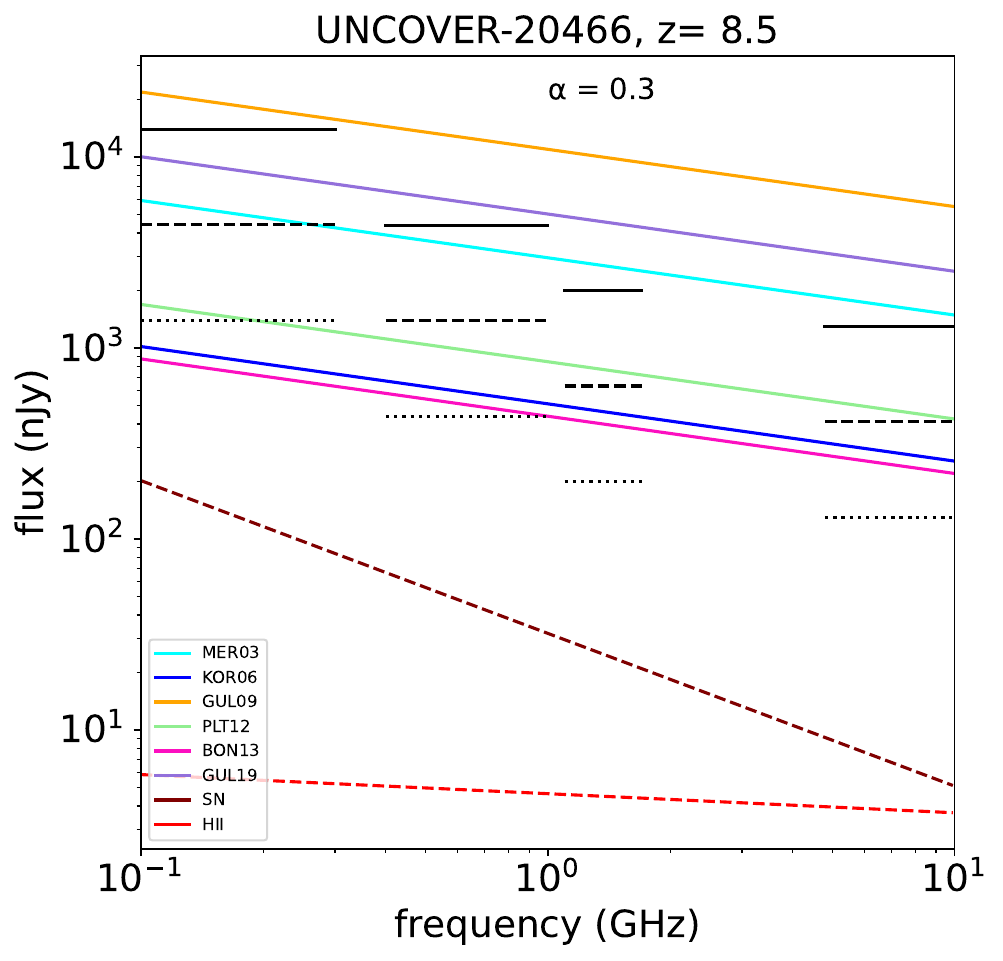}
\end{center}
\vspace{-0.1cm}
\caption{Radio flux densities for AGNs observed in CEERS, JADES and UNCOVER for $\alpha$ = 0.3 with detection limits for ngVLA (left column) and SKA (right column). The dotted, dashed and dot-dashed red and brown lines are H II region and SN flux densities for SFRs of 1, 3 and 10 \Ms\ yr$^{-1}$, respectively. The black solid, dashed and dotted horizontal bars show ngVLA and SKA limits for integration times of 1, 10 and 100 hr, respectively.}
\label{fig:f3}
\end{figure*}

\begin{figure*} 
\begin{center}
\includegraphics[scale=0.45]{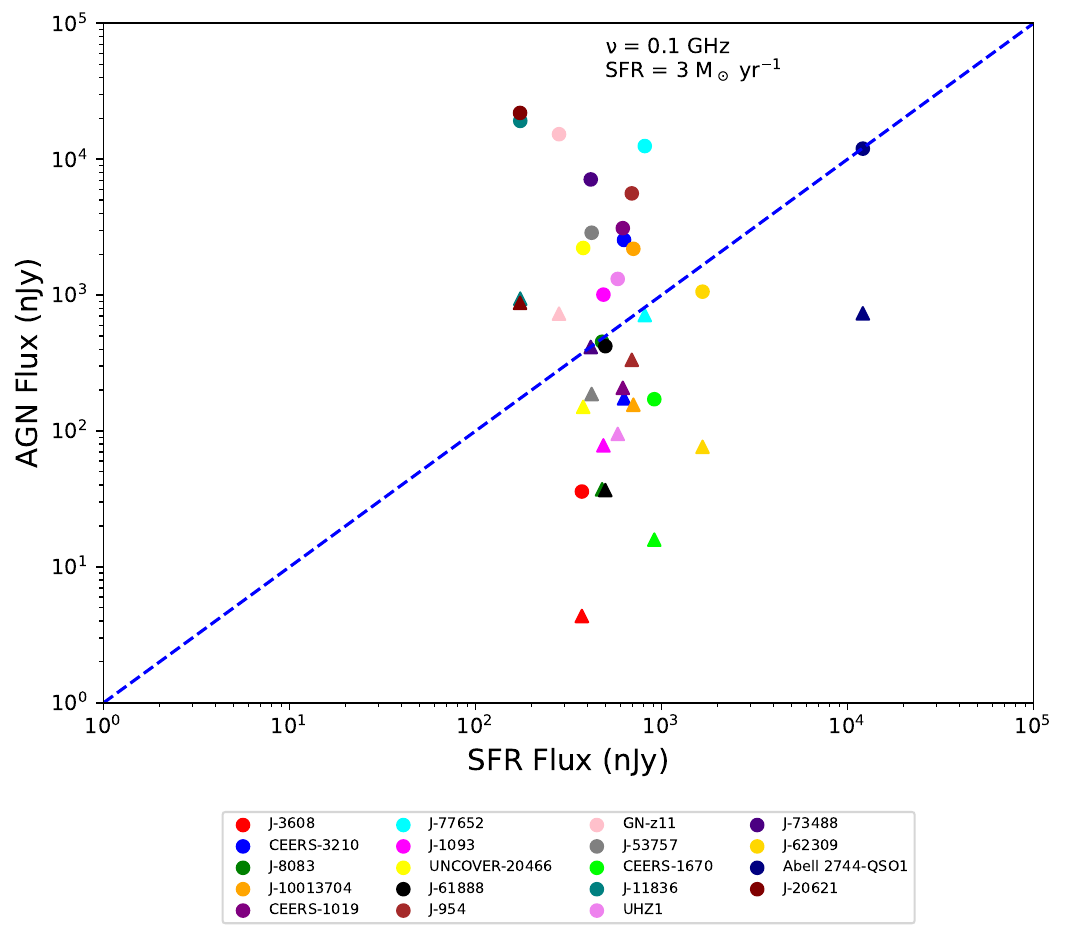}
\includegraphics[scale=0.45]{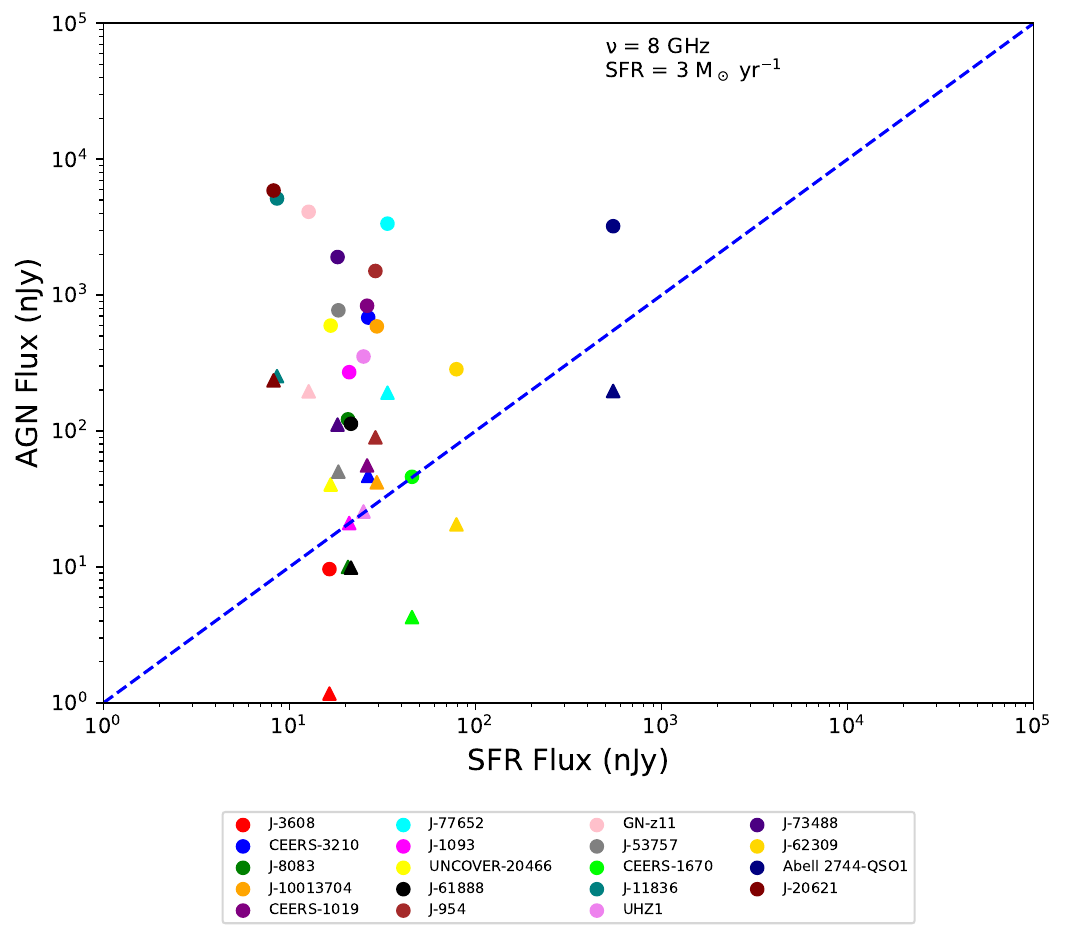} 
\includegraphics[scale=0.45]{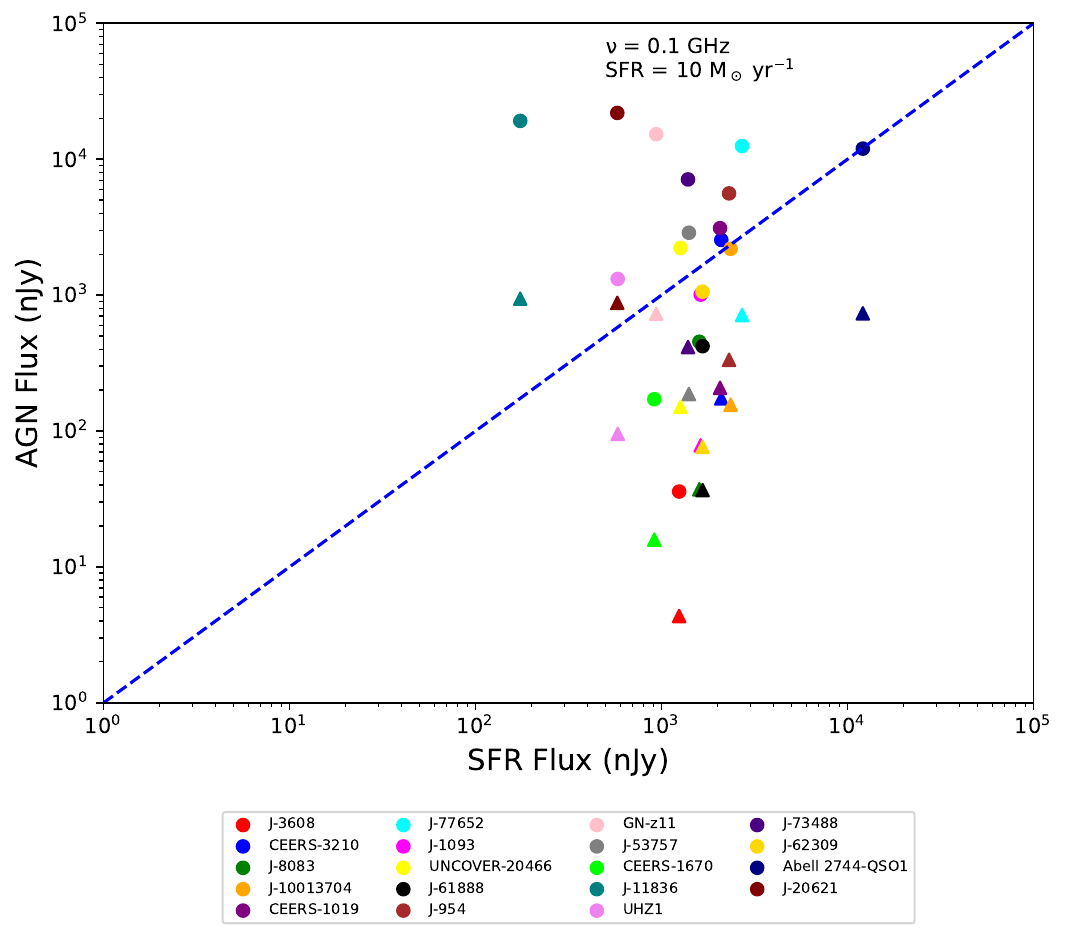}
\includegraphics[scale=0.45]{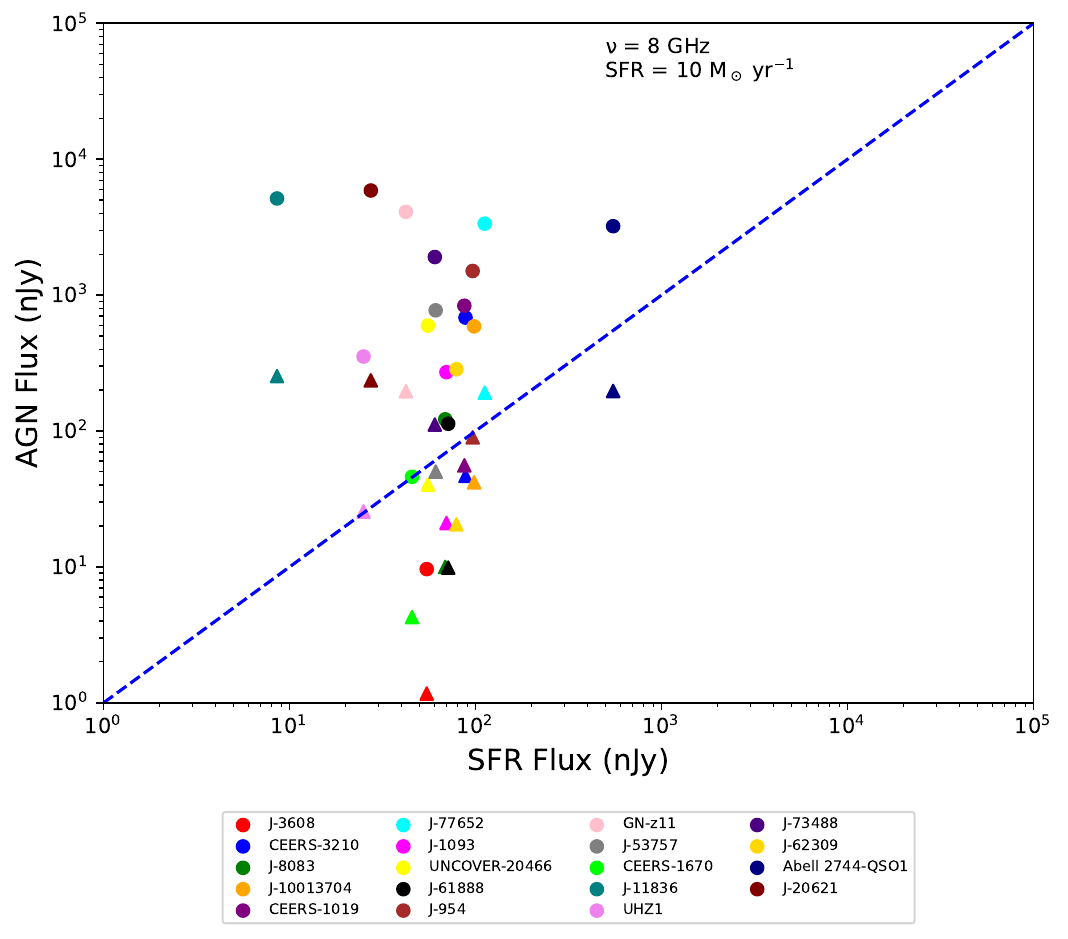}
\end{center}
\vspace{-0.1cm}
\caption{The AGN radio flux vs the expected flux from the host galaxy (which includes both contributions from HII regions and SN remnants) for $\rm \alpha=0.3$. The dot symbols show the highest flux from FPs and triangles represent the lowest flux expected from FPs at given frequencies of 0.1 GHz \& 8 GHz. The sources for which SFRs are unknown, we assume fixed values of 3 \& 10 $\rm \Ms/yr$ while for other sources SFRs are listed in table ~\ref{tbl:tbl1}.}
\label{fig:s3}
\end{figure*}

We show radio flux densities for three of the AGNs in Figures~\ref{fig:f1} - \ref{fig:f3} for $\alpha =$ 0.3. The other $\alpha =$ 0.3 fluxes are shown in the Appendix in Figures~\ref{fig:f4} -  \ref{fig:f7} and the $\alpha =$ 0.7 fluxes are shown in Figures~\ref{fig:f7} - \ref{fig:f12}.  The BHs in the JADES survey have masses of $ 10^{5.65} - 10^{7.9}~\Ms$ and bolometric luminosities of $\rm 10^{43.8} - 10^{45.6}~L_{\odot}$. Typical BH flux densities vary from a few hundred to a few thousand nJy at 0.1 GHz and about 20 to a few hundred nJy at 10 GHz. J-954 and J-73488 have the highest flux densities, up to $\sim 10^4$ nJy because of their large BH masses, high bolometric luminosities, and greater Eddington ratios. J-10013704 has the lowest radio flux, about 40 nJy at 0.1 GHz, because it has the smallest BH mass. This is the only source that is accreting at the Eddington limit, as the others grow at rates of 10 - 40\% Eddington. 

Our estimates show that BH radio flux densities for most of the sources can be detected with ngVLA with integration times of 10 - 100 hr except for J-10013704 and J-3608.  Radio emission from J-73488 and J-954 can be detected with integration times of only 1 hr at 1 - 10 GHz.  On the other hand, SKA would require integration times of 100 hr or more at frequencies $> 1$ GHz to detect these sources except for J-10013704, J-3608 and J-62309, which require even longer times.  J-954 and J-73488, the brightest of these AGNs, can be observed with SKA with 10 - 100 hr integration times. 

Radio emission from H II regions is below that of BHs in almost all cases with SFRs of up to 10 \Ms\ yr$^{-1}$ except for J-10013704, where it exceeds the most pessimistic of the FPs at 1 - 10 GHz.  They in general are at least factor of a few lower than the least of the FP flux densities.  While radio emission from SNe is generally below that of BHs for SFRs $<$ 1 \Ms\ yr$^{-1}$, they become comparable to some FPs for SFRs above 3 \Ms\ yr$^{-1}$.  For J-3608, J-62309, J-77652 and J-10013704, the flux densities from SNe rival those of the BHs even at SFRs of 1 \Ms\ yr$^{-1}$.  Synchrotron flux from SN remnants falls off much faster in frequency than BH flux, so SFRs can be inferred for host galaxies of AGNs whose radio emission decreases more quickly with frequency than predicted for a BH alone.  In this respect, ngVLA is more suitable in discriminating between SN and BH emission because of its greater sensitivity than SKA at high frequencies.

The BHs in the CEERS, UNCOVER and Abell 2744 lensing surveys have masses of $10^{6.95} - 10^{8.16}~\Ms$ and bolometric luminosities of $10^{44.4} - 10^{45.8}~L_{\odot}$.  The BH flux density estimates for Abell 2744-QSO1 range from 700 - 12,000 nJy at 0.1 GHz and 200 - 3000 nJy at 10 GHz. Its H II region flux densities are an order of magnitude lower than those of the BH.  Radio emission from SNe is also below that of the BHs above 1 GHz but becomes comparable to some FPs at lower frequencies.  NgVLA and SKA can detect this AGN with integration times of 1 hr and 10 - 100 hr, respectively.  The radio flux density from the BH in CEERS-1019 is 80 - 1000 nJy at 0.1 GHz and 2 - 300 nJy at 10 GHz.  Its host galaxy has an SFR of 30 \Ms\ yr$^{-1}$ for which the expected flux from H II regions is about 40 nJy, still less than that of the BH.  However, radio emission from SNe below 1 GHz is on par with that of the BH and even exceeds it at 0.1 GHz.  NgVLA can detect CEERS-1019 with integration times of 100 hr but SKA would require longer times.  The BH flux density from CEERS-1670 is about factor of 1.4 higher than CEERS-1019 due to differences in redshifts and bolometric luminosities. SFRs in the host galaxy of CEERS-1670 are a factor of 9 lower than in CEERS-1019 so flux densities from H II regions are factor of a few lower than those of the BH. The SNe radio emission is below that of the BH above 1 GHz but becomes comparable to radio flux from its BH below 1 GHz. NgVLA can detect CEERS-1670 with integration times of 100 hr but SKA would require longer times.

The estimated BH mass in CEERS-3210 is 0.9 - 4.7 $\times 10^7$ \Ms\ depending on the degree of dust obscuration. At the upper limit in mass shown in Figures~\ref{fig:f6} and \ref{fig:f12}, the expected flux density from the BH is factor of 5 higher than for CEERS-1670 and CEERS-1019 at 0.1 - 10 GHz.  SFRs are not available for CEERS-3210 but we expect fluxes from the BH to be higher than those from the H II regions and SNe at the SFRs in the other two CEERS sources.  CEERS-3210 can detected by ngVLA with 10 hr integration times and by SKA with about 100 hr pointings.  GN-z11, the most distant BH candidate observed to date at $z =10.6$, has a $1.5 \times 10^6$ \Ms\ BH that can be marginally detected by ngVLA with 100 hr integration times but only with longer times by SKA. Radio emission from its H II regions becomes equal to the flux from some FPs due to the 25 \Ms\ yr$^{-1}$ SFR in the host galaxy.  Below 3 GHz, flux from SNe dominates emission from the BH.  The flux density for the 1.45 $\times 10^8$ \Ms\ BH in UNCOVER-20466 is 1000 - 20,000 nJy at 0.1 GHz and 200 - 5500 nJy at 10 GHz.  We again expect that emission from the BH will be dominant for moderate SFRs ($<$ 30 \Ms\ yr$^{-1}$).  Radio emission from the BH in this source can be detected with ngVLA with just a 1 hr integration time but SKA will need 10 - 100 hr.

UHZ1 was recently spectroscopically confirmed to be at $z =$ 10.1 by \cite{Gould23}, who found through stellar synthesis modeling that the SFR is about 1.1 \Ms\ yr$^{-1}$, a factor of four lower than reported by \citet{Bod23}.  \citet{W23} calculated flux densities for UHZ1 but we plot them in Figure~\ref{fig:f13} for the new SFR and redshift.  They confirm that emission from H II regions and SNe is well below that of the BH.  NgVLA can detect UHZ1 in a 1 hr pointing and SKA could detect it with a 10 - 100 hr exposure.  We find that almost all the sources can be detected with ngVLA with 10 - 100 hr integration times except GN-z11, which may require longer times.

To further elucidate the contribution to AGN flux by SF, we show the flux of the BH versus that of the host galaxy, which is the sum of the H II region and SNR fluxes, in Figure~\ref{fig:s3}. We select the FPs with the highest and lowest fluxes at 0.1 GHz and 8 GHz as limiting cases. In general, the BH flux from the highest FP dominates that of the host galaxy at 10 GHz for SFRs of 3 \Ms\ yr$^{-1}$ and 10 \Ms\ yr$^{-1}$ with the exception of J-3608, which has the lowest Eddington ratio. For CEERS-1670, the BH and its host galaxy have comparable fluxes. For SFRs of 3 \Ms\ yr$^{-1}$, even flux from the lowest FP at 10 GHz is higher than SF flux in the host galaxies for 12 of the 19 sources but at 10 \Ms\ yr$^{-1}$ flux from host galaxy dominates in 14 of the 19 sources. At 0.1 GHz, for 3 \Ms\ yr$^{-1}$ the BH flux from the highest FP is still higher for most of the sources except J-3608 and CEERS-170, and even at 10 \Ms\ yr$^{-1}$ BH flux dominates that of the host galaxies for 10 sources. For the lowest FP the host galaxy flux exceeds that of the BH for most of the sources. Overall, our results suggest that BH flux dominates that of their host galaxies at 10 GHz.

The AGN flux densities for $\alpha$ = 0.7 are about a factor of 2 - 3 higher at 0.1 GHz and 2 - 3 times lower at 10 GHz in comparison to $\alpha$ = 0.3.  BH fluxes for $\alpha$ = 0.7 in general are lower in the sensitivity range of ngVLA (1-10 GHz) so longer integration times of 10 -100 hr would be required to observe them depending on the BH mass and redshift except for GN-z11 and J-10013704, which still require more than 100 hr integration times. SKA is better at detecting some $\alpha$ = 0.7 sources because of its sensitivity at lower frequencies (0.1 GHz) for integration times of at least 100 hrs and longer.  However, overall, ngVLA is more suitable for detection of high redshift AGNs because of its shorter integration times.

\section{Discussion and Conclusion}

Radio observations by ngVLA and SKA will be able to probe the properties of low-luminosity high-redshift AGNs recently found by the {\em JWST} JADES, CEERS and UNCOVER surveys in the NIR and similar objects in future surveys.  Their greatest utility will be in targeted searches because the flux densities of these less massive BHs may be too low to be found in general surveys with large areas but lower sensitivities.  Typical radio flux densities for these AGNs are a few hundred to a few thousand nJy at 0.1 GHz for $\alpha = 0.3$ but reach a few 10$^4$ nJy in some cases, and a few tens to a few hundred nJy at 10 GHz.  For $\alpha$ = 0.7, the AGN fluxes are about a factor of 2 - 3 higher at 0.1 GHz and a factor of 2 - 3 lower at 10 GHz. Radio emission from H II regions for SFRs up to 10 \Ms\ yr$^{-1}$ are generally lower than BH emission except for a few sources. However, radio flux densities from SN remnants become comparable to those of the BH for some FPs above 3 \Ms\ yr$^{-1}$ and could be used to constrain SFRs in the host galaxies because they can be distinguished from BH fluxes below $\sim$ 1 GHz. In general, we find that most of these sources can be detected by ngVLA with 10 - 100 hr integration times (and a few with only 1 hr) but SKA would require at least 10 - 100 hr, even with its greater sensitivity at lower frequencies.

Our flux estimates are subject to uncertainties in BH masses, Eddington ratios and SFRs for the host galaxies. Although current observations favor an AGN in GN-z11 because of the detection of higher ionization lines and higher gas densities like those in broad line region (BLR) AGNs, other scenarios such as Pop III star clusters can not be fully ruled out \citep[see the detailed discussion in][]{Maio23a}.

\begin{acknowledgments}
%\acknowledgments

MAL thanks the UAEU for funding via UPAR grant No. 12S111. We thank the referee, whose comments improved the quality of this paper. 

\end{acknowledgments}

\bibliography{ref.bib}

\begin{thebibliography}{}
\expandafter\ifx\csname natexlab\endcsname\relax\def\natexlab#1{#1}\fi
\providecommand{\url}[1]{\href{#1}{#1}}
\providecommand{\dodoi}[1]{doi:~\href{http://doi.org/#1}{\nolinkurl{#1}}}
\providecommand{\doeprint}[1]{\href{http://ascl.net/#1}{\nolinkurl{http://ascl.net/#1}}}
\providecommand{\doarXiv}[1]{\href{https://arxiv.org/abs/#1}{\nolinkurl{https://arxiv.org/abs/#1}}}

\bibitem[{{Ba{\~n}ados} {et~al.}(2021){Ba{\~n}ados}, {Mazzucchelli}, {Momjian},
  {Eilers}, {Wang}, {Schindler}, {Connor}, {Andika}, {Barth}, {Carilli},
  {Davies}, {Decarli}, {Fan}, {Farina}, {Hennawi}, {Pensabene}, {Stern},
  {Venemans}, {Wenzl}, \& {Yang}}]{ban21}
{Ba{\~n}ados}, E., {Mazzucchelli}, C., {Momjian}, E., {et~al.} 2021, \apj, 909,
  80, \dodoi{10.3847/1538-4357/abe239}

\bibitem[{{Belladitta} {et~al.}(2020){Belladitta}, {Moretti}, {Caccianiga},
  {Spingola}, {Severgnini}, {Della Ceca}, {Ghisellini}, {Dallacasa},
  {Sbarrato}, {Cicone}, {Cassar{\`a}}, \& {Pedani}}]{bd20}
{Belladitta}, S., {Moretti}, A., {Caccianiga}, A., {et~al.} 2020, \aap, 635,
  L7, \dodoi{10.1051/0004-6361/201937395}

\bibitem[{{Bogd{\'a}n} {et~al.}(2023){Bogd{\'a}n}, {Goulding}, {Natarajan},
  {Kov{\'a}cs}, {Tremblay}, {Chadayammuri}, {Volonteri}, {Kraft}, {Forman},
  {Jones}, {Churazov}, \& {Zhuravleva}}]{Bod23}
{Bogd{\'a}n}, {\'A}., {Goulding}, A.~D., {Natarajan}, P., {et~al.} 2023, Nature
  Astronomy, \dodoi{10.1038/s41550-023-02111-9}

\bibitem[{{Bonchi} {et~al.}(2013){Bonchi}, {La Franca}, {Melini}, {Bongiorno},
  \& {Fiore}}]{bonchi13}
{Bonchi}, A., {La Franca}, F., {Melini}, G., {Bongiorno}, A., \& {Fiore}, F.
  2013, \mnras, 429, 1970, \dodoi{10.1093/mnras/sts456}

\bibitem[{{Condon}(1992)}]{con92}
{Condon}, J.~J. 1992, \araa, 30, 575,
  \dodoi{10.1146/annurev.aa.30.090192.003043}

\bibitem[{{Condon} {et~al.}(2002){Condon}, {Cotton}, \& {Broderick}}]{ccb02}
{Condon}, J.~J., {Cotton}, W.~D., \& {Broderick}, J.~J. 2002, \aj, 124, 675,
  \dodoi{10.1086/341650}

\bibitem[{{Fan} {et~al.}(2023){Fan}, {Ba{\~n}ados}, \& {Simcoe}}]{fbs23}
{Fan}, X., {Ba{\~n}ados}, E., \& {Simcoe}, R.~A. 2023, \araa, 61, 373,
  \dodoi{10.1146/annurev-astro-052920-102455}

\bibitem[{{Furtak} {et~al.}(2023){Furtak}, {Labb{\'e}}, {Zitrin}, {Greene},
  {Dayal}, {Chemerynska}, {Kokorev}, {Miller}, {Goulding}, {Bezanson},
  {Brammer}, {Cutler}, {Leja}, {Pan}, {Price}, {Wang}, {Weaver}, {Whitaker},
  {Atek}, {Bogd{\'a}n}, {Charlot}, {Curtis-Lake}, {van Dokkum}, {Endsley},
  {Fudamoto}, {Fujimoto}, {de Graaff}, {Glazebrook}, {Juneau}, {Marchesini},
  {Maseda}, {Nelson}, {Oesch}, {Plat}, {Setton}, {Stark}, \&
  {Williams}}]{Furt23}
{Furtak}, L.~J., {Labb{\'e}}, I., {Zitrin}, A., {et~al.} 2023, arXiv e-prints,
  arXiv:2308.05735, \dodoi{10.48550/arXiv.2308.05735}

\bibitem[{{Gloudemans} {et~al.}(2021){Gloudemans}, {Duncan}, {R{\"o}ttgering},
  {Shimwell}, {Venemans}, {Best}, {Br{\"u}ggen}, {Calistro Rivera}, {Drabent},
  {Hardcastle}, {Miley}, {Schwarz}, {Saxena}, {Smith}, \& {Williams}}]{glou21}
{Gloudemans}, A.~J., {Duncan}, K.~J., {R{\"o}ttgering}, H.~J.~A., {et~al.}
  2021, \aap, 656, A137, \dodoi{10.1051/0004-6361/202141722}

\bibitem[{{Goulding} {et~al.}(2023){Goulding}, {Greene}, {Setton}, {Labbe},
  {Bezanson}, {Miller}, {Atek}, {Bogd{\'a}n}, {Brammer}, {Chemerynska},
  {Cutler}, {Dayal}, {Fudamoto}, {Fujimoto}, {Furtak}, {Kokorev}, {Khullar},
  {Leja}, {Marchesini}, {Natarajan}, {Nelson}, {Oesch}, {Pan}, {Papovich},
  {Price}, {van Dokkum}, {Wang}, {Weaver}, {Whitaker}, \& {Zitrin}}]{Gould23}
{Goulding}, A.~D., {Greene}, J.~E., {Setton}, D.~J., {et~al.} 2023, \apjl, 955,
  L24, \dodoi{10.3847/2041-8213/acf7c5}

\bibitem[{{G{\"u}ltekin} {et~al.}(2014){G{\"u}ltekin}, {Cackett}, {King},
  {Miller}, \& {Pinkney}}]{gul14}
{G{\"u}ltekin}, K., {Cackett}, E.~M., {King}, A.~L., {Miller}, J.~M., \&
  {Pinkney}, J. 2014, \apjl, 788, L22, \dodoi{10.1088/2041-8205/788/2/L22}

\bibitem[{{G{\"u}ltekin} {et~al.}(2009){G{\"u}ltekin}, {Cackett}, {Miller}, {Di
  Matteo}, {Markoff}, \& {Richstone}}]{gul09}
{G{\"u}ltekin}, K., {Cackett}, E.~M., {Miller}, J.~M., {et~al.} 2009, \apj,
  706, 404, \dodoi{10.1088/0004-637X/706/1/404}

\bibitem[{{G{\"u}ltekin} {et~al.}(2019){G{\"u}ltekin}, {King}, {Cackett},
  {Nyland}, {Miller}, {Di Matteo}, {Markoff}, \& {Rupen}}]{gul19}
{G{\"u}ltekin}, K., {King}, A.~L., {Cackett}, E.~M., {et~al.} 2019, \apj, 871,
  80, \dodoi{10.3847/1538-4357/aaf6b9}

\bibitem[{{Herrington} {et~al.}(2023){Herrington}, {Whalen}, \&
  {Woods}}]{herr23a}
{Herrington}, N.~P., {Whalen}, D.~J., \& {Woods}, T.~E. 2023, \mnras, 521, 463,
  \dodoi{10.1093/mnras/stad572}

\bibitem[{{Hosokawa} {et~al.}(2013){Hosokawa}, {Yorke}, {Inayoshi}, {Omukai},
  \& {Yoshida}}]{hos13}
{Hosokawa}, T., {Yorke}, H.~W., {Inayoshi}, K., {Omukai}, K., \& {Yoshida}, N.
  2013, \apj, 778, 178, \dodoi{10.1088/0004-637X/778/2/178}

\bibitem[{{Inayoshi} {et~al.}(2019){Inayoshi}, {Visbal}, \& {Haiman}}]{ivh20}
{Inayoshi}, K., {Visbal}, E., \& {Haiman}, Z. 2019, arXiv:1911.05791,
  arXiv:1911.05791.
\newblock \doarXiv{1911.05791}

\bibitem[{{Joggerst} {et~al.}(2010){Joggerst}, {Almgren}, {Bell}, {Heger},
  {Whalen}, \& {Woosley}}]{jet09b}
{Joggerst}, C.~C., {Almgren}, A., {Bell}, J., {et~al.} 2010, \apj, 709, 11,
  \dodoi{10.1088/0004-637X/709/1/11}

\bibitem[{{Johnson} {et~al.}(2013){Johnson}, {Whalen}, {Li}, \& {Holz}}]{jet13}
{Johnson}, J.~L., {Whalen}, D.~J., {Li}, H., \& {Holz}, D.~E. 2013, \apj, 771,
  116, \dodoi{10.1088/0004-637X/771/2/116}

\bibitem[{{Kennicutt}(1998)}]{ken98}
{Kennicutt}, Robert~C., J. 1998, \araa, 36, 189,
  \dodoi{10.1146/annurev.astro.36.1.189}

\bibitem[{{Kocevski} {et~al.}(2023){Kocevski}, {Onoue}, {Inayoshi}, {Trump},
  {Arrabal Haro}, {Grazian}, {Dickinson}, {Finkelstein}, {Kartaltepe},
  {Hirschmann}, {Aird}, {Holwerda}, {Fujimoto}, {Juneau}, {Amor{\'\i}n},
  {Backhaus}, {Bagley}, {Barro}, {Bell}, {Bisigello}, {Calabr{\`o}}, {Cleri},
  {Cooper}, {Ding}, {Grogin}, {Ho}, {Hutchison}, {Inoue}, {Jiang}, {Jones},
  {Koekemoer}, {Li}, {Li}, {McGrath}, {Molina}, {Papovich},
  {P{\'e}rez-Gonz{\'a}lez}, {Pirzkal}, {Wilkins}, {Yang}, \& {Yung}}]{Koc23}
{Kocevski}, D.~D., {Onoue}, M., {Inayoshi}, K., {et~al.} 2023, \apjl, 954, L4,
  \dodoi{10.3847/2041-8213/ace5a0}

\bibitem[{{Kokorev} {et~al.}(2023){Kokorev}, {Fujimoto}, {Labbe}, {Greene},
  {Bezanson}, {Dayal}, {Nelson}, {Atek}, {Brammer}, {Caputi}, {Chemerynska},
  {Cutler}, {Feldmann}, {Fudamoto}, {Furtak}, {Goulding}, {de Graaff}, {Leja},
  {Marchesini}, {Miller}, {Nanayakkara}, {Oesch}, {Pan}, {Price}, {Setton},
  {Smit}, {Stefanon}, {Wang}, {Weaver}, {Whitaker}, {Williams}, \&
  {Zitrin}}]{Kok23}
{Kokorev}, V., {Fujimoto}, S., {Labbe}, I., {et~al.} 2023, \apjl, 957, L7,
  \dodoi{10.3847/2041-8213/ad037a}

\bibitem[{{K{\"o}rding} {et~al.}(2006){K{\"o}rding}, {Falcke}, \&
  {Corbel}}]{kord06}
{K{\"o}rding}, E., {Falcke}, H., \& {Corbel}, S. 2006, \aap, 456, 439,
  \dodoi{10.1051/0004-6361:20054144}

\bibitem[{{Lambrides} {et~al.}(2023){Lambrides}, {Chiaberge}, {Long}, {Liu},
  {Akins}, {Ptak}, {Taufik Andika}, {Capetti}, {Casey}, {Champagne},
  {Chworowsky}, {Cooper}, {Ding}, {Faisst}, {Franco}, {Gillman}, {Gozaliasl},
  {Hall}, {Harish}, {Hayward}, {Hirschmann}, {Hutchison}, {Jahnke}, {Jin},
  {Kartaltepe}, {Koekemoer}, {Kokorev}, {Manning}, {Martin}, {McKinney},
  {Norman}, {Onoue}, {Robertson}, {Shuntov}, {Silverman}, {Stiavelli},
  {Trakhtenbrot}, {Vardoulaki}, {Zavala}, {Allen}, {Ilbert}, {McCracken},
  {Paquereau}, {Rhodes}, \& {Toft}}]{Lamb23}
{Lambrides}, E., {Chiaberge}, M., {Long}, A., {et~al.} 2023, arXiv e-prints,
  arXiv:2308.12823, \dodoi{10.48550/arXiv.2308.12823}

\bibitem[{{Larson} {et~al.}(2023){Larson}, {Finkelstein}, {Kocevski},
  {Hutchison}, {Trump}, {Arrabal Haro}, {Bromm}, {Cleri}, {Dickinson},
  {Fujimoto}, {Kartaltepe}, {Koekemoer}, {Papovich}, {Pirzkal}, {Tacchella},
  {Zavala}, {Bagley}, {Behroozi}, {Champagne}, {Cole}, {Jung}, {Morales},
  {Yang}, {Zhang}, {Zitrin}, {Amor{\'\i}n}, {Burgarella}, {Casey}, {Ch{\'a}vez
  Ortiz}, {Cox}, {Chworowsky}, {Fontana}, {Gawiser}, {Grazian}, {Grogin},
  {Harish}, {Hathi}, {Hirschmann}, {Holwerda}, {Juneau}, {Leung}, {Lucas},
  {McGrath}, {P{\'e}rez-Gonz{\'a}lez}, {Rigby}, {Seill{\'e}}, {Simons}, {de La
  Vega}, {Weiner}, {Wilkins}, {Yung}, \& {Ceers Team}}]{Lars23}
{Larson}, R.~L., {Finkelstein}, S.~L., {Kocevski}, D.~D., {et~al.} 2023, \apjl,
  953, L29, \dodoi{10.3847/2041-8213/ace619}

\bibitem[{{Latif} \& {Ferrara}(2016)}]{lf16}
{Latif}, M.~A., \& {Ferrara}, A. 2016, \pasa, 33, e051,
  \dodoi{10.1017/pasa.2016.41}

\bibitem[{{Latif} \& {Khochfar}(2020)}]{L20}
{Latif}, M.~A., \& {Khochfar}, S. 2020, \mnras, 497, 3761,
  \dodoi{10.1093/mnras/staa2218}

\bibitem[{{Latif} {et~al.}(2018){Latif}, {Volonteri}, \& {Wise}}]{L18}
{Latif}, M.~A., {Volonteri}, M., \& {Wise}, J.~H. 2018, \mnras, 476, 5016,
  \dodoi{10.1093/mnras/sty622}

\bibitem[{{Latif} {et~al.}(2022{\natexlab{a}}){Latif}, {Whalen}, \&
  {Khochfar}}]{latif22a}
{Latif}, M.~A., {Whalen}, D., \& {Khochfar}, S. 2022{\natexlab{a}}, \apj, 925,
  28, \dodoi{10.3847/1538-4357/ac3916}

\bibitem[{{Latif} {et~al.}(2022{\natexlab{b}}){Latif}, {Whalen}, {Khochfar},
  {Herrington}, \& {Woods}}]{latif22b}
{Latif}, M.~A., {Whalen}, D.~J., {Khochfar}, S., {Herrington}, N.~P., \&
  {Woods}, T.~E. 2022{\natexlab{b}}, \nat, 607, 48,
  \dodoi{10.1038/s41586-022-04813-y}

\bibitem[{{Latif} {et~al.}(2024){Latif}, {Whalen}, \& {Mezcua}}]{L24}
{Latif}, M.~A., {Whalen}, D.~J., \& {Mezcua}, M. 2024, \mnras, 527, L37,
  \dodoi{10.1093/mnrasl/slad102}

\bibitem[{{Maiolino} {et~al.}(2023{\natexlab{a}}){Maiolino}, {Scholtz},
  {Curtis-Lake}, {Carniani}, {Baker}, {de Graaff}, {Tacchella}, {{\"U}bler},
  {D'Eugenio}, {Witstok}, {Curti}, {Arribas}, {Bunker}, {Charlot},
  {Chevallard}, {Eisenstein}, {Egami}, {Ji}, {Jones}, {Lyu}, {Rawle},
  {Robertson}, {Rujopakarn}, {Perna}, {Sun}, {Venturi}, {Williams}, \&
  {Willott}}]{Maio23}
{Maiolino}, R., {Scholtz}, J., {Curtis-Lake}, E., {et~al.} 2023{\natexlab{a}},
  arXiv e-prints, arXiv:2308.01230, \dodoi{10.48550/arXiv.2308.01230}

\bibitem[{{Maiolino} {et~al.}(2023{\natexlab{b}}){Maiolino}, {Scholtz},
  {Witstok}, {Carniani}, {D'Eugenio}, {de Graaff}, {Uebler}, {Tacchella},
  {Curtis-Lake}, {Arribas}, {Bunker}, {Charlot}, {Chevallard}, {Curti},
  {Looser}, {Maseda}, {Rawle}, {Rodriguez Del Pino}, {Willott}, {Egami},
  {Eisenstein}, {Hainline}, {Robertson}, {Williams}, {Willmer}, {Baker},
  {Boyett}, {DeCoursey}, {Fabian}, {Helton}, {Ji}, {Jones}, {Kumari},
  {Laporte}, {Nelson}, {Perna}, {Sandles}, {Shivaei}, \& {Sun}}]{Maio23a}
{Maiolino}, R., {Scholtz}, J., {Witstok}, J., {et~al.} 2023{\natexlab{b}},
  arXiv e-prints, arXiv:2305.12492, \dodoi{10.48550/arXiv.2305.12492}

\bibitem[{{Marconi} {et~al.}(2004){Marconi}, {Risaliti}, {Gilli}, {Hunt},
  {Maiolino}, \& {Salvati}}]{marc04}
{Marconi}, A., {Risaliti}, G., {Gilli}, R., {et~al.} 2004, \mnras, 351, 169,
  \dodoi{10.1111/j.1365-2966.2004.07765.x}

\bibitem[{{Mayer} \& {Bonoli}(2019)}]{may19}
{Mayer}, L., \& {Bonoli}, S. 2019, Reports on Progress in Physics, 82, 016901,
  \dodoi{10.1088/1361-6633/aad6a5}

\bibitem[{{McGreer} {et~al.}(2006){McGreer}, {Becker}, {Helfand}, \&
  {White}}]{mcg06}
{McGreer}, I.~D., {Becker}, R.~H., {Helfand}, D.~J., \& {White}, R.~L. 2006,
  \apj, 652, 157, \dodoi{10.1086/507767}

\bibitem[{{Meiksin} \& {Whalen}(2013)}]{mw12}
{Meiksin}, A., \& {Whalen}, D.~J. 2013, \mnras, 430, 2854,
  \dodoi{10.1093/mnras/stt089}

\bibitem[{{Merloni} {et~al.}(2003){Merloni}, {Heinz}, \& {di Matteo}}]{merl03}
{Merloni}, A., {Heinz}, S., \& {di Matteo}, T. 2003, \mnras, 345, 1057,
  \dodoi{10.1046/j.1365-2966.2003.07017.x}

\bibitem[{{Nath} {et~al.}(2023){Nath}, {Vasiliev}, {Drozdov}, \&
  {Shchekinov}}]{Nath23}
{Nath}, B.~B., {Vasiliev}, E.~O., {Drozdov}, S.~A., \& {Shchekinov}, Y.~A.
  2023, \mnras, 521, 662, \dodoi{10.1093/mnras/stad505}

\bibitem[{{Patrick} {et~al.}(2023){Patrick}, {Whalen}, {Latif}, \&
  {Elford}}]{pat23}
{Patrick}, S.~J., {Whalen}, D.~J., {Latif}, M.~A., \& {Elford}, J.~S. 2023,
  \mnras, 522, 3795, \dodoi{10.1093/mnras/stad1179}

\bibitem[{{Planck Collaboration} {et~al.}(2016){Planck Collaboration}, {Ade},
  {Aghanim}, {Arnaud}, {Ashdown}, {Aumont}, {Baccigalupi}, {Banday},
  {Barreiro}, {Bartlett}, \& et~al.}]{planck2}
{Planck Collaboration}, {Ade}, P.~A.~R., {Aghanim}, N., {et~al.} 2016, \aap,
  594, A13, \dodoi{10.1051/0004-6361/201525830}

\bibitem[{{Plotkin} {et~al.}(2012){Plotkin}, {Markoff}, {Kelly}, {K{\"o}rding},
  \& {Anderson}}]{plot12}
{Plotkin}, R.~M., {Markoff}, S., {Kelly}, B.~C., {K{\"o}rding}, E., \&
  {Anderson}, S.~F. 2012, \mnras, 419, 267,
  \dodoi{10.1111/j.1365-2966.2011.19689.x}

\bibitem[{{Schneider} {et~al.}(2023){Schneider}, {Valiante}, {Trinca},
  {Graziani}, {Volonteri}, \& {Maiolino}}]{Sch23}
{Schneider}, R., {Valiante}, R., {Trinca}, A., {et~al.} 2023, \mnras, 526,
  3250, \dodoi{10.1093/mnras/stad2503}

\bibitem[{{Smidt} {et~al.}(2018){Smidt}, {Whalen}, {Johnson}, {Surace}, \&
  {Li}}]{smidt18}
{Smidt}, J., {Whalen}, D.~J., {Johnson}, J.~L., {Surace}, M., \& {Li}, H. 2018,
  \apj, 865, 126, \dodoi{10.3847/1538-4357/aad7b8}

\bibitem[{{Volonteri} {et~al.}(2023){Volonteri}, {Habouzit}, \&
  {Colpi}}]{Vol23}
{Volonteri}, M., {Habouzit}, M., \& {Colpi}, M. 2023, \mnras, 521, 241,
  \dodoi{10.1093/mnras/stad499}

\bibitem[{{Wang} {et~al.}(2021){Wang}, {Yang}, {Fan}, {Hennawi}, {Barth},
  {Banados}, {Bian}, {Boutsia}, {Connor}, {Davies}, {Decarli}, {Eilers},
  {Farina}, {Green}, {Jiang}, {Li}, {Mazzucchelli}, {Nanni}, {Schindler},
  {Venemans}, {Walter}, {Wu}, \& {Yue}}]{wang21}
{Wang}, F., {Yang}, J., {Fan}, X., {et~al.} 2021, \apjl, 907, L1,
  \dodoi{10.3847/2041-8213/abd8c6}

\bibitem[{{Whalen} {et~al.}(2004){Whalen}, {Abel}, \& {Norman}}]{wan04}
{Whalen}, D., {Abel}, T., \& {Norman}, M.~L. 2004, \apj, 610, 14,
  \dodoi{10.1086/421548}

\bibitem[{{Whalen} {et~al.}(2008){Whalen}, {van Veelen}, {O'Shea}, \&
  {Norman}}]{wet08a}
{Whalen}, D., {van Veelen}, B., {O'Shea}, B.~W., \& {Norman}, M.~L. 2008, \apj,
  682, 49, \dodoi{10.1086/589643}

\bibitem[{{Whalen} \& {Fryer}(2012)}]{wf12}
{Whalen}, D.~J., \& {Fryer}, C.~L. 2012, \apjl, 756, L19,
  \dodoi{10.1088/2041-8205/756/1/L19}

\bibitem[{{Whalen} {et~al.}(2023){Whalen}, {Latif}, \& {Mezcua}}]{W23}
{Whalen}, D.~J., {Latif}, M.~A., \& {Mezcua}, M. 2023, \apj, 956, 133,
  \dodoi{10.3847/1538-4357/acf92c}

\bibitem[{{Whalen} {et~al.}(2020){Whalen}, {Mezcua}, {Meiksin}, {Hartwig}, \&
  {Latif}}]{wet20a}
{Whalen}, D.~J., {Mezcua}, M., {Meiksin}, A., {Hartwig}, T., \& {Latif}, M.~A.
  2020, \apjl, 896, L45, \dodoi{10.3847/2041-8213/ab9a30}

\bibitem[{{Whalen} {et~al.}(2021){Whalen}, {Mezcua}, {Patrick}, {Meiksin}, \&
  {Latif}}]{wet21a}
{Whalen}, D.~J., {Mezcua}, M., {Patrick}, S.~J., {Meiksin}, A., \& {Latif},
  M.~A. 2021, \apjl, 922, L39, \dodoi{10.3847/2041-8213/ac35e6}

\bibitem[{{Willott} {et~al.}(2010){Willott}, {Delorme}, {Reyl{\'e}}, {Albert},
  {Bergeron}, {Crampton}, {Delfosse}, {Forveille}, {Hutchings}, {McLure},
  {Omont}, \& {Schade}}]{wil10}
{Willott}, C.~J., {Delorme}, P., {Reyl{\'e}}, C., {et~al.} 2010, \aj, 139, 906,
  \dodoi{10.1088/0004-6256/139/3/906}

\bibitem[{{Woods} {et~al.}(2017){Woods}, {Heger}, {Whalen}, {Haemmerl{\'e}}, \&
  {Klessen}}]{tyr17}
{Woods}, T.~E., {Heger}, A., {Whalen}, D.~J., {Haemmerl{\'e}}, L., \&
  {Klessen}, R.~S. 2017, \apjl, 842, L6, \dodoi{10.3847/2041-8213/aa7412}

\bibitem[{{Woods} {et~al.}(2019){Woods}, {Agarwal}, {Bromm}, {Bunker}, {Chen},
  {Chon}, {Ferrara}, {Glover}, {Haemmerl{\'e}}, {Haiman}, {Hartwig}, {Heger},
  {Hirano}, {Hosokawa}, {Inayoshi}, {Klessen}, {Kobayashi}, {Koliopanos},
  {Latif}, {Li}, {Mayer}, {Mezcua}, {Natarajan}, {Pacucci}, {Rees}, {Regan},
  {Sakurai}, {Salvadori}, {Schneider}, {Surace}, {Tanaka}, {Whalen}, \&
  {Yoshida}}]{titans}
{Woods}, T.~E., {Agarwal}, B., {Bromm}, V., {et~al.} 2019, Publications of the
  Astronomical Society of Australia, 36, e027, \dodoi{10.1017/pasa.2019.14}

\bibitem[{{Yang} {et~al.}(2020){Yang}, {Wang}, {Fan}, {Hennawi}, {Davies},
  {Yue}, {Banados}, {Wu}, {Venemans}, {Barth}, {Bian}, {Boutsia}, {Decarli},
  {Farina}, {Green}, {Jiang}, {Li}, {Mazzucchelli}, \& {Walter}}]{yang20}
{Yang}, J., {Wang}, F., {Fan}, X., {et~al.} 2020, \apjl, 897, L14,
  \dodoi{10.3847/2041-8213/ab9c26}

\bibitem[{{Zhang} {et~al.}(2022){Zhang}, {An}, {Wang}, {Frey}, {Gurvits},
  {Gab{\'a}nyi}, {Perger}, \& {Paragi}}]{zng22}
{Zhang}, Y., {An}, T., {Wang}, A., {et~al.} 2022, \aap, 662, L2,
  \dodoi{10.1051/0004-6361/202243785}

\bibitem[{{Zhu} {et~al.}(2022){Zhu}, {Li}, {Li}, {Maji}, {Yajima}, {Schneider},
  \& {Hernquist}}]{zhu22}
{Zhu}, Q., {Li}, Y., {Li}, Y., {et~al.} 2022, \mnras, 514, 5583,
  \dodoi{10.1093/mnras/stac1556}

\end{thebibliography}
\bibliographystyle{aasjournal}
\newpage
\appendix
\setcounter{figure}{0} 
\counterwithin{figure}{section}

\section{Figures}

\begin{figure*} 
\begin{center}
\includegraphics[scale=0.45]{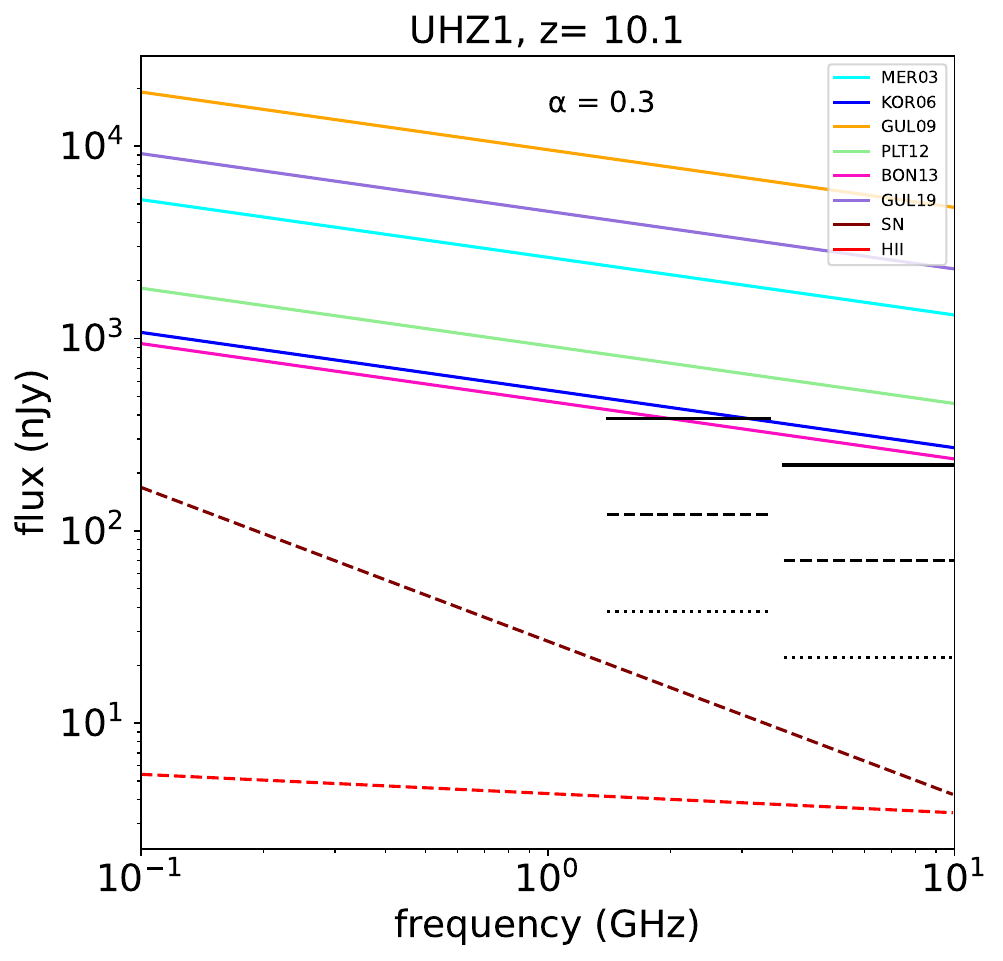}
\includegraphics[scale=0.45]{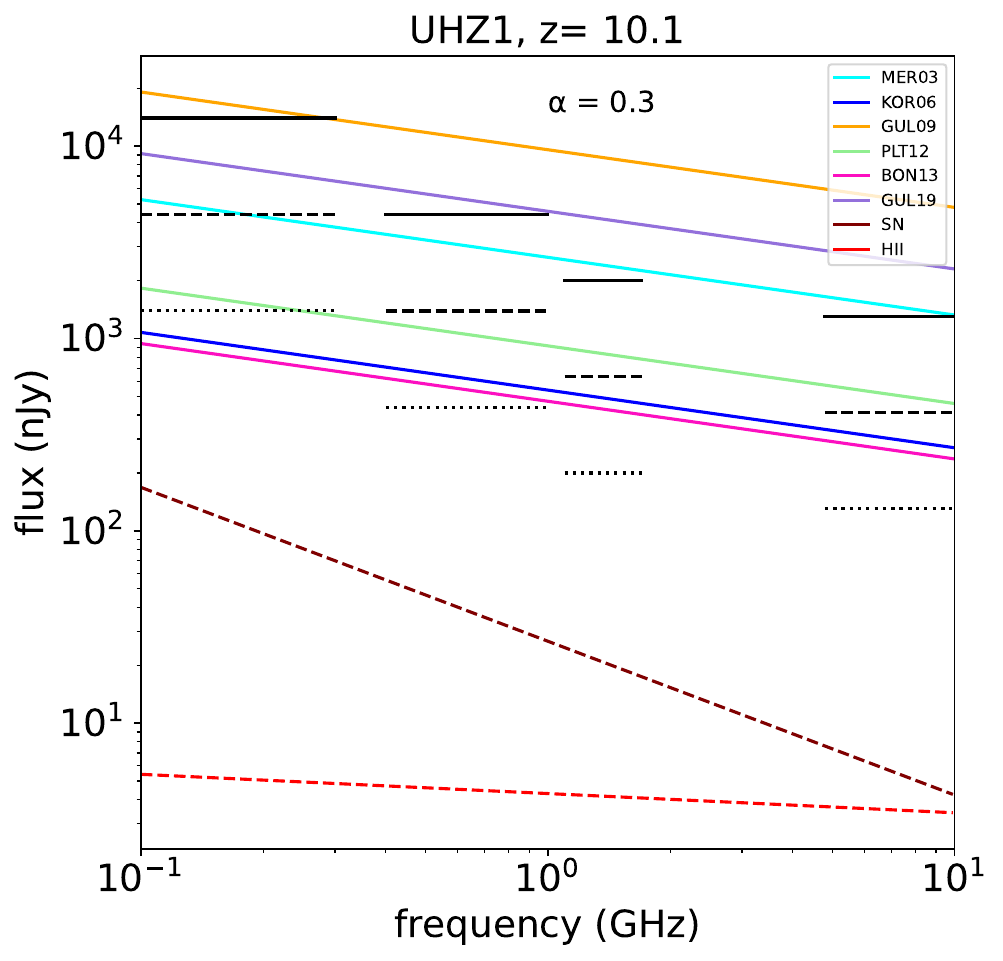}
\includegraphics[scale=0.45]{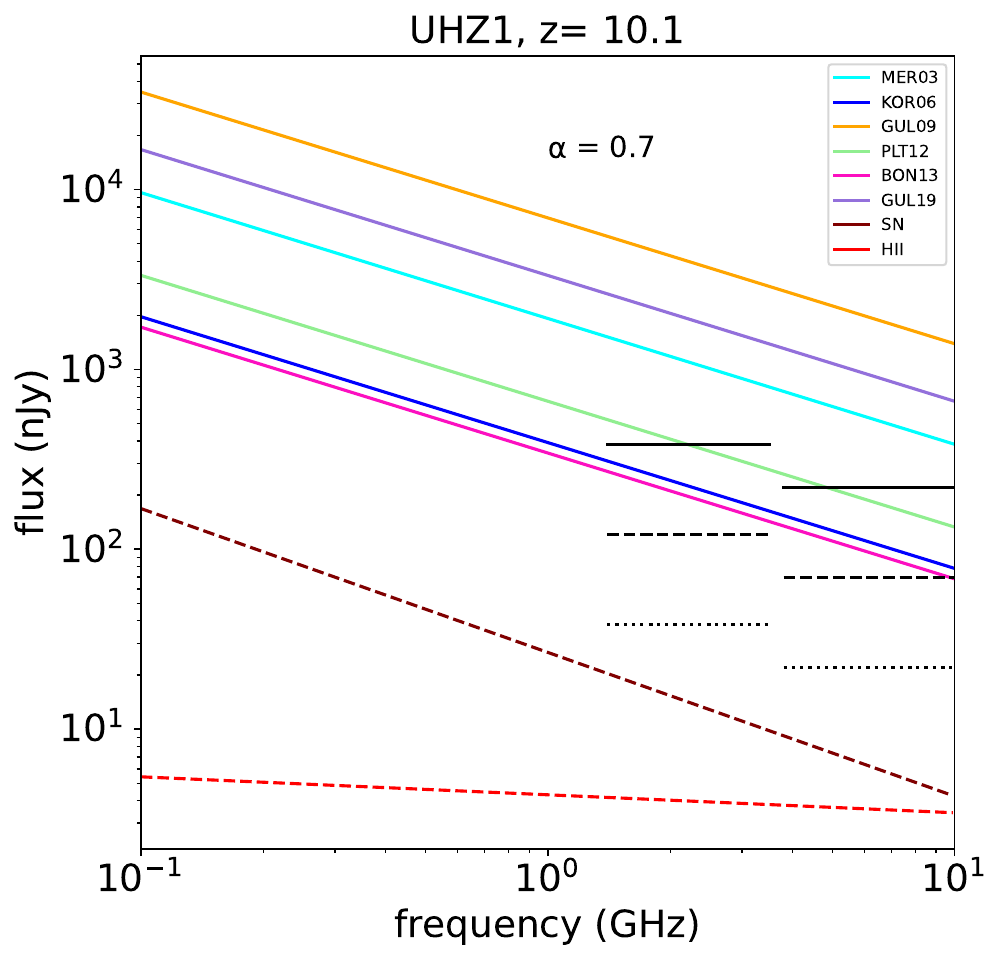}
\includegraphics[scale=0.45]{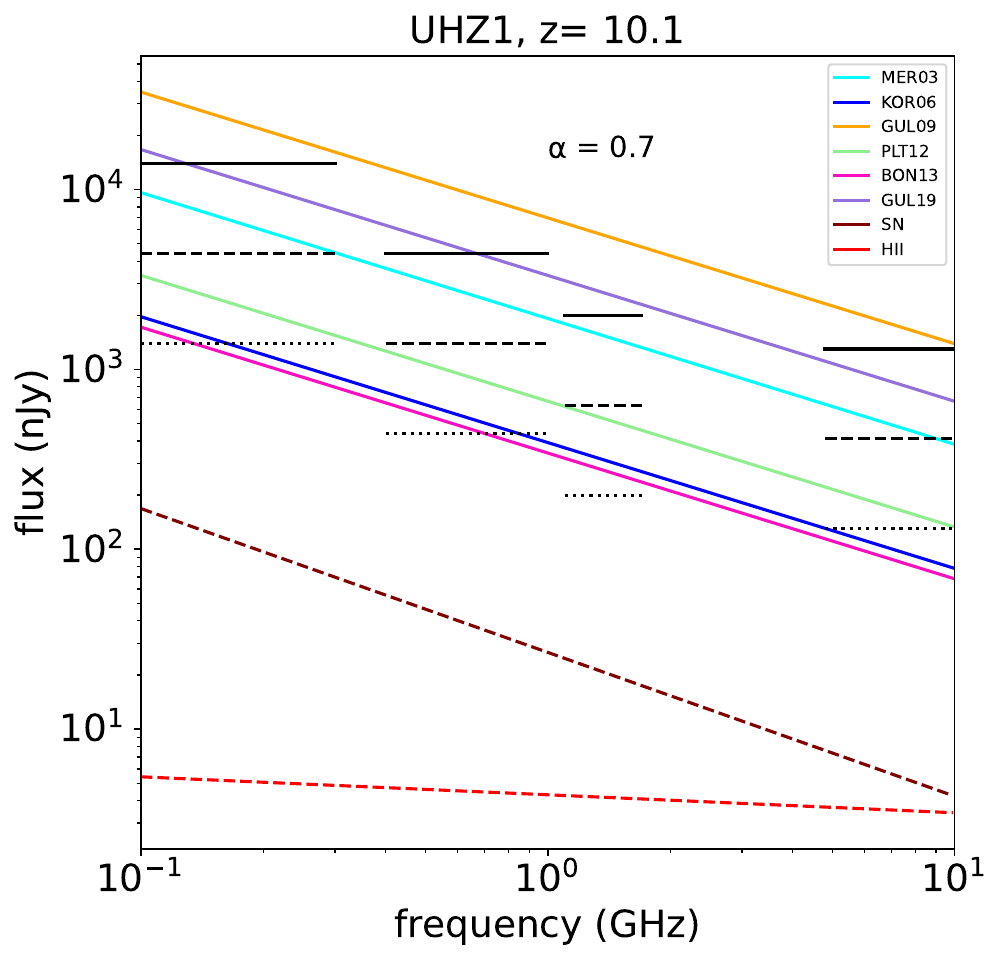}
\end{center}
\vspace{-0.1cm}
\caption{Radio flux densities for UHZ1 for $\alpha =$ 0.3 (top row) and 0.7 (bottom row) with detection limits  for ngVLA (left column) and SKA (right column). The dotted, dashed and dot-dashed red and brown lines are H II region and SN flux densities for SFRs of 1, 3 and 10 \Ms\ yr$^{-1}$, respectively. The black solid, dashed and dotted horizontal bars are H II region and SN flux densities for SKA and ngVLA limits for integration times of 1, 10 and 100 hr, respectively.}
\label{fig:f13}
\end{figure*}

\begin{figure*} 
\begin{center}
\includegraphics[scale=0.45]{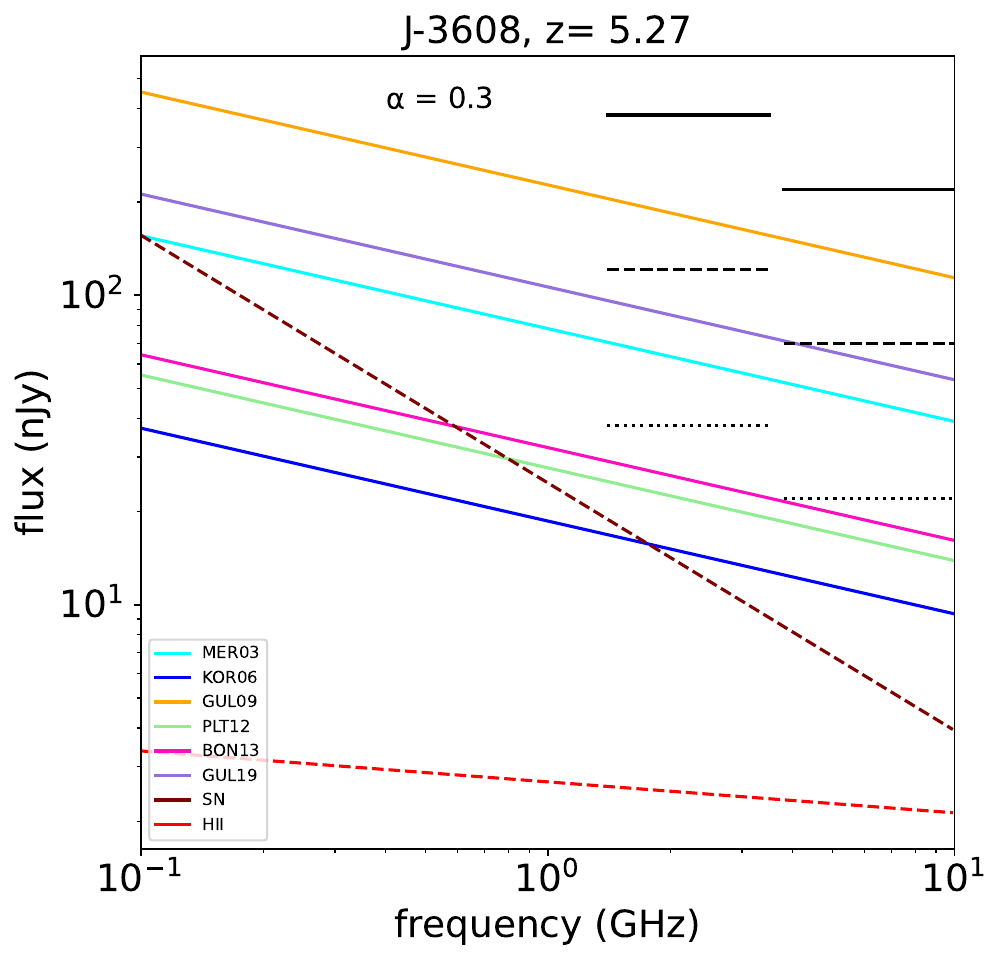}
\includegraphics[scale=0.45]{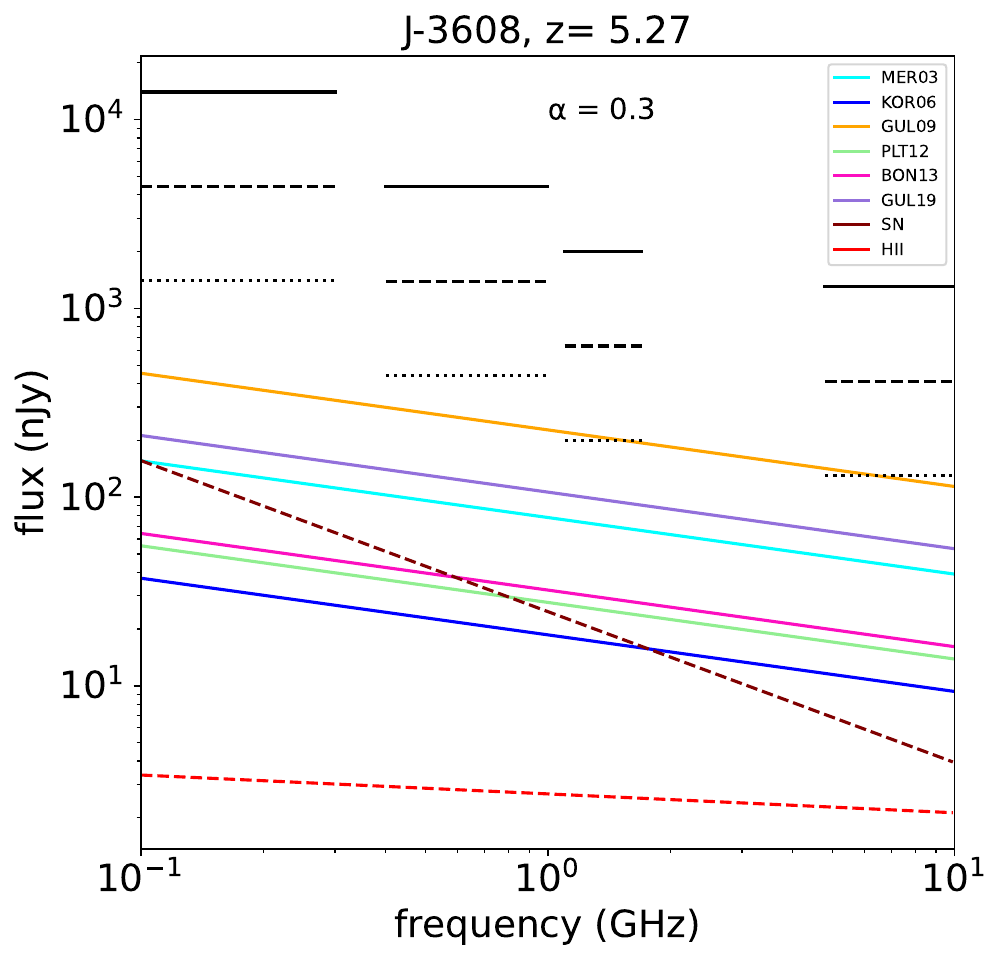}
\includegraphics[scale=0.45]{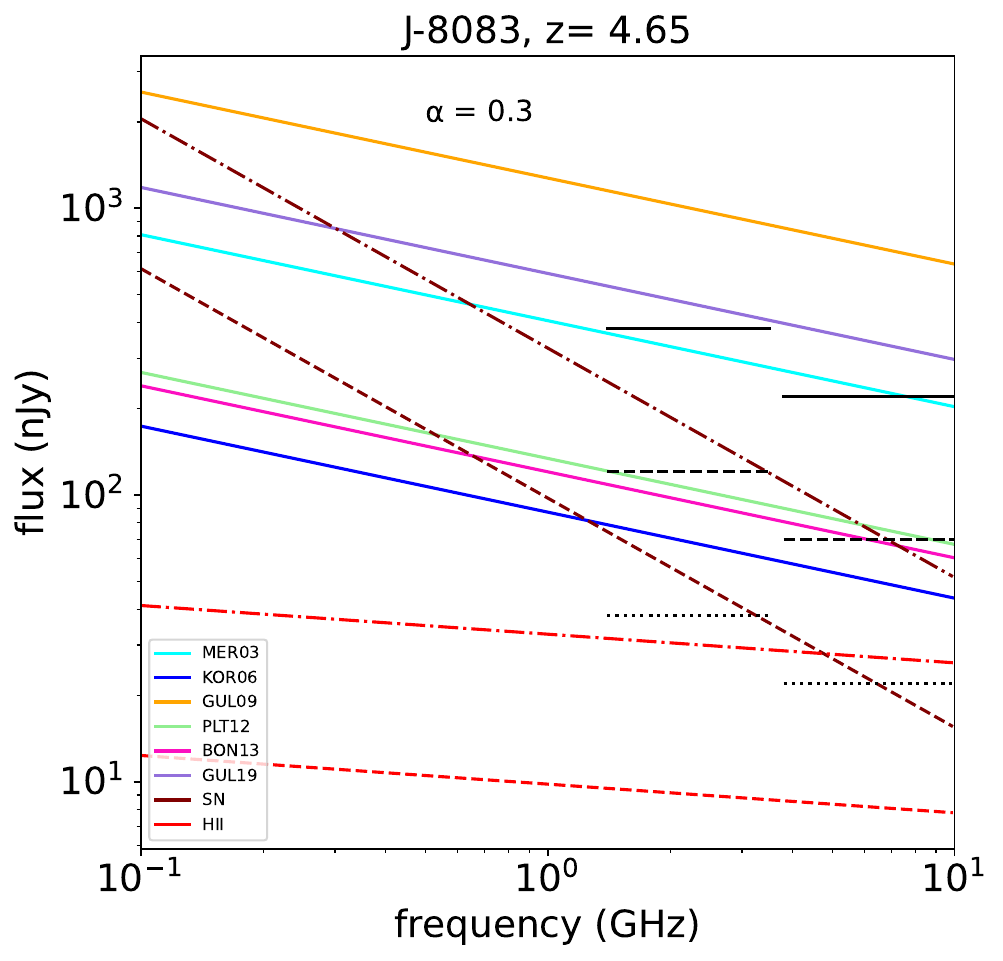} 
\includegraphics[scale=0.45]{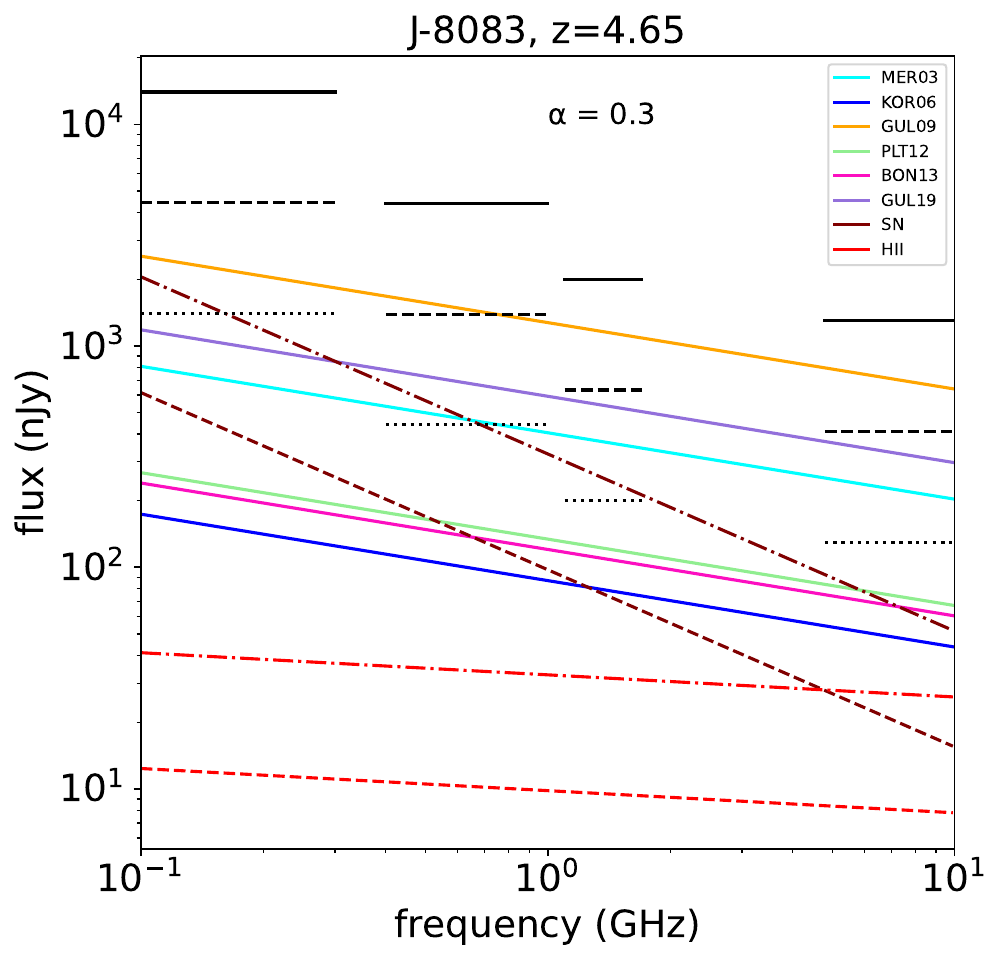}
\includegraphics[scale=0.45]{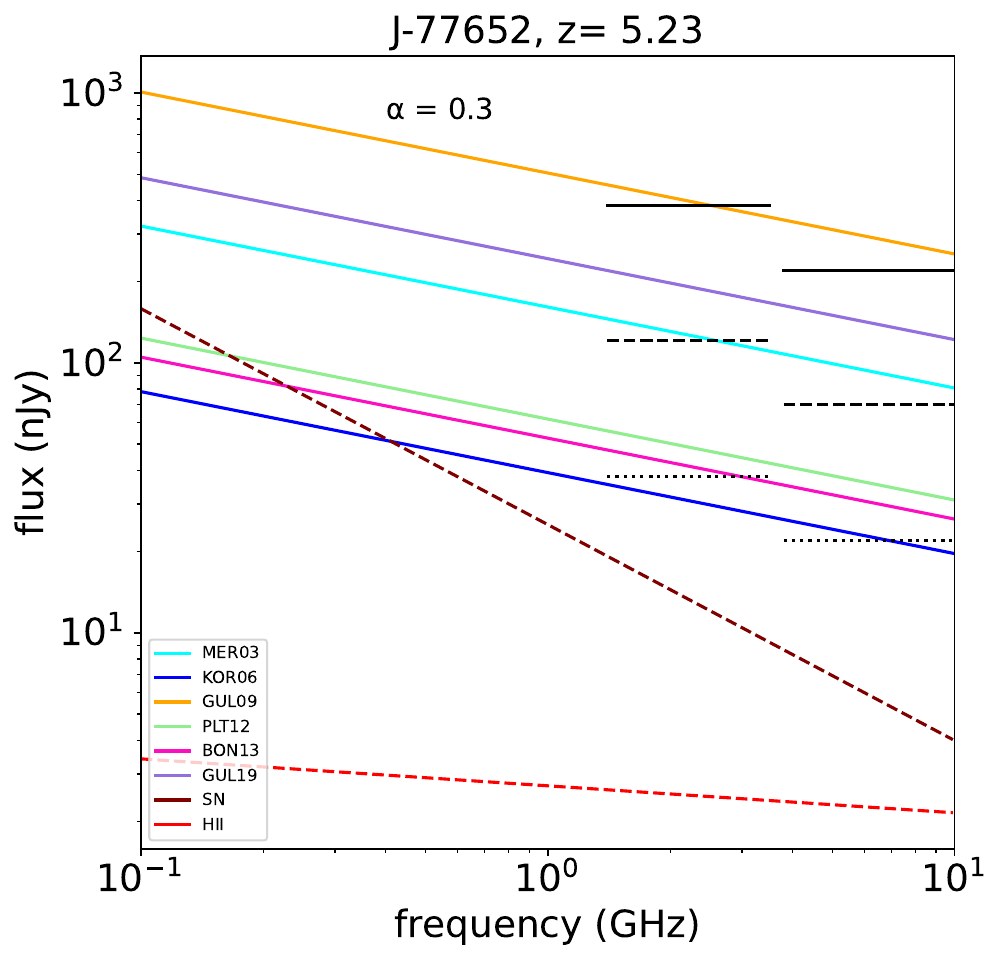}
\includegraphics[scale=0.45]{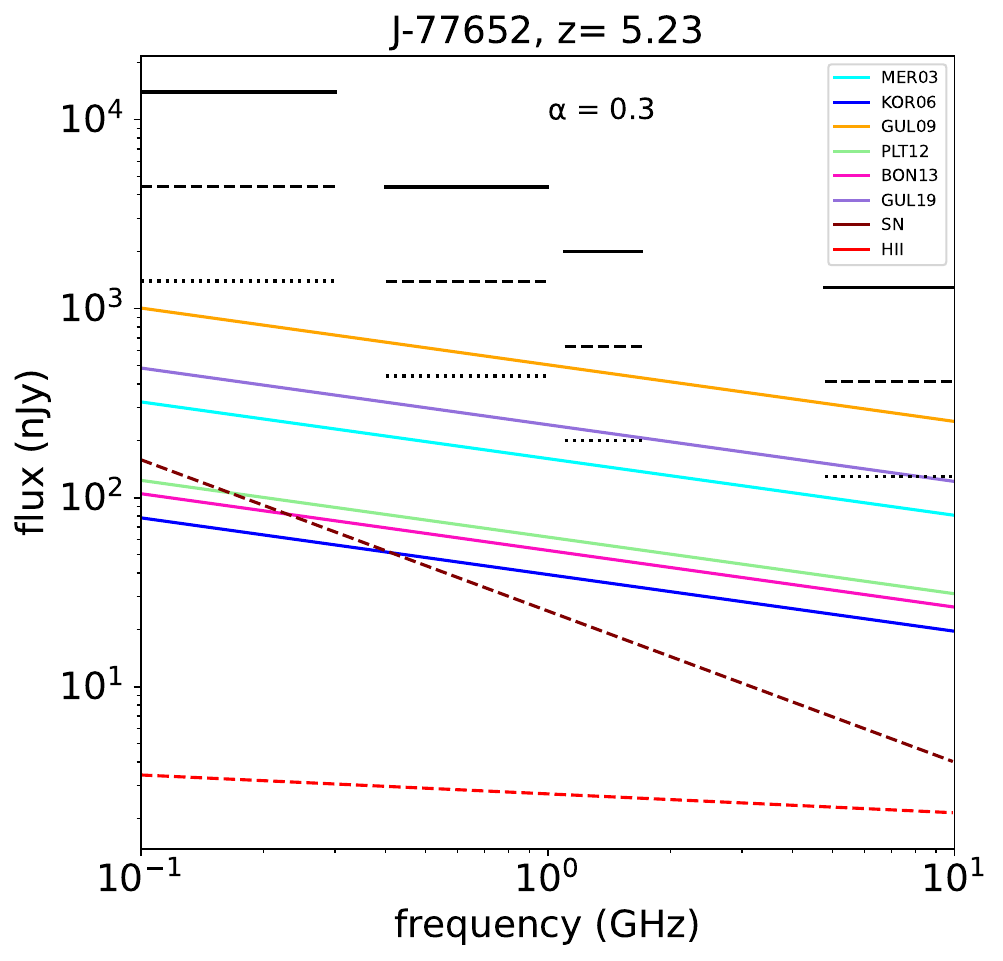}
\end{center}
\vspace{-0.1cm}
\caption{Radio flux densities for AGNs observed in JADES for $\alpha$ = 0.3 with detection limits for ngVLA (left column) and SKA (right column). The dotted, dashed and dot-dashed red and brown lines are H II region and SN flux densities for SFRs of 1, 3 and 10 \Ms\ yr$^{-1}$, respectively. The black solid, dashed and dotted horizontal bars show ngVLA and SKA detection limits for integration times of 1, 10 and 100 hr, respectively.}
\label{fig:f4}
\end{figure*}

\begin{figure*} 
\begin{center}
\includegraphics[scale=0.45]{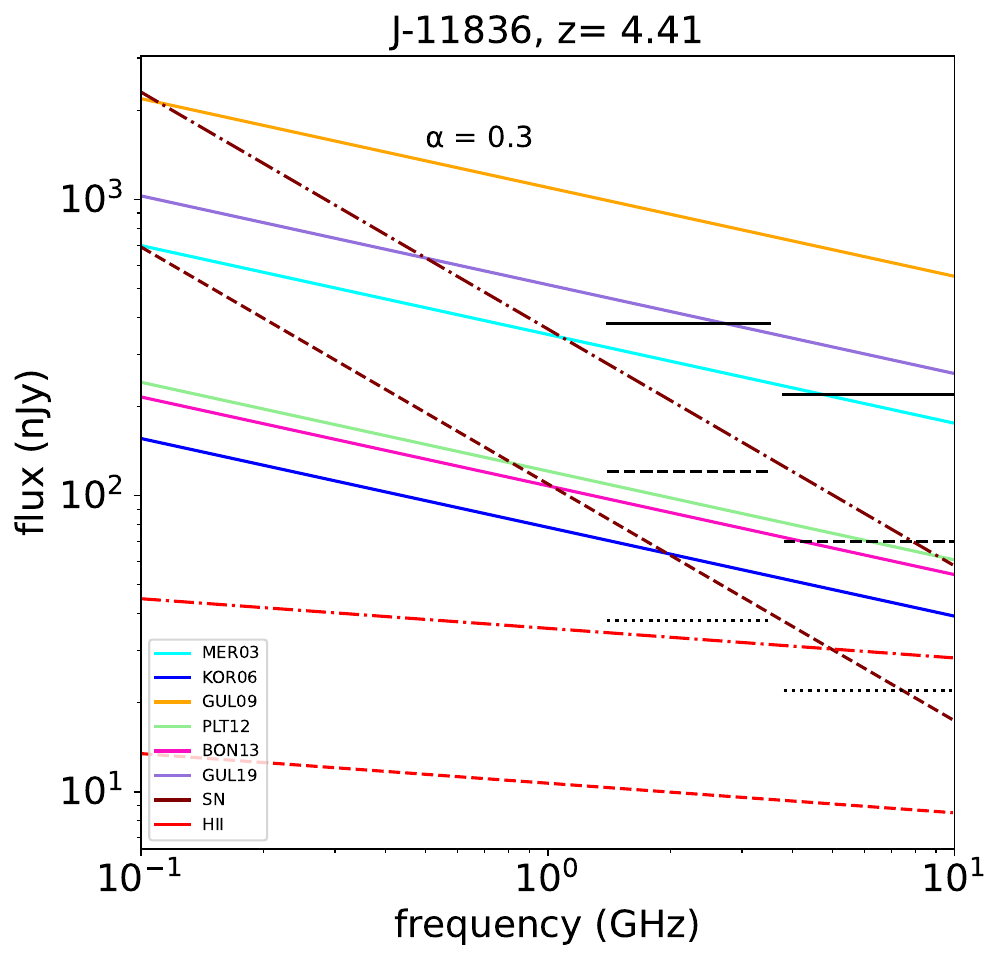} 
\includegraphics[scale=0.45]{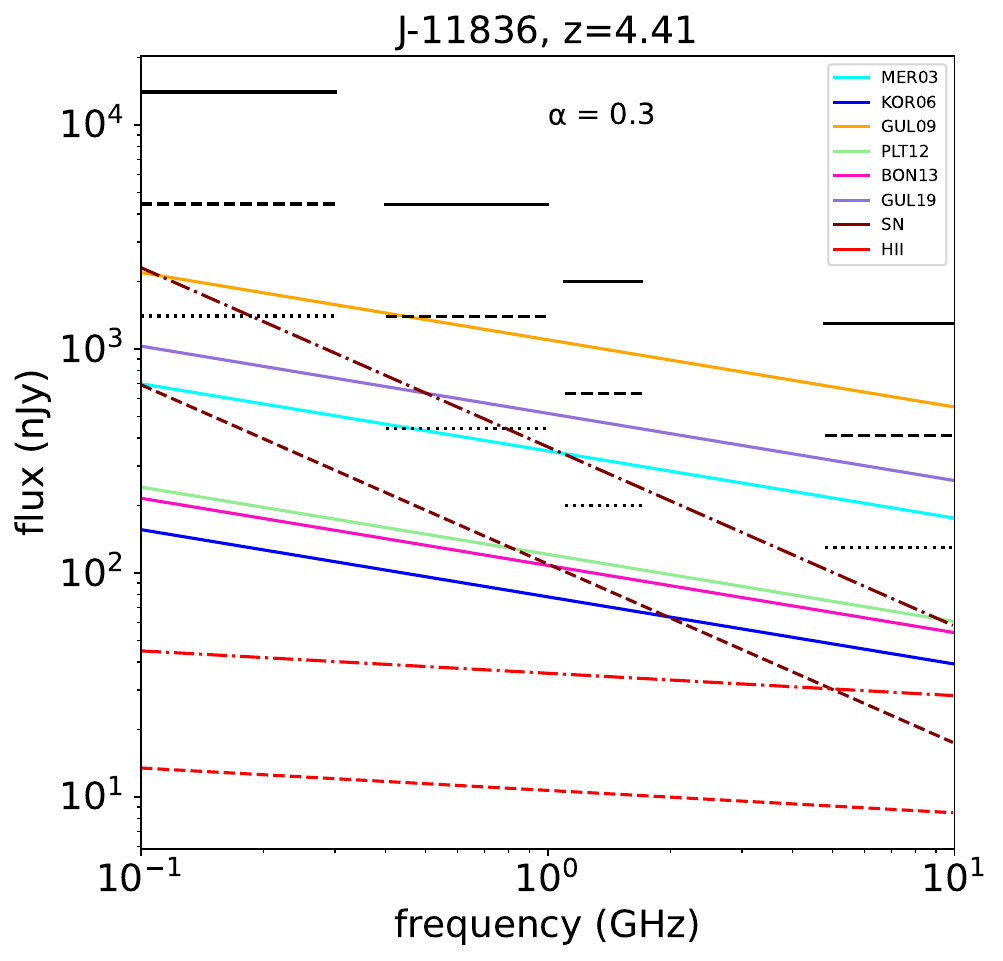} 
\includegraphics[scale=0.45]{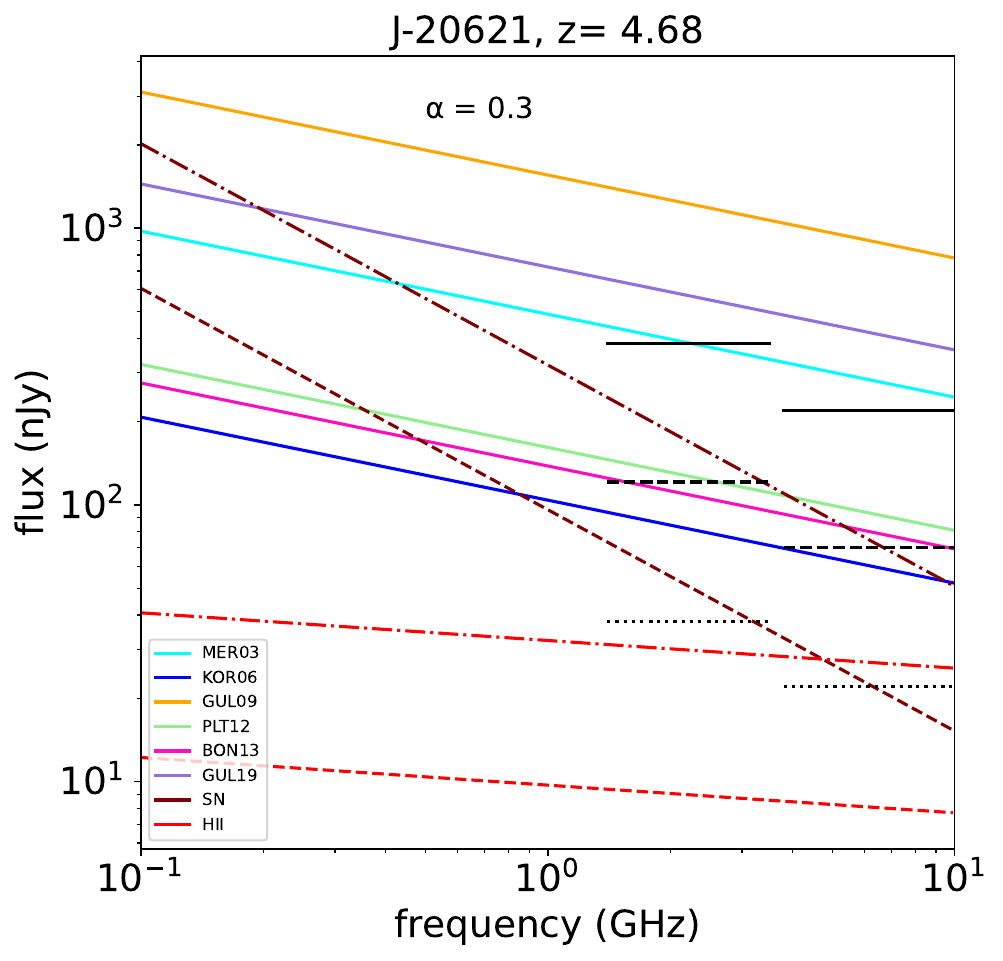} 
\includegraphics[scale=0.45]{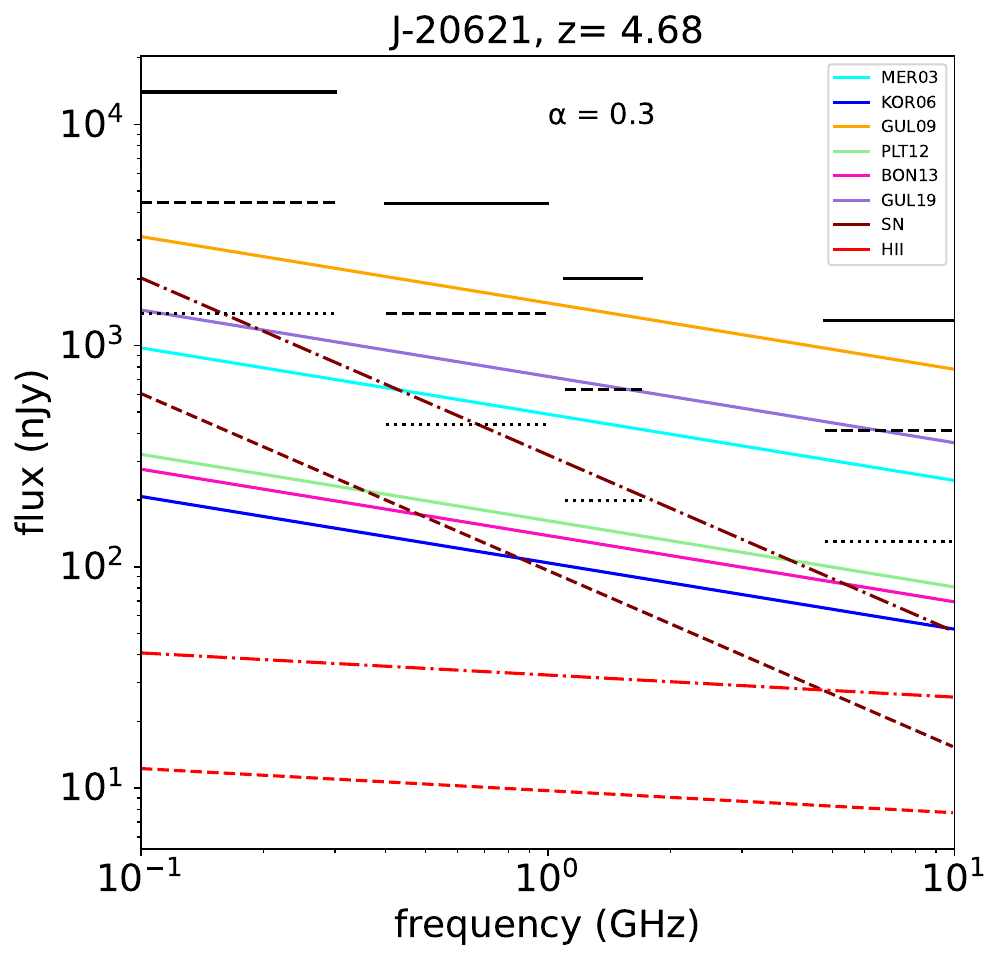}
\includegraphics[scale=0.45]{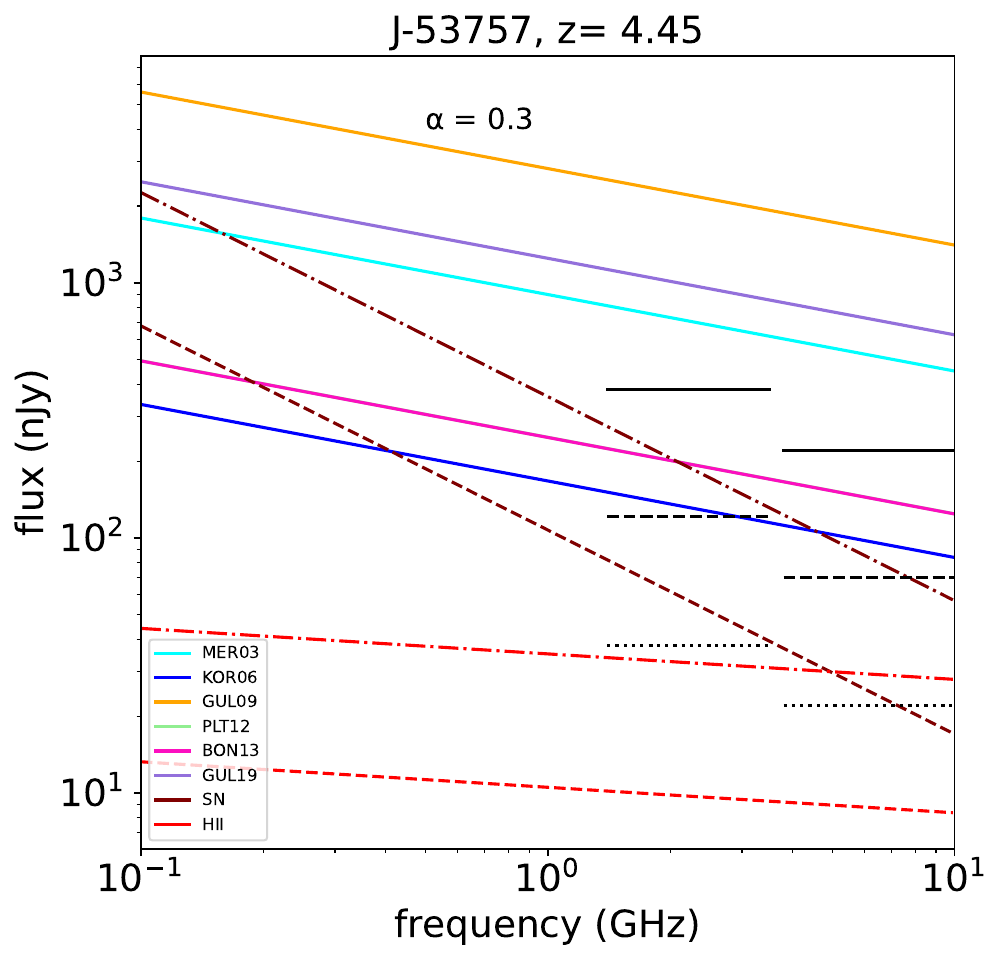} 
\includegraphics[scale=0.45]{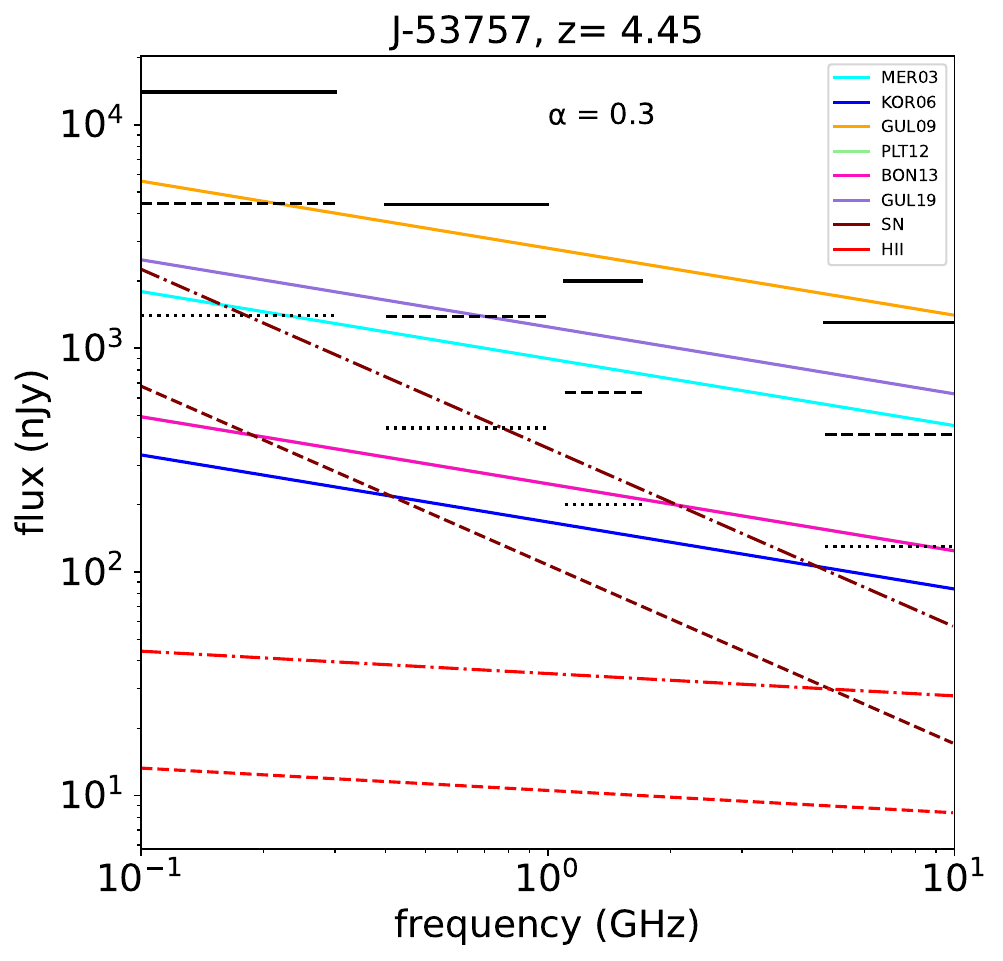} 
\end{center}
\vspace{-0.1cm}
\caption{Radio flux densities for AGNs observed in JADES for $\alpha$ = 0.3 with detection limits for ngVLA (left column) and SKA (right column). The dotted, dashed and dot-dashed red and brown lines are H II region and SN flux densities for SFRs of 1, 3 and 10 \Ms\ yr$^{-1}$, respectively. The black solid, dashed and dotted horizontal bars show ngVLA and SKA detection limits for integration times of 1, 10 and 100 hr, respectively.}
\label{fig:f5}
\end{figure*}

\begin{figure*} 
\begin{center}
\includegraphics[scale=0.45]{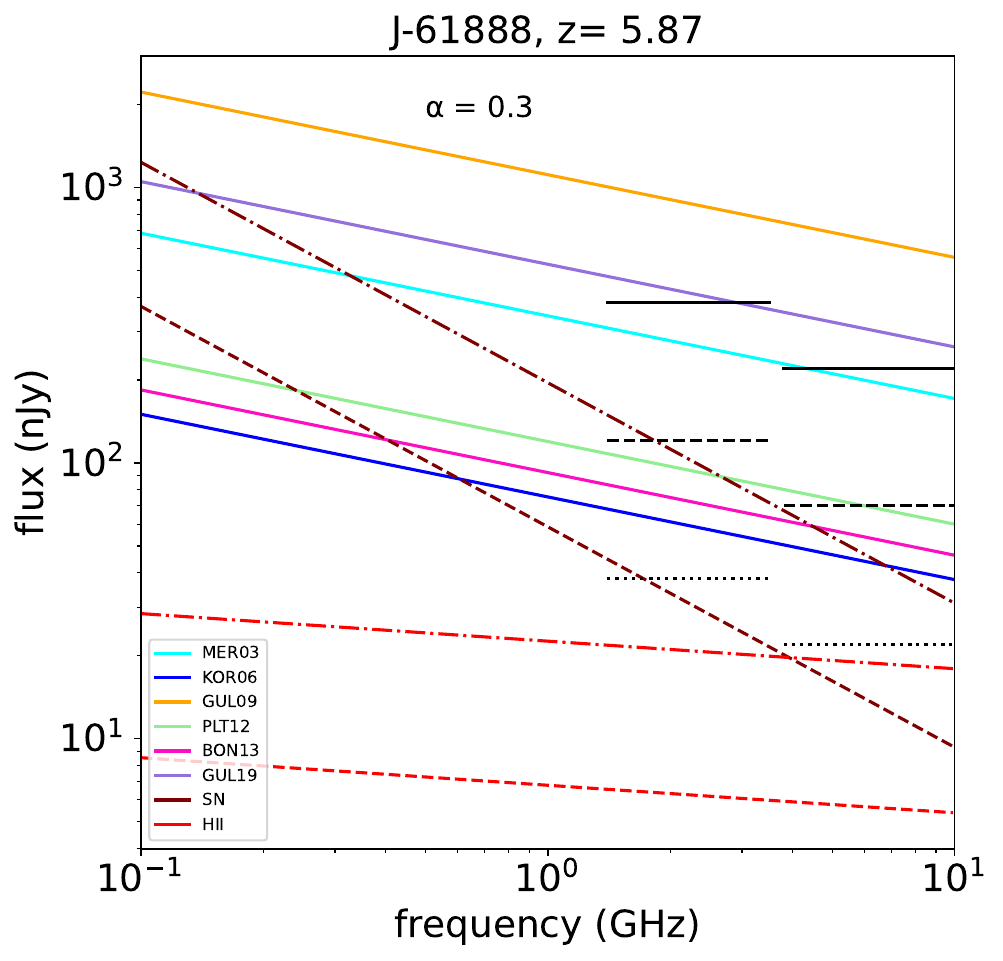}
\includegraphics[scale=0.45]{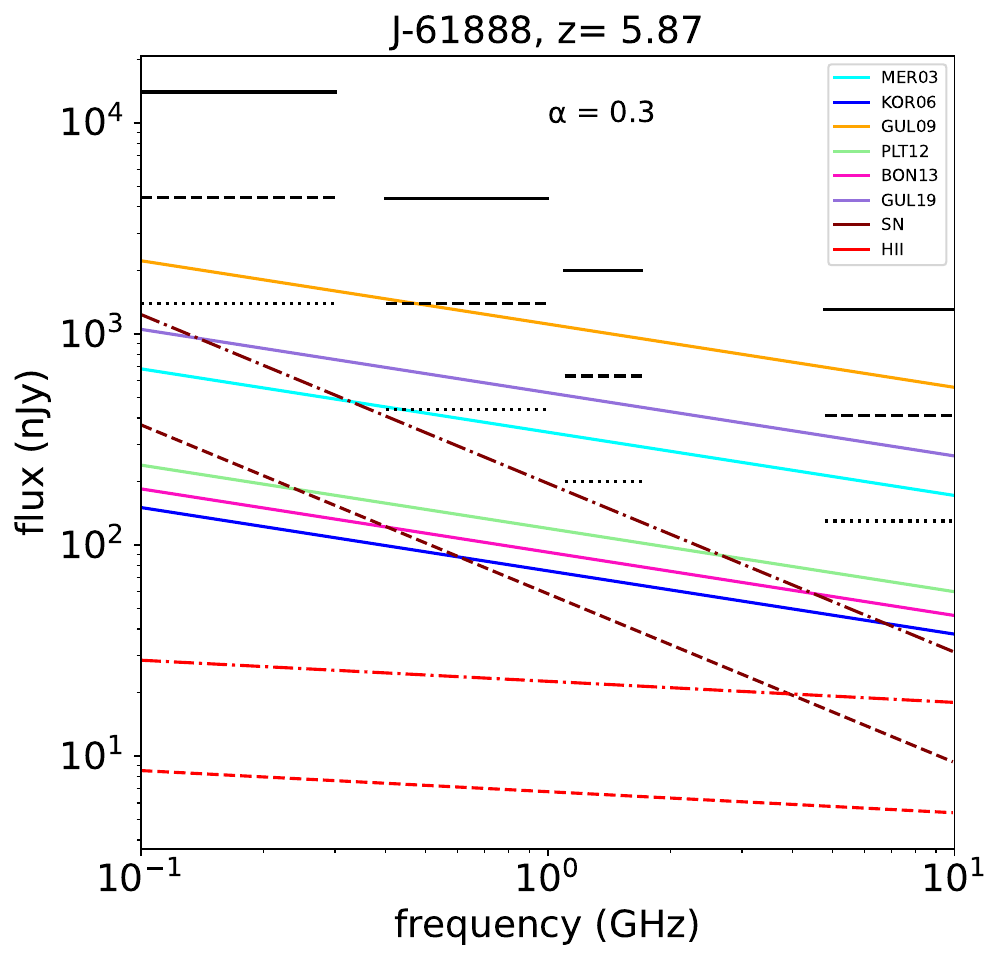}
\includegraphics[scale=0.45]{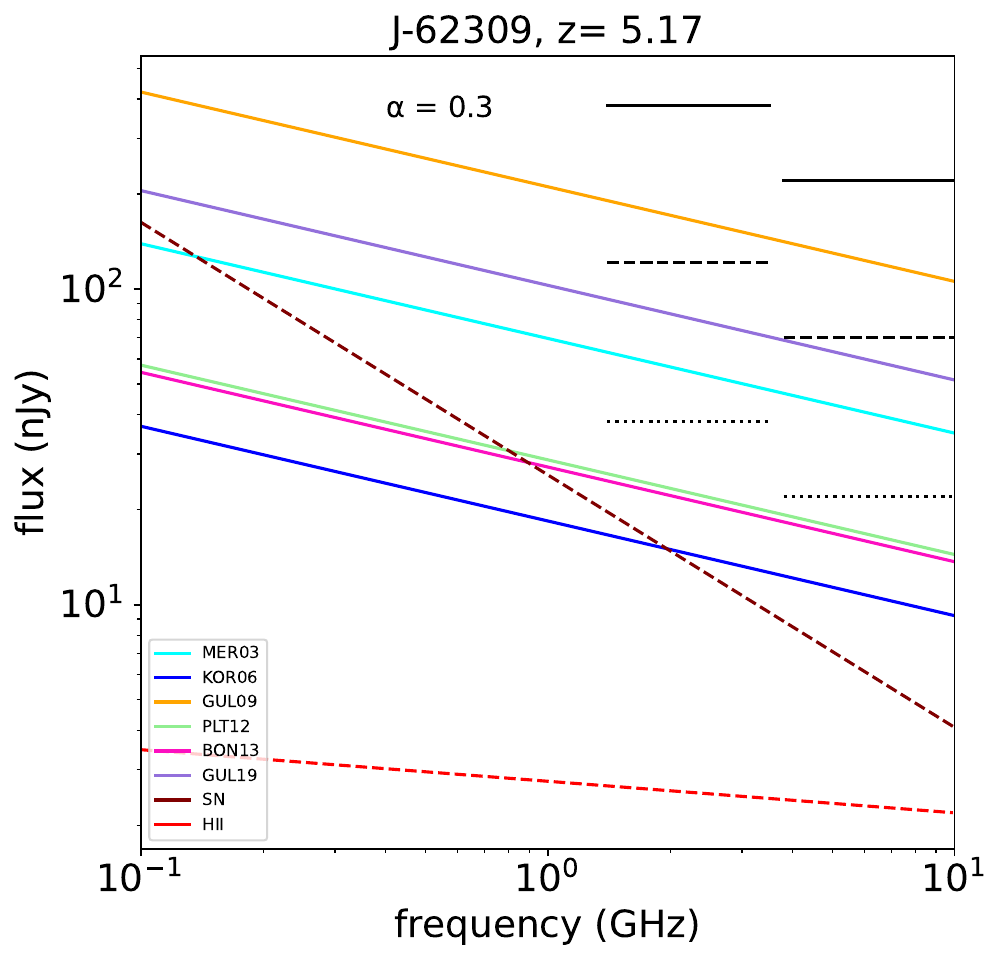}
\includegraphics[scale=0.45]{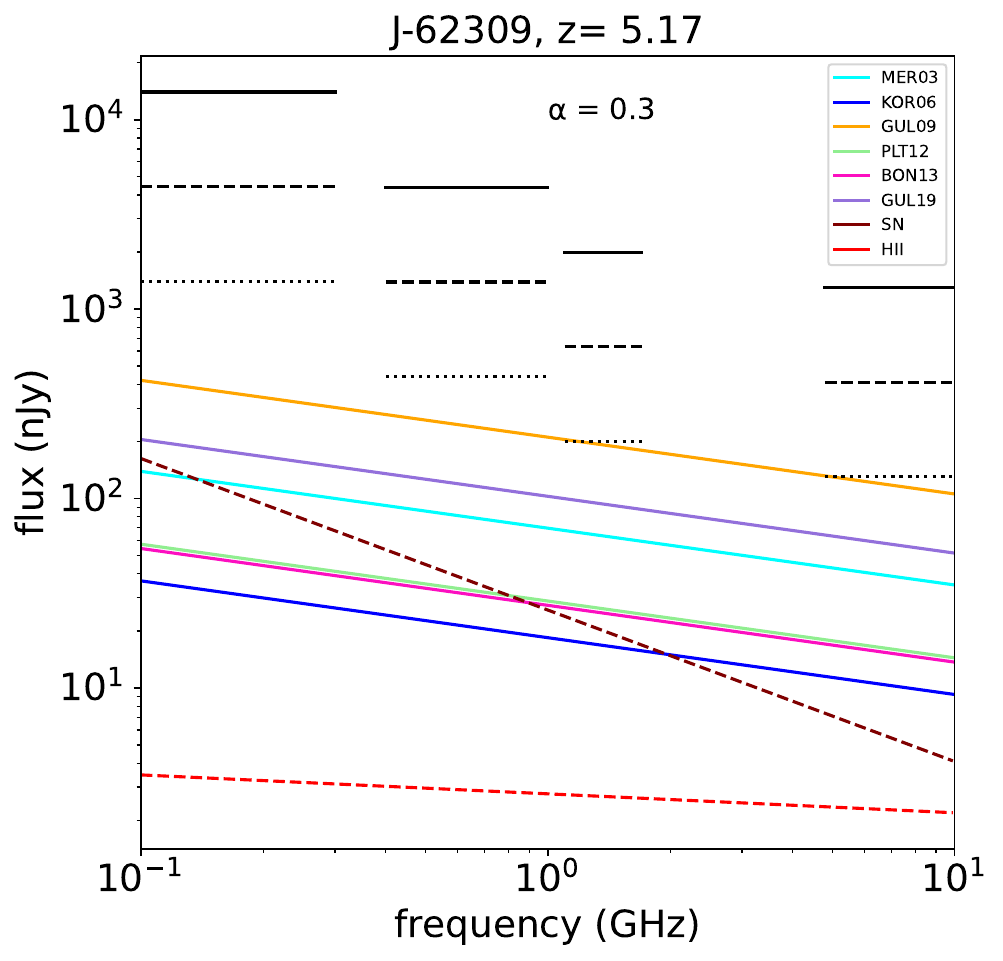}
\includegraphics[scale=0.45]{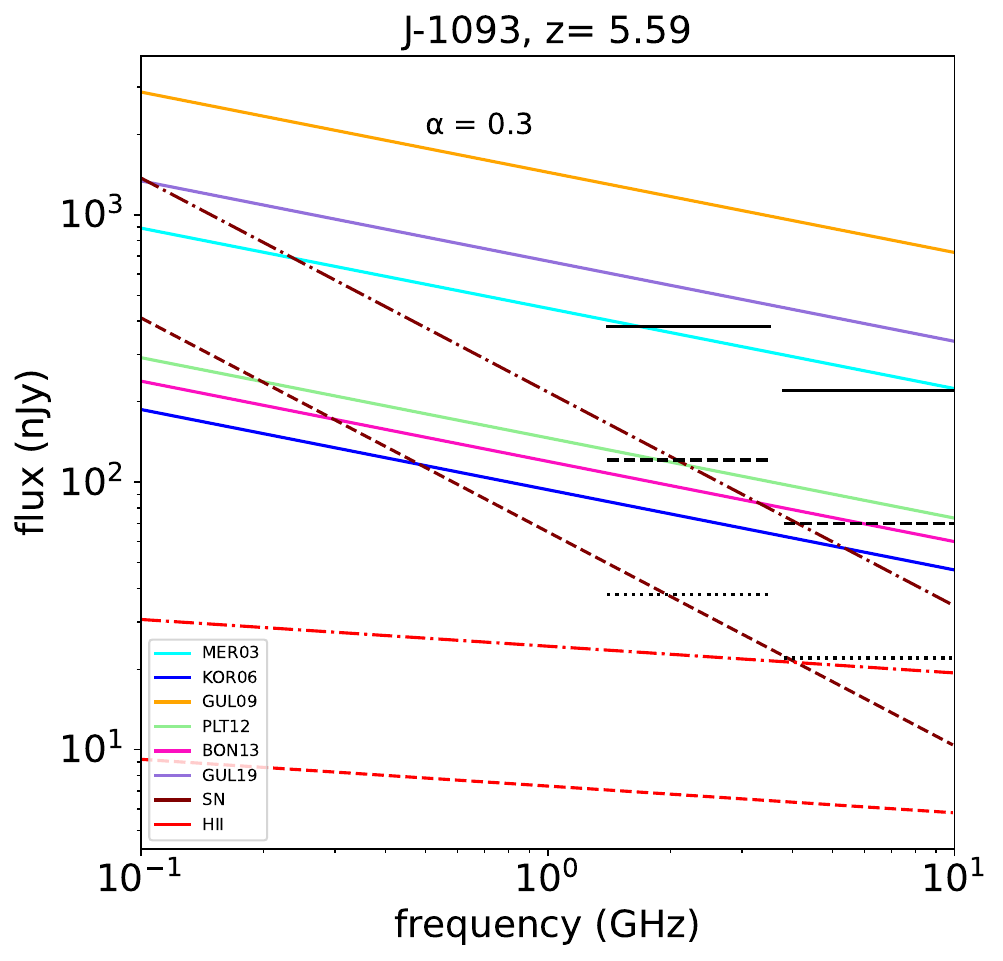}
\includegraphics[scale=0.45]{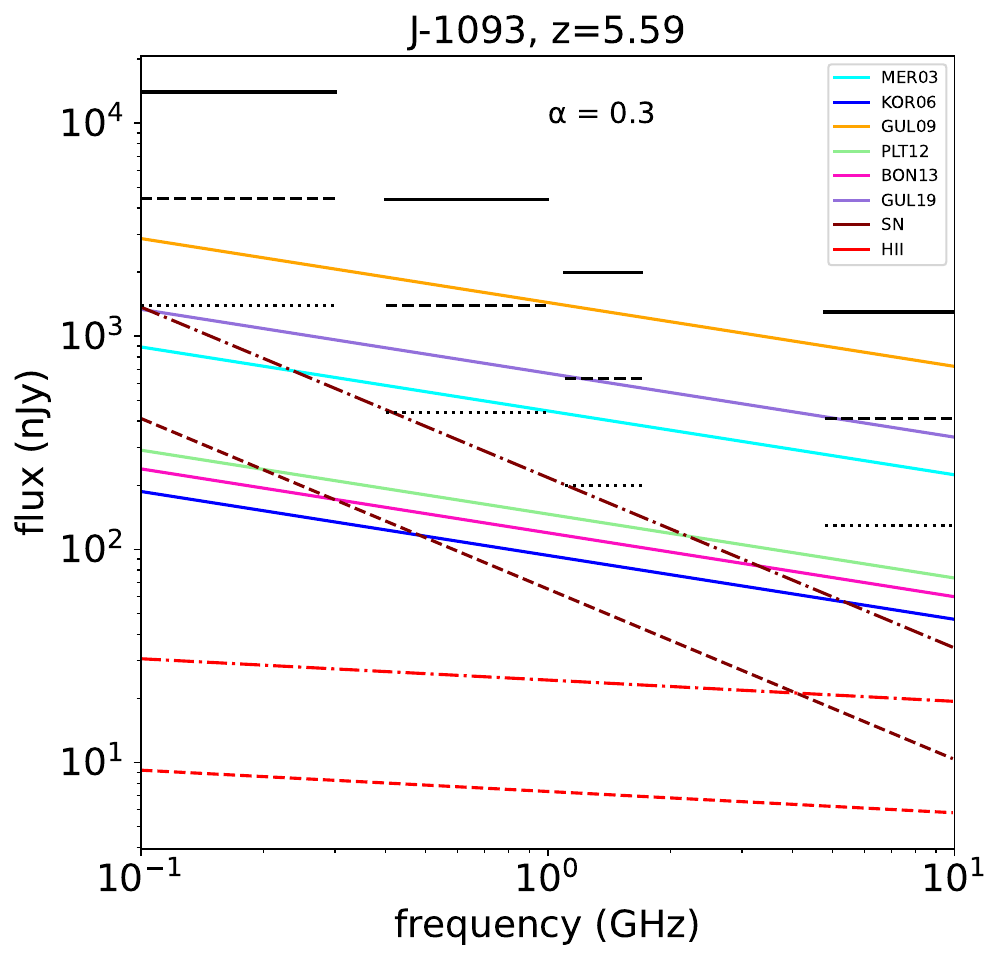}
\end{center}
\vspace{-0.1cm}
\caption{Radio flux densities for AGNs observed in JADES for $\alpha$ = 0.3 with detection limits for ngVLA (left column) and SKA (right column). The dotted, dashed and dot-dashed red and brown lines are H II region and SN flux densities for SFRs of 1, 3 and 10 \Ms\ yr$^{-1}$, respectively. The black solid, dashed and dotted horizontal bars show ngVLA and SKA detection limits for integration times of 1, 10 and 100 hr, respectively.}
\label{fig:f6}
\end{figure*}

\begin{figure*} 
\begin{center}
\includegraphics[scale=0.45]{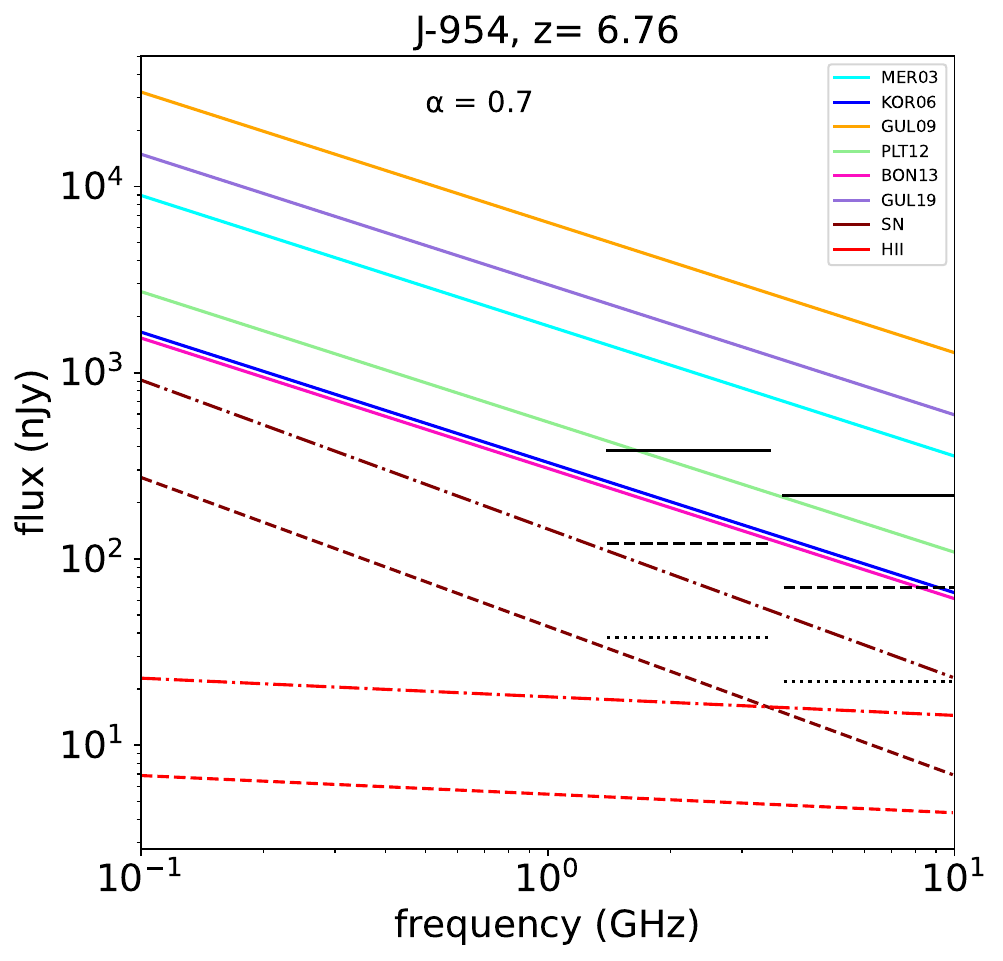}
\includegraphics[scale=0.45]{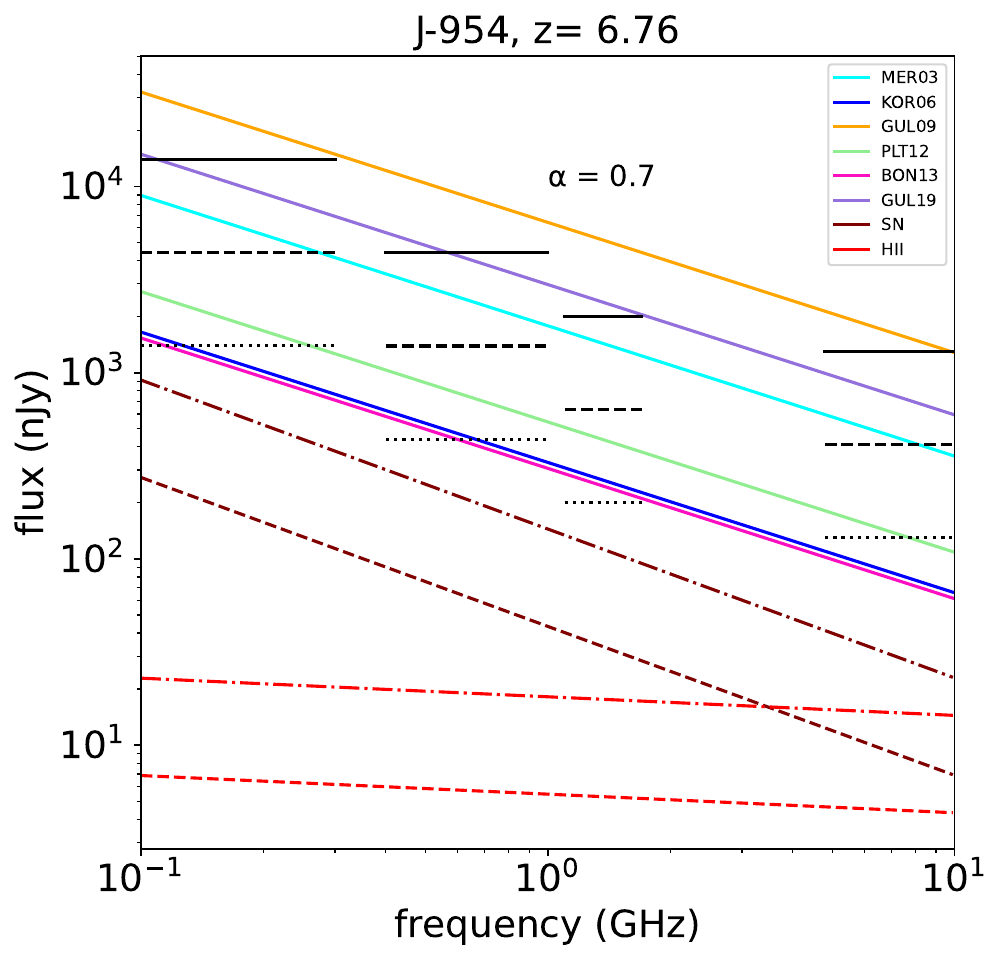}
\includegraphics[scale=0.45]{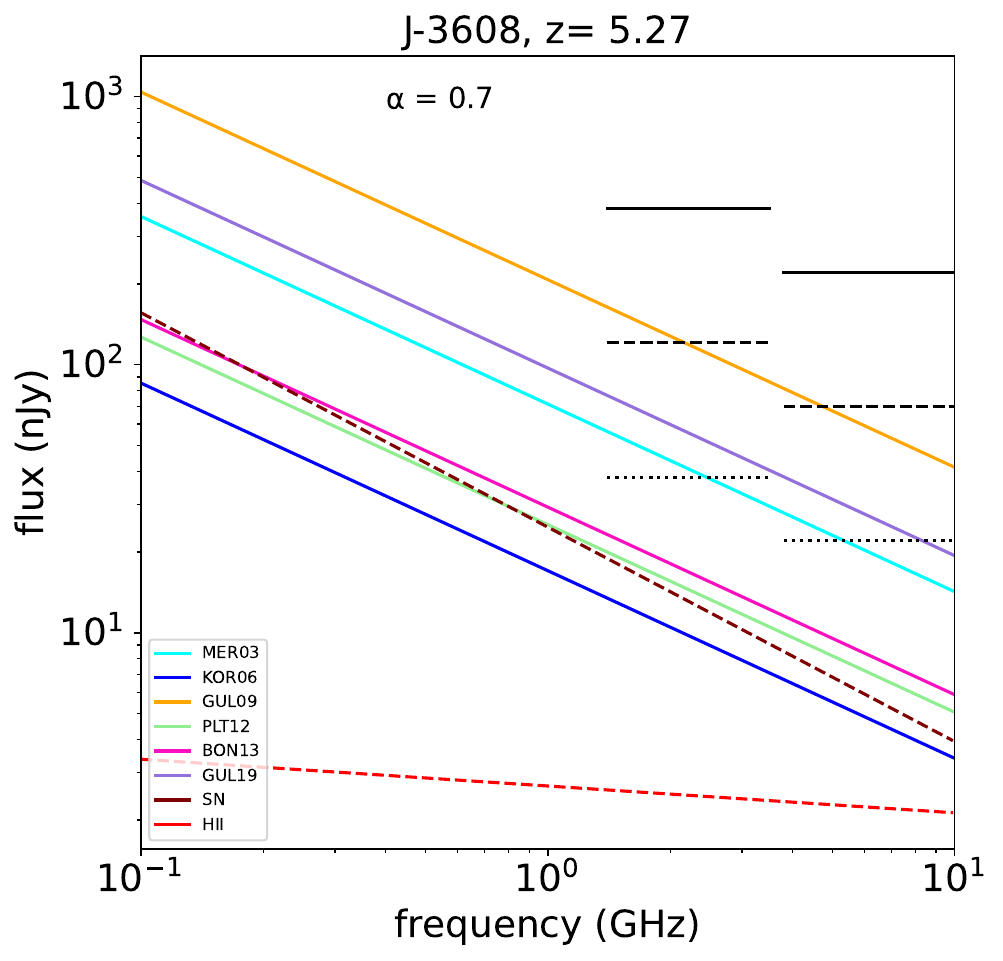}
\includegraphics[scale=0.45]{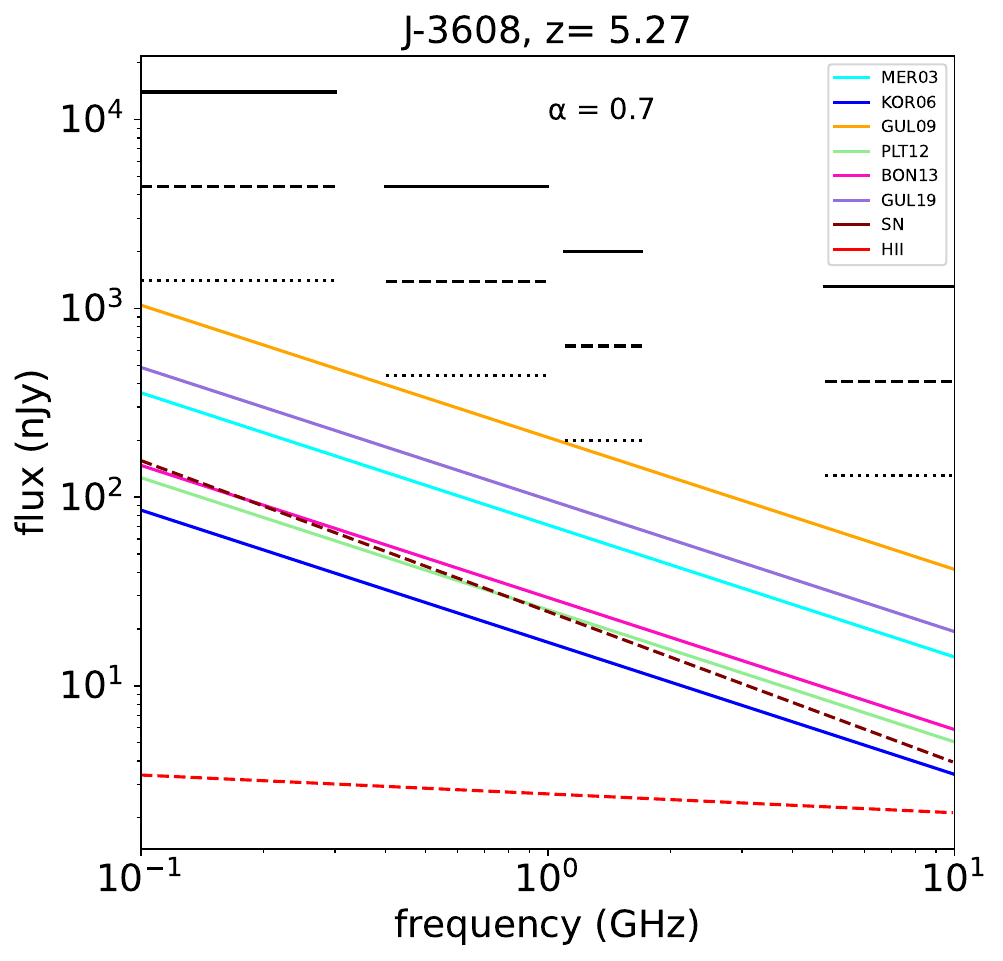}
\includegraphics[scale=0.45]{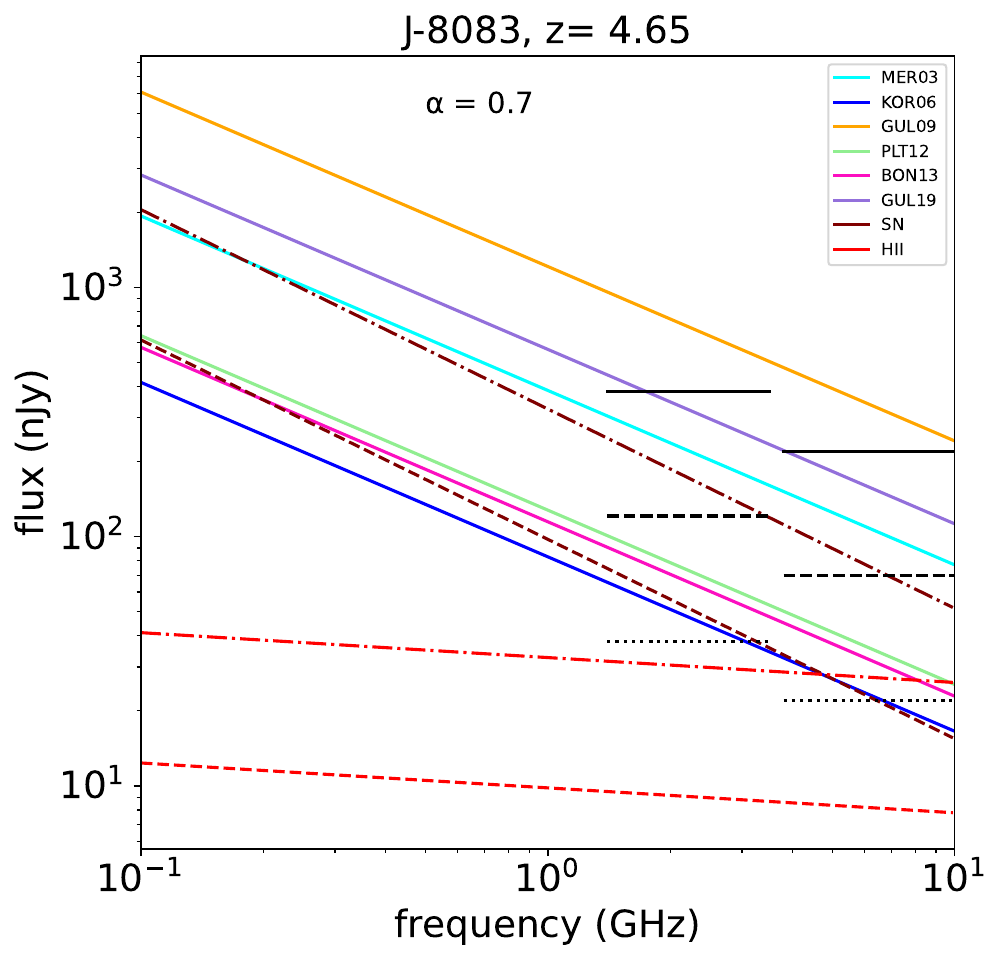}
\includegraphics[scale=0.45]{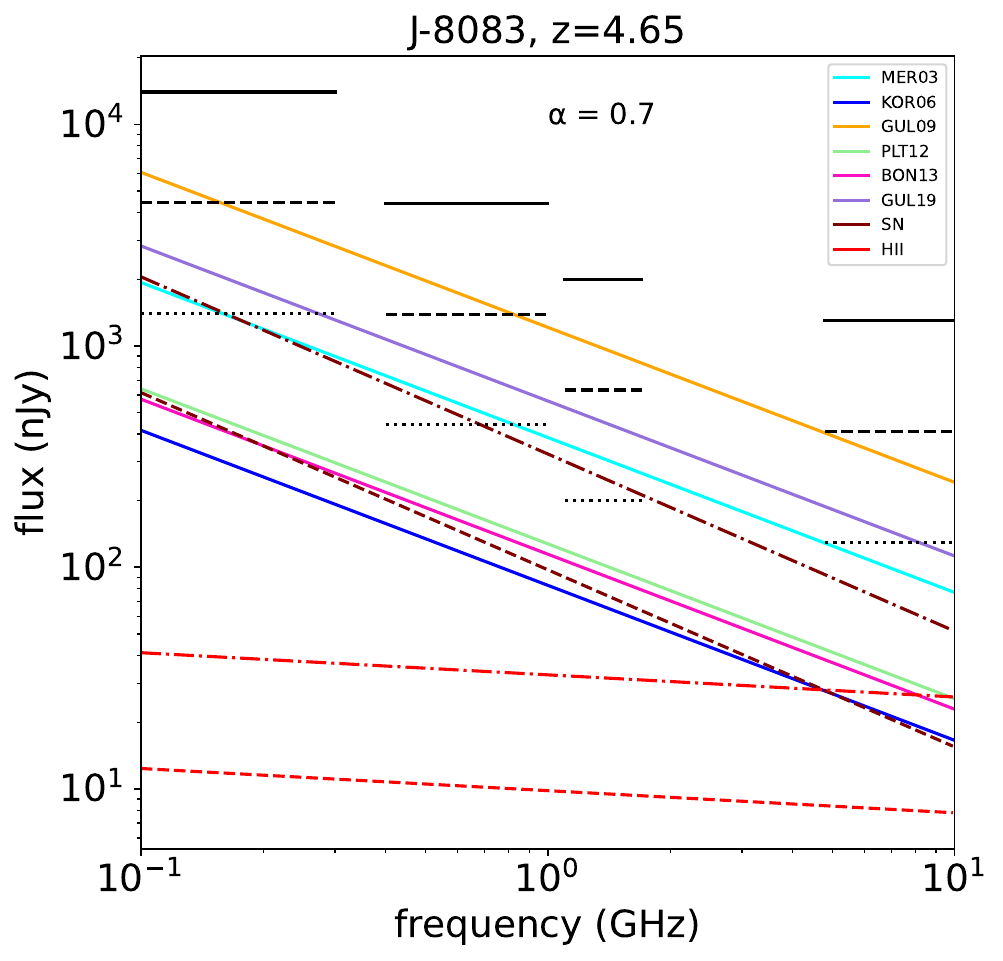} 
\end{center}
\vspace{-0.1cm}
\caption{Radio flux densities for AGNs observed in JADES for $\alpha$ = 0.7 with detection limits for ngVLA (left column) and SKA (right column). The dotted, dashed and dot-dashed red and brown lines are H II region and SN flux densities for SFRs of 1, 3 and 10 \Ms\ yr$^{-1}$, respectively. The black solid, dashed and dotted horizontal bars show SKA limits for integration times of 1, 10 and 100 hr, respectively.}
\label{fig:f7}
\end{figure*}

\begin{figure*} 
\begin{center}
\includegraphics[scale=0.45]{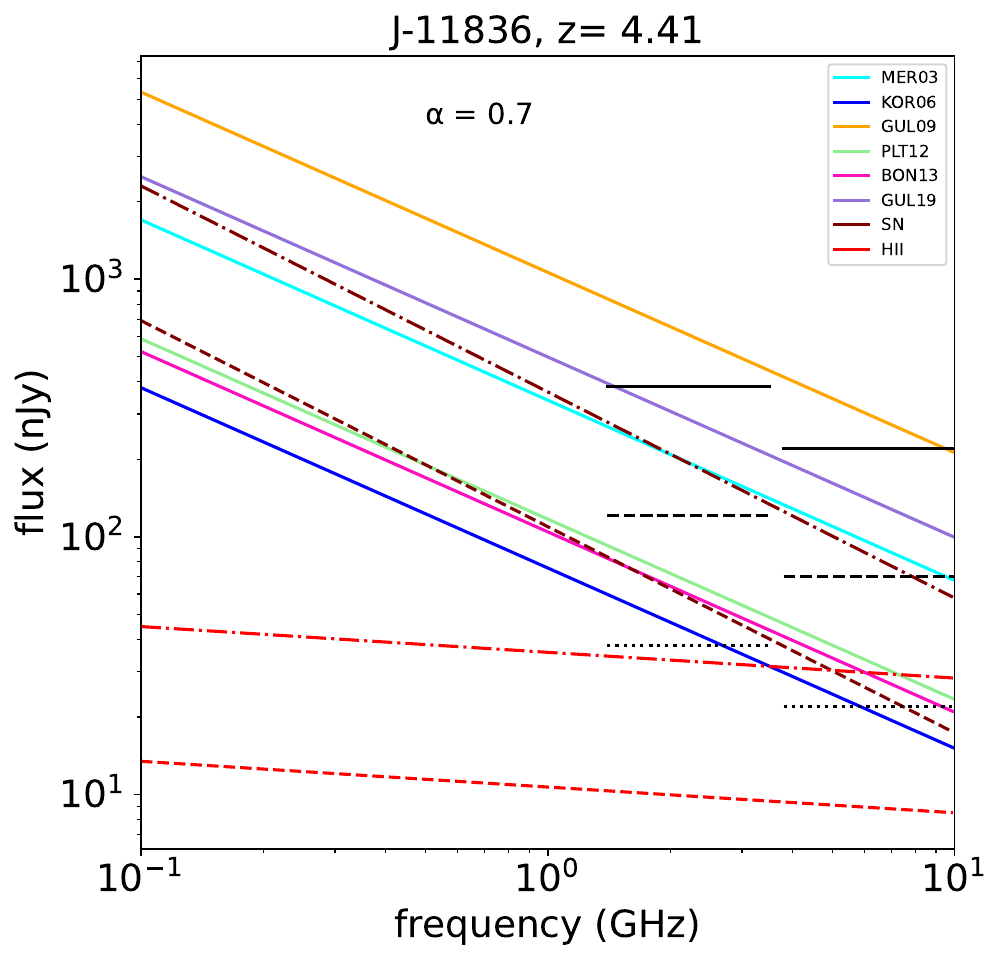} 
\includegraphics[scale=0.45]{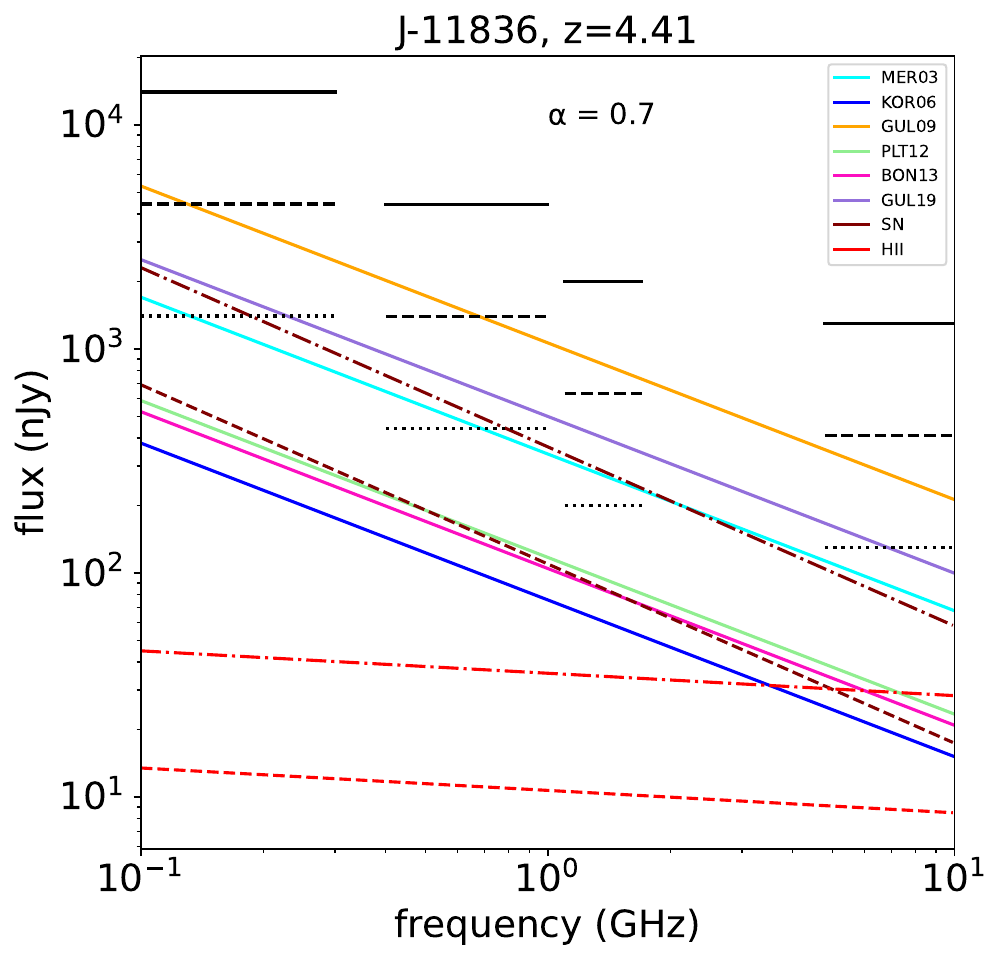} 
\includegraphics[scale=0.45]{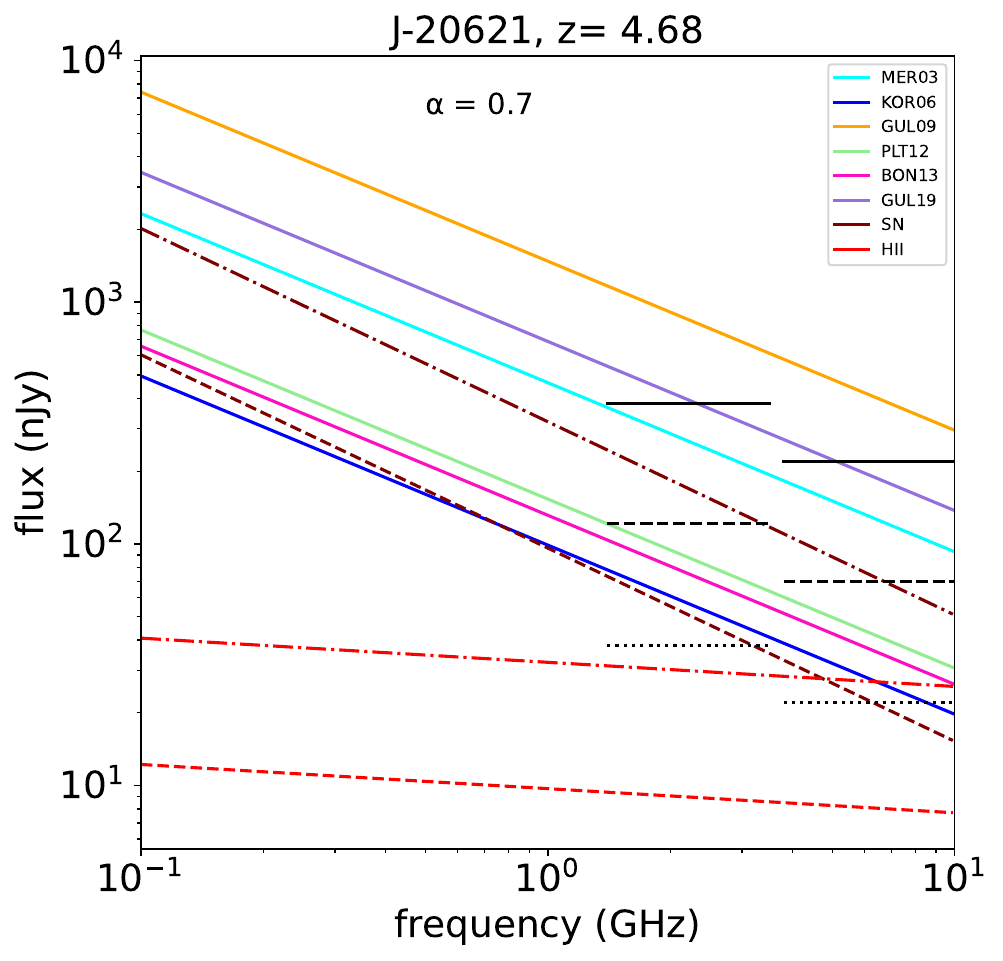} 
\includegraphics[scale=0.45]{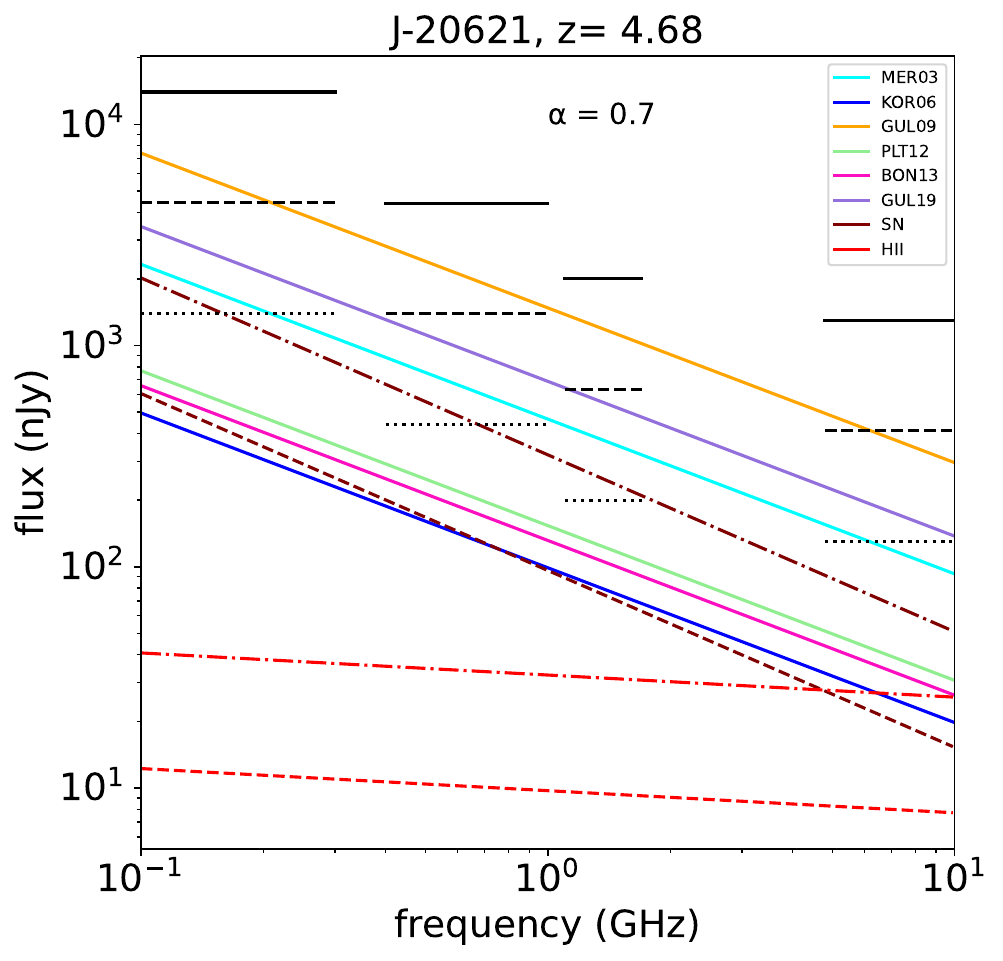} 
\includegraphics[scale=0.45]{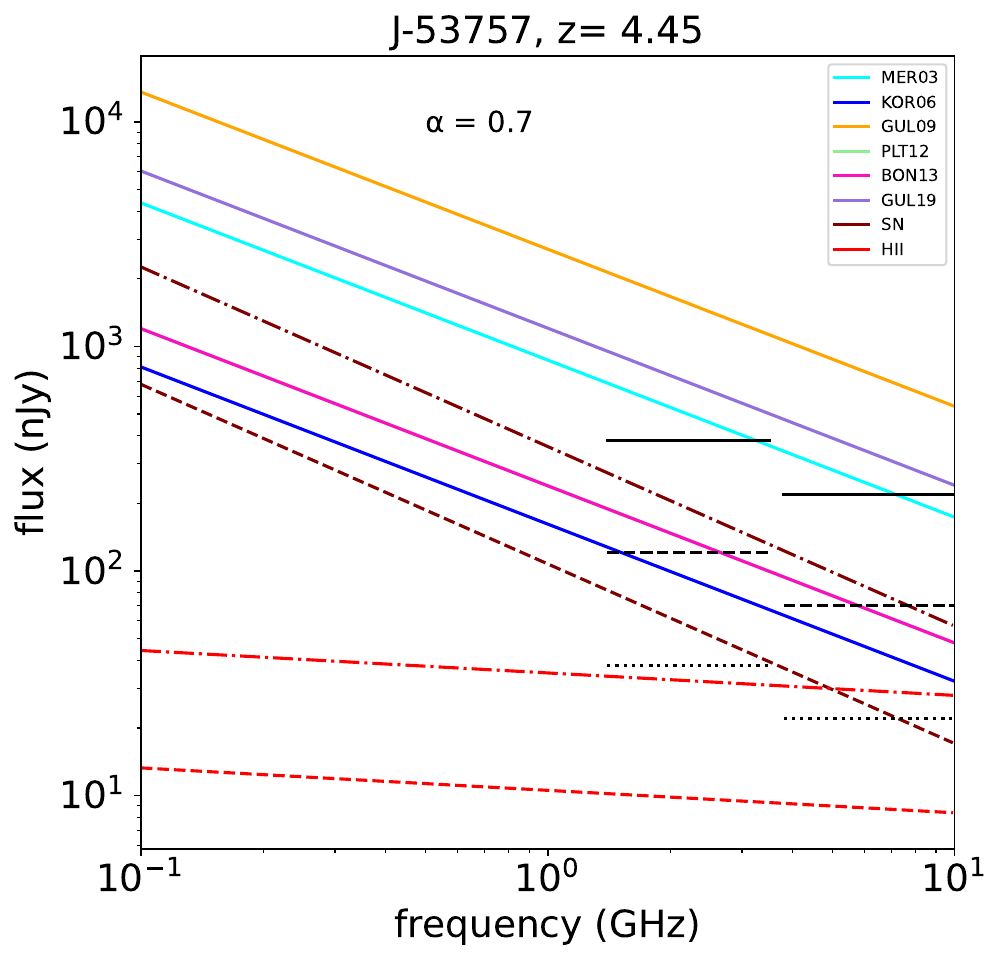} 
\includegraphics[scale=0.45]{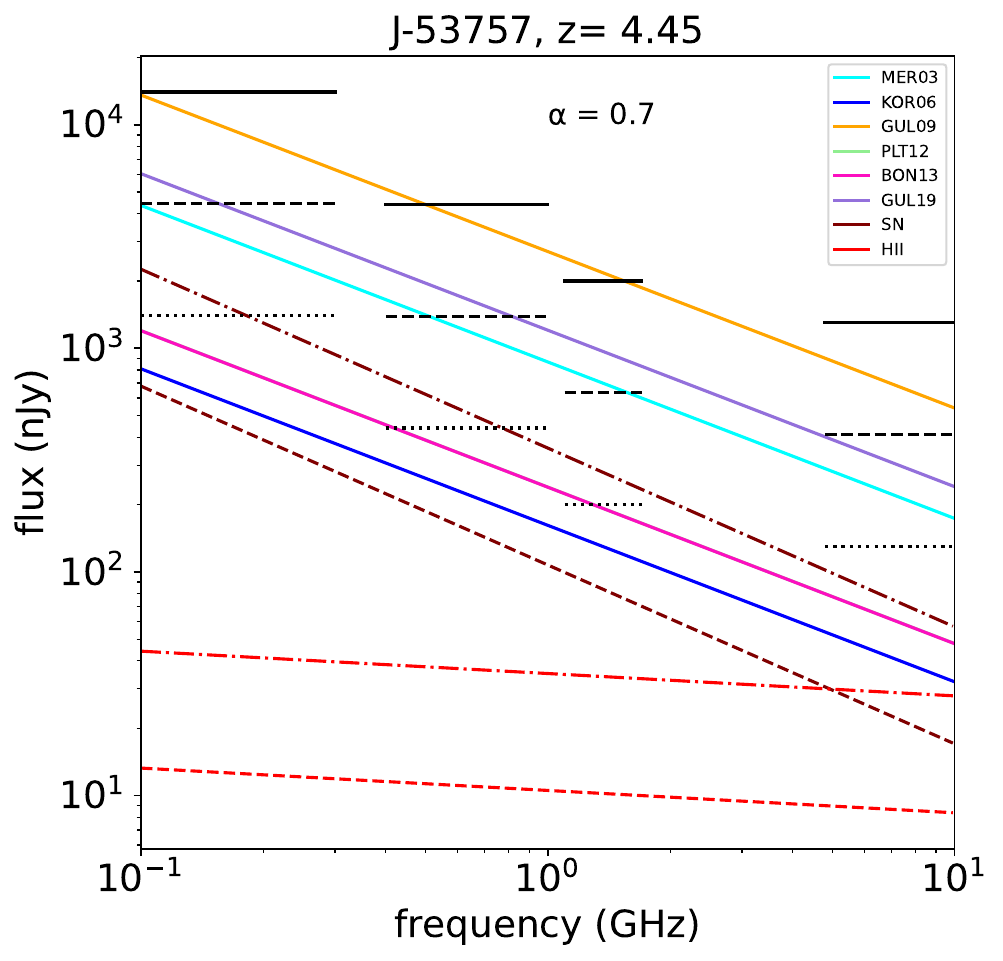} 
\end{center}
\vspace{-0.1cm}
\caption{Radio flux densities for AGNs observed in JADES  for $\alpha$ = 0.7 with detection limits for ngVLA (left column) and SKA (right column). The dotted, dashed and dot-dashed red and brown lines are H II region and SN flux densities for SFRs of 1, 3 and 10 \Ms\ yr$^{-1}$, respectively. The black solid, dashed and dotted horizontal bars show SKA limits for integration times of 1, 10 and 100 hr, respectively.}
\label{fig:f8}
\end{figure*}

\begin{figure*} 
\begin{center}
\includegraphics[scale=0.45]{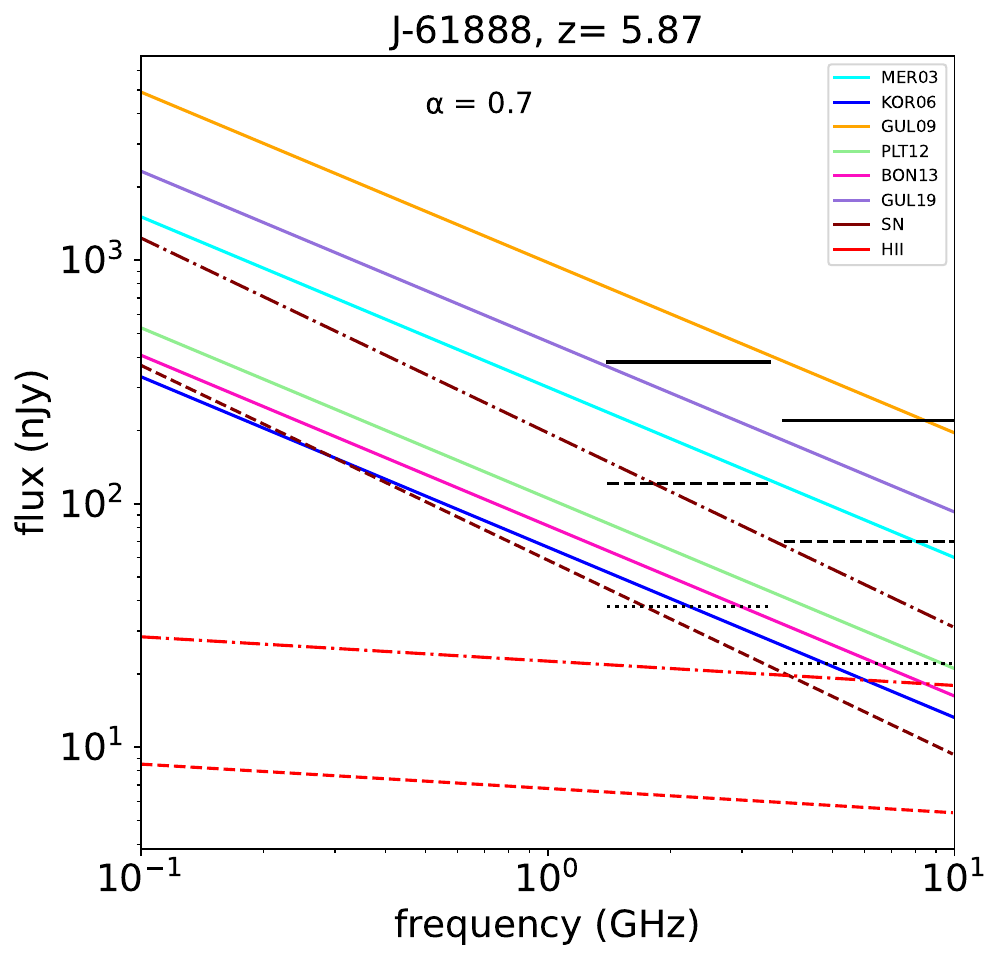}
\includegraphics[scale=0.45]{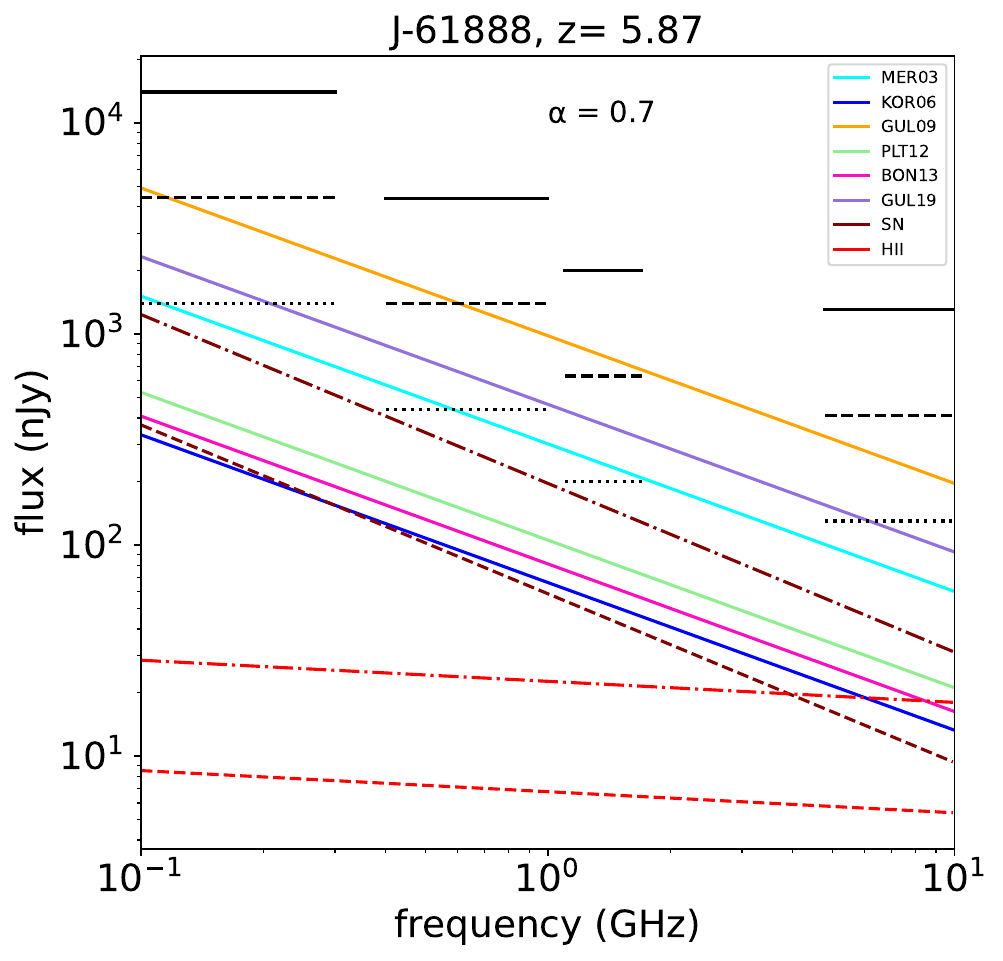}
\includegraphics[scale=0.45]{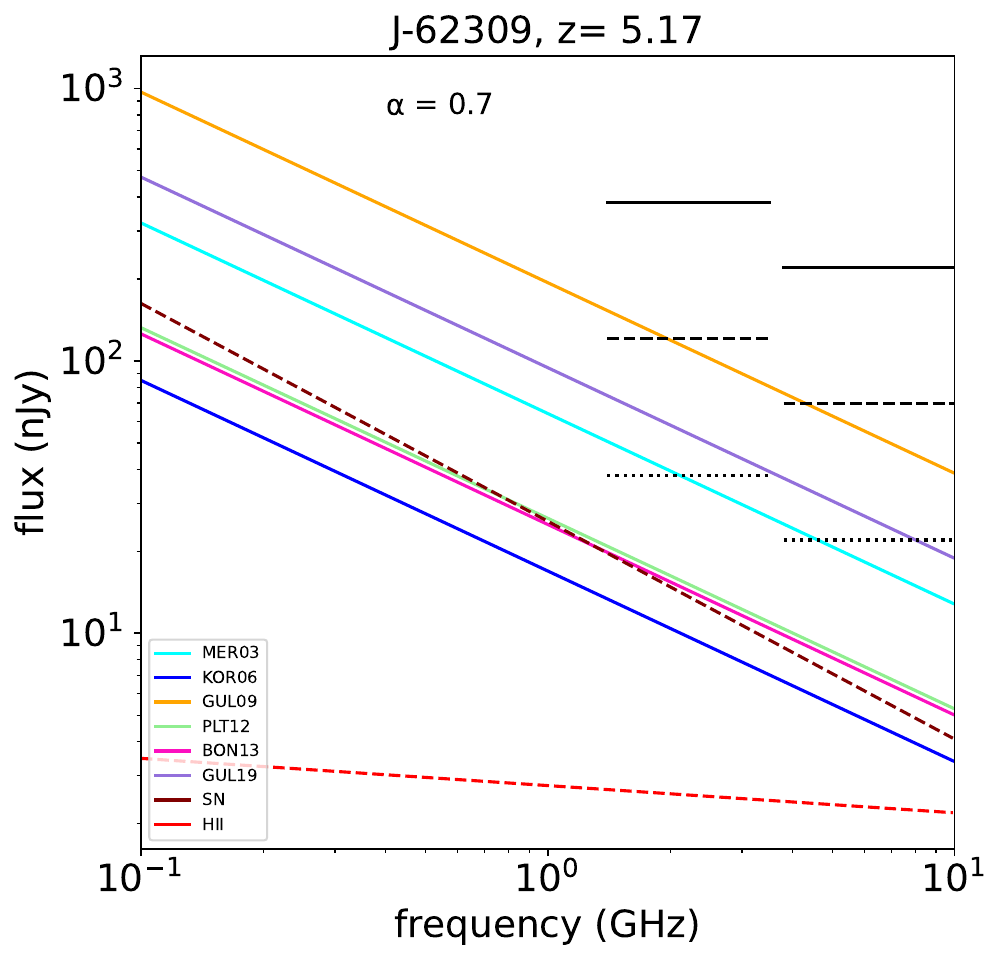}
\includegraphics[scale=0.45]{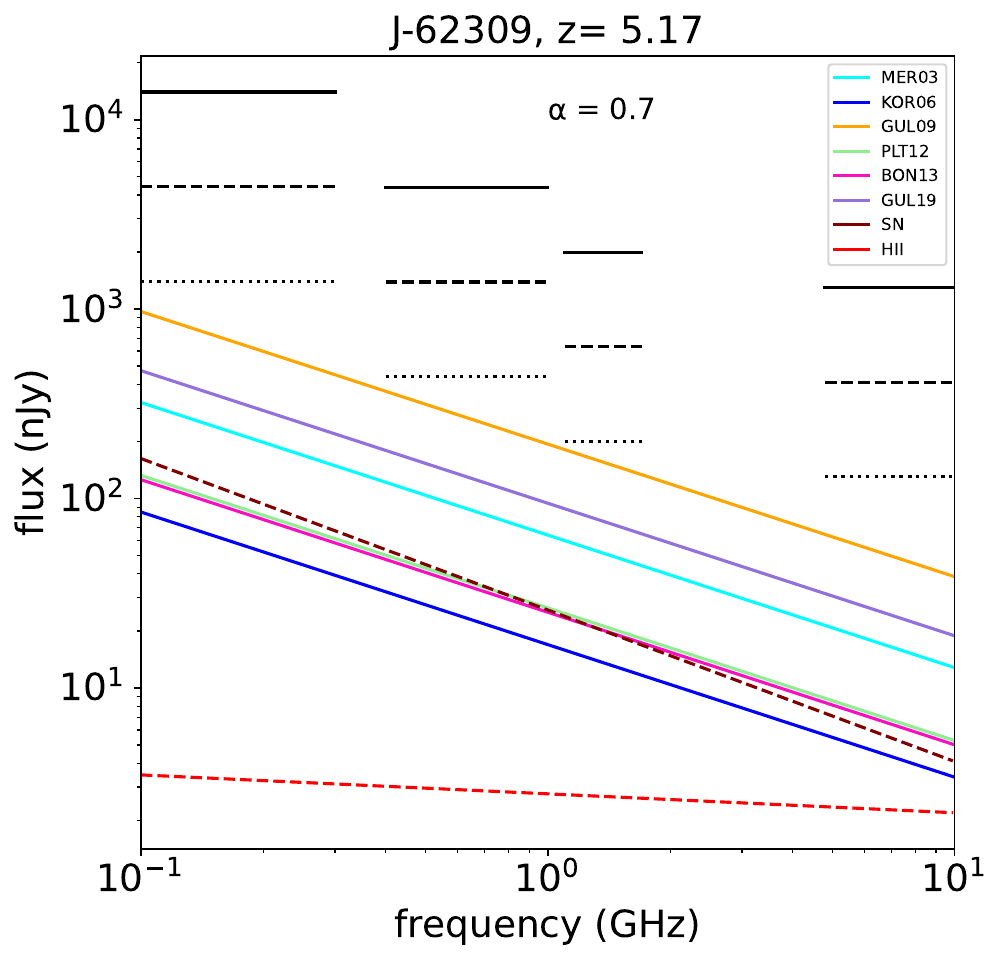}
\includegraphics[scale=0.45]{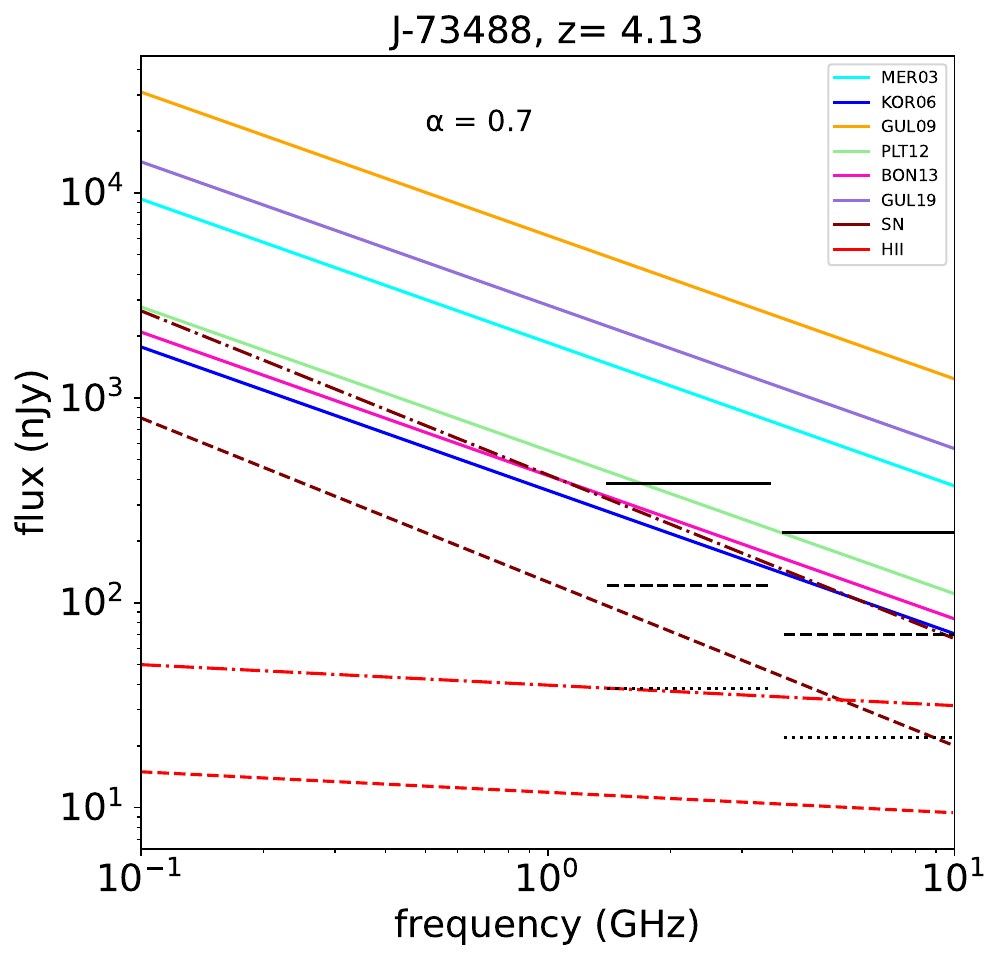}
\includegraphics[scale=0.45]{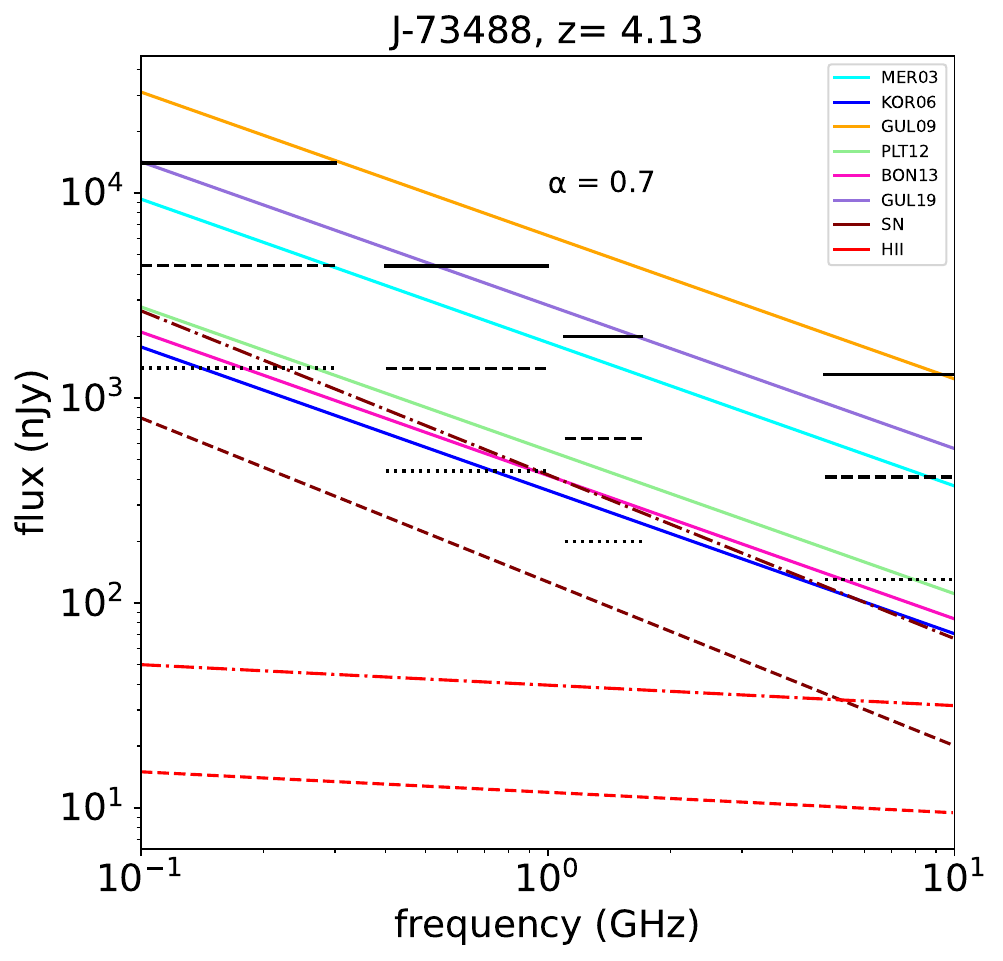}
\end{center}
\vspace{-0.1cm}
\caption{Radio flux densities for AGNs observed in JADES  for $\alpha$ = 0.7 with detection limits for ngVLA (left column) and SKA (right column). The dotted, dashed and dot-dashed red and brown lines are H II region and SN flux densities for SFRs of 1, 3 and 10 \Ms\ yr$^{-1}$, respectively. The black solid, dashed and dotted horizontal bars show SKA limits for integration times of 1, 10 and 100 hr, respectively.}
\label{fig:f9}
\end{figure*}

\begin{figure*} 
\begin{center}
\includegraphics[scale=0.45]{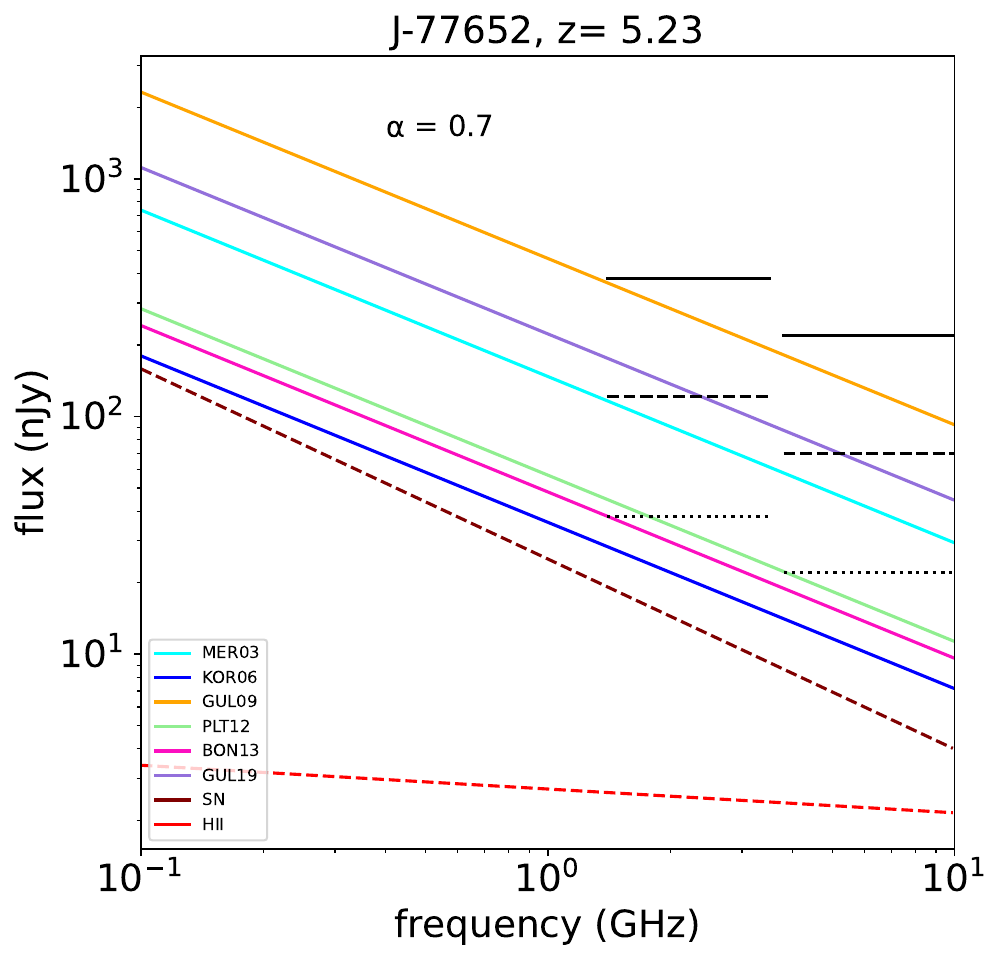}
\includegraphics[scale=0.45]{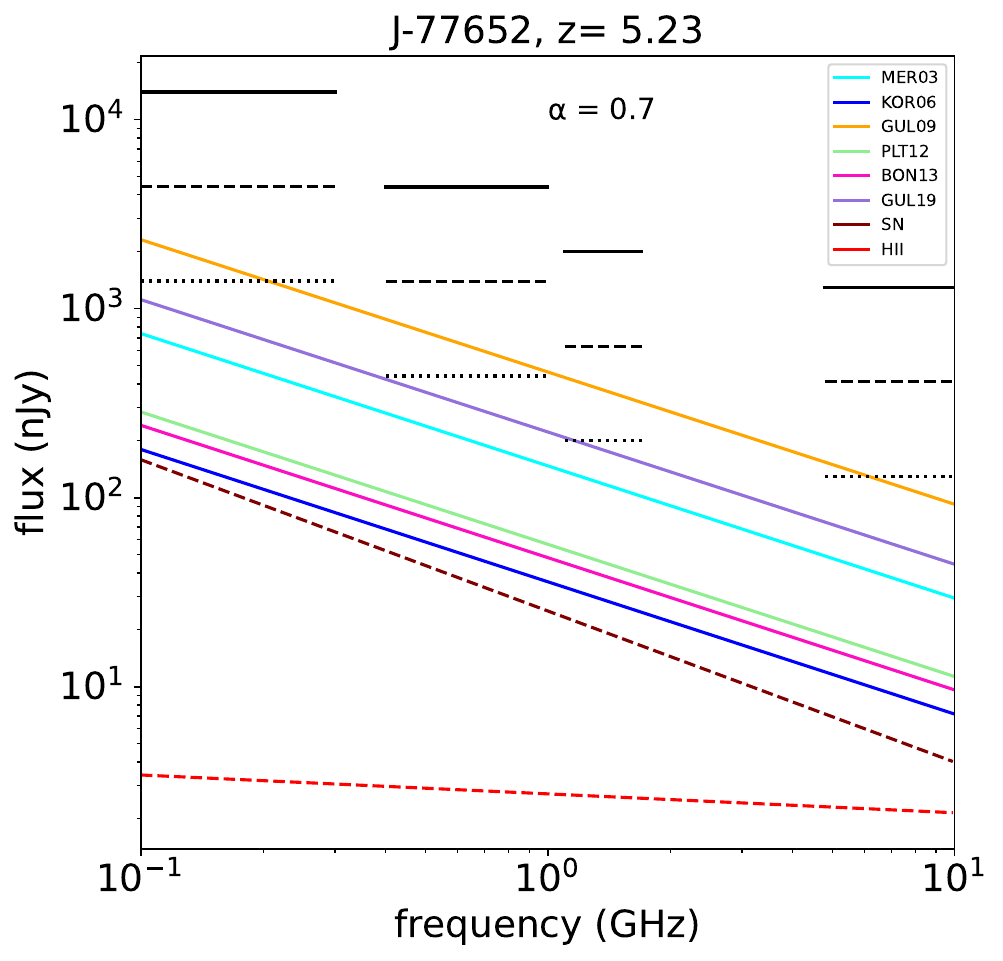}
\includegraphics[scale=0.45]{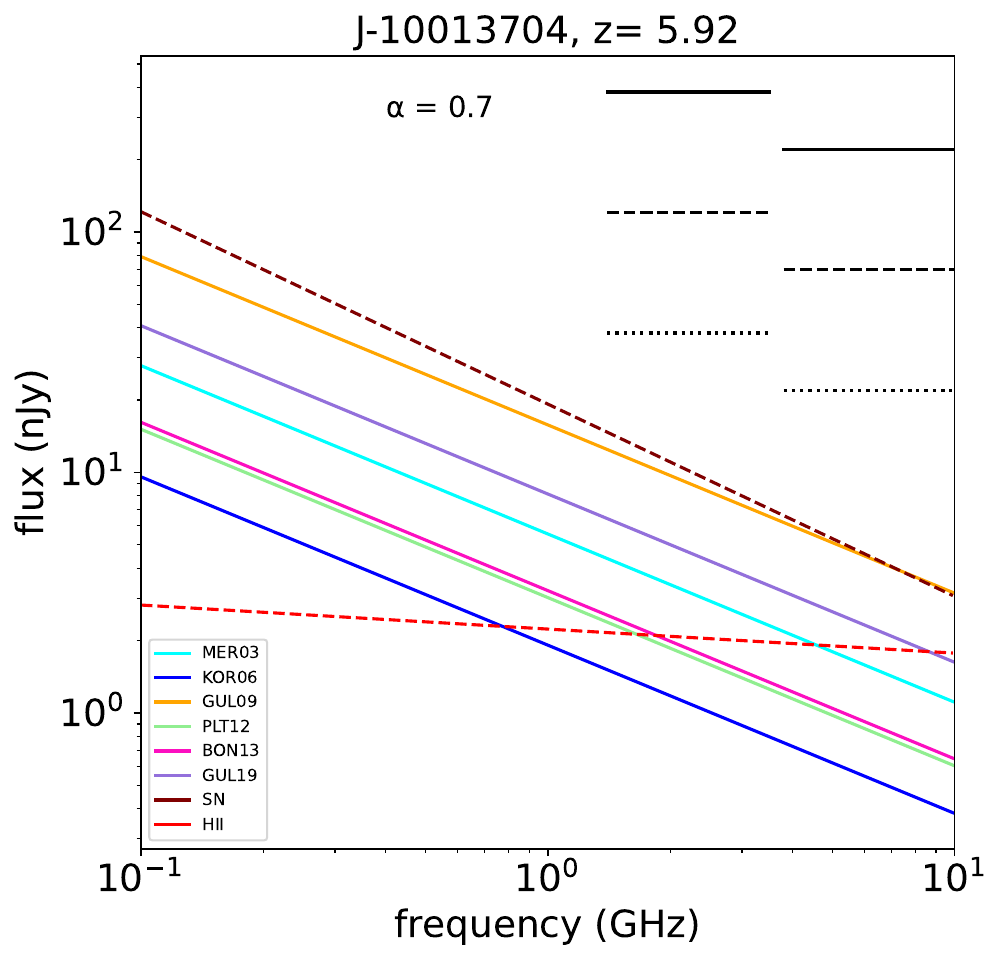}
\includegraphics[scale=0.45]{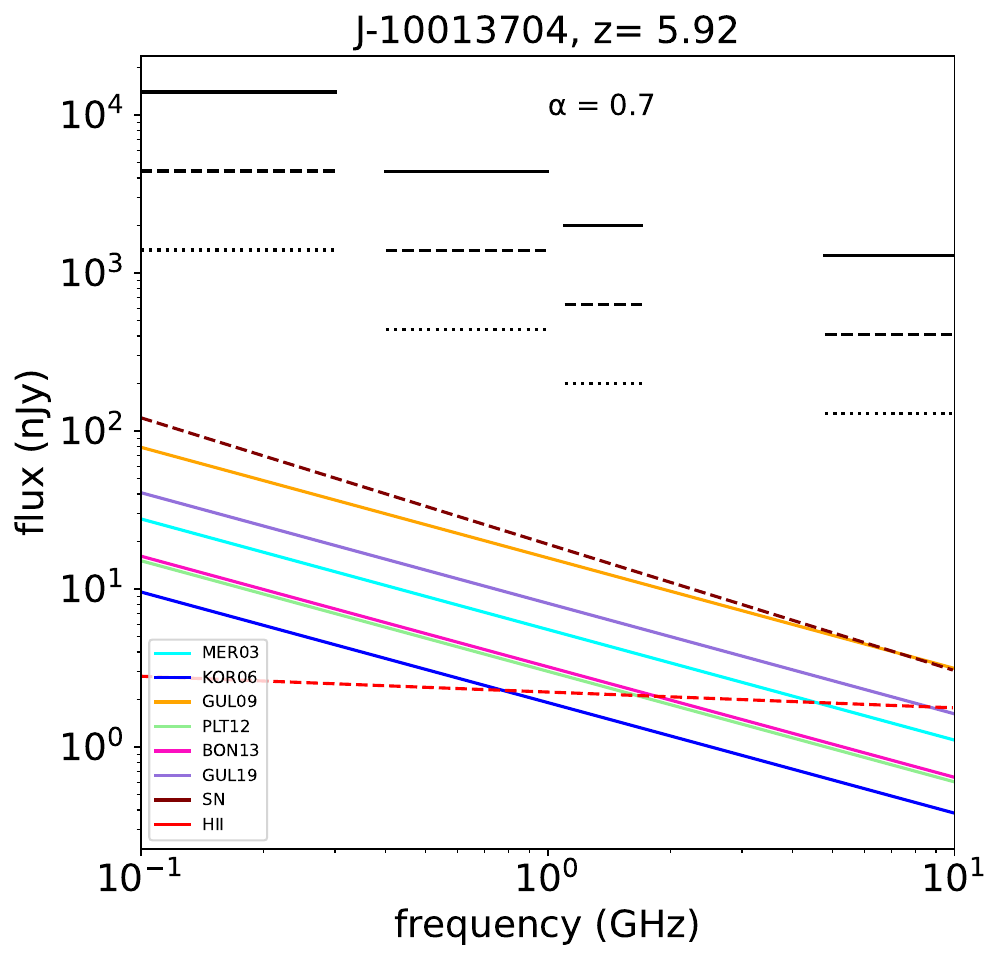}
\includegraphics[scale=0.45]{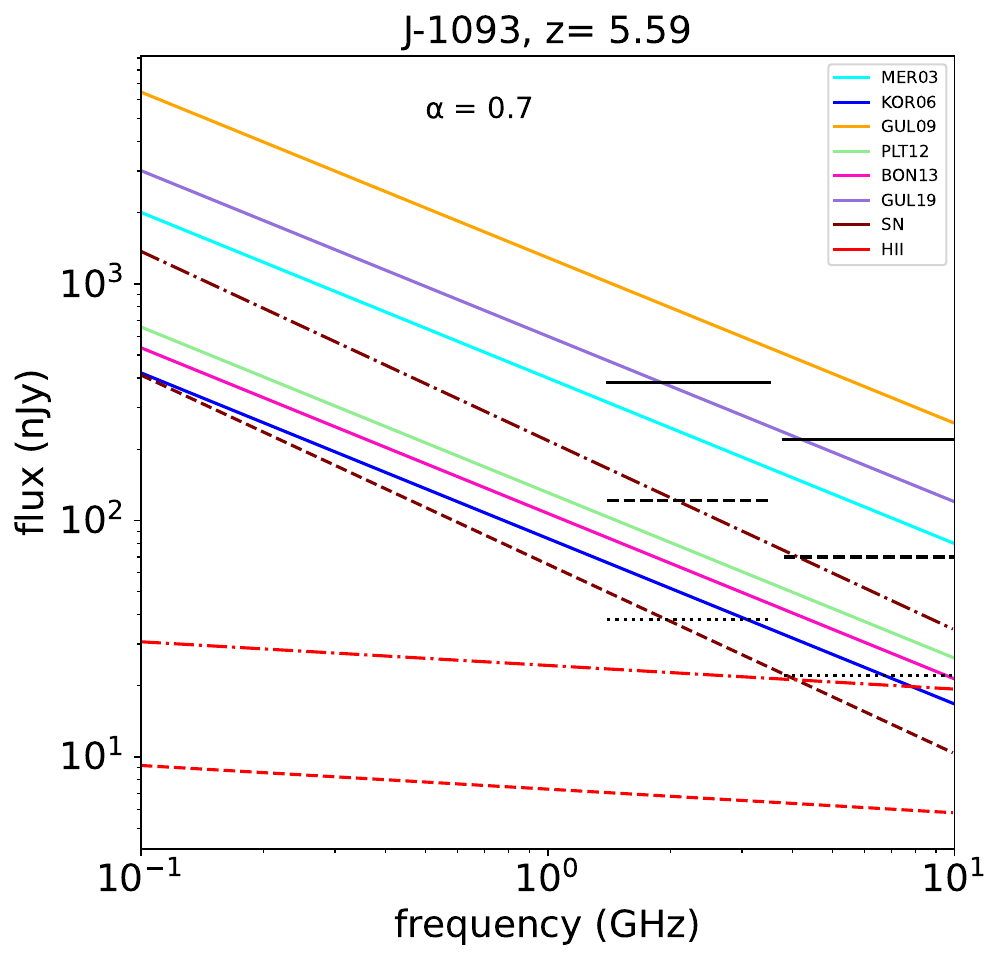}
\includegraphics[scale=0.45]{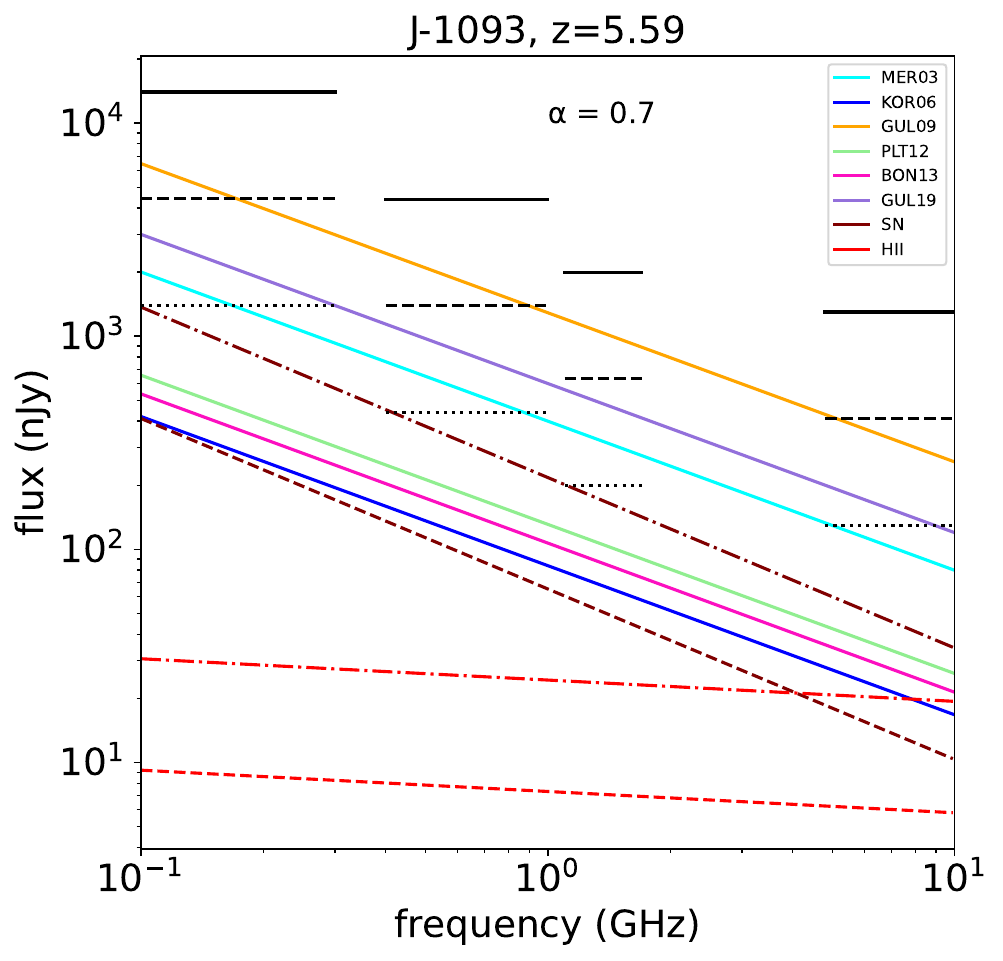}
\end{center}
\vspace{-0.1cm}
\caption{Radio flux densities for AGNs observed in JADES  for $\alpha$ = 0.7 with detection limits for ngVLA (left column) and SKA (right column). The dotted, dashed and dot-dashed red and brown lines are H II region and SN flux densities for SFRs of 1, 3 and 10 \Ms\ yr$^{-1}$, respectively. The black solid, dashed and dotted horizontal bars show SKA limits for integration times of 1, 10 and 100 hr, respectively.}
\label{fig:f10}
\end{figure*}

\begin{figure*} 
\begin{center}
\includegraphics[scale=0.45]{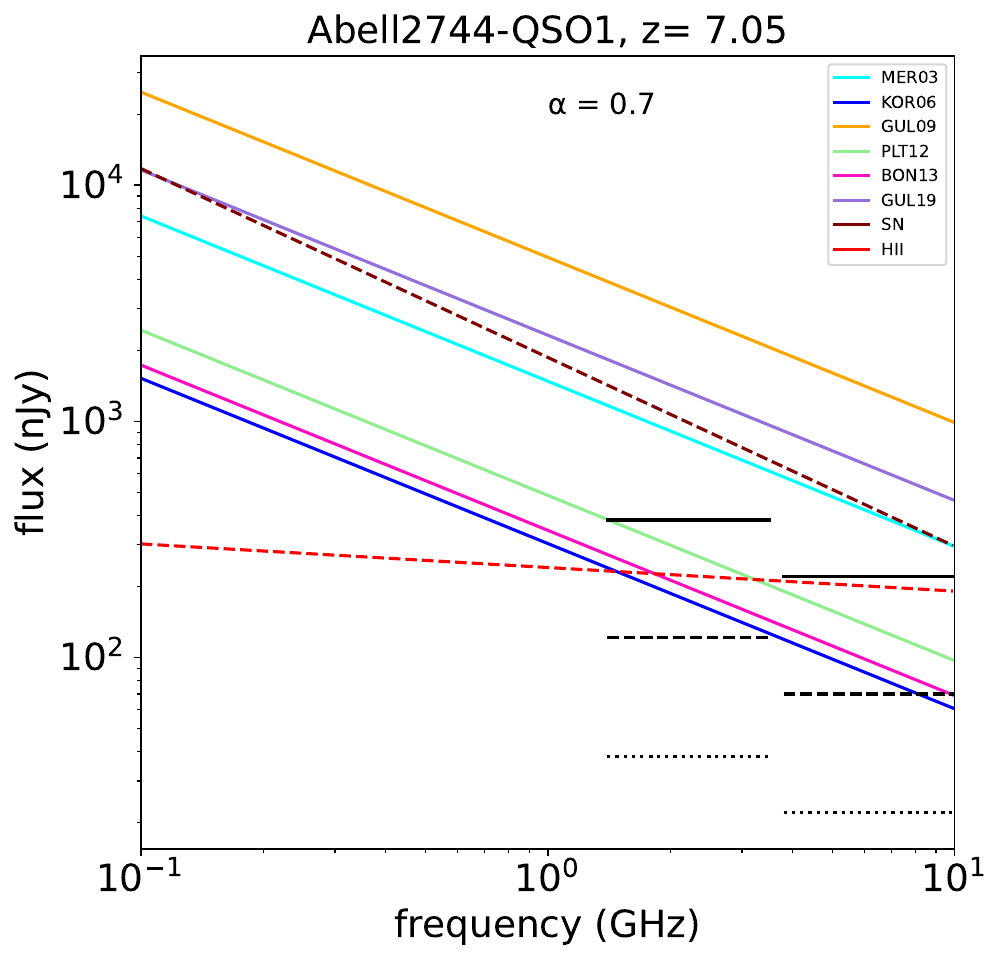}
\includegraphics[scale=0.45]{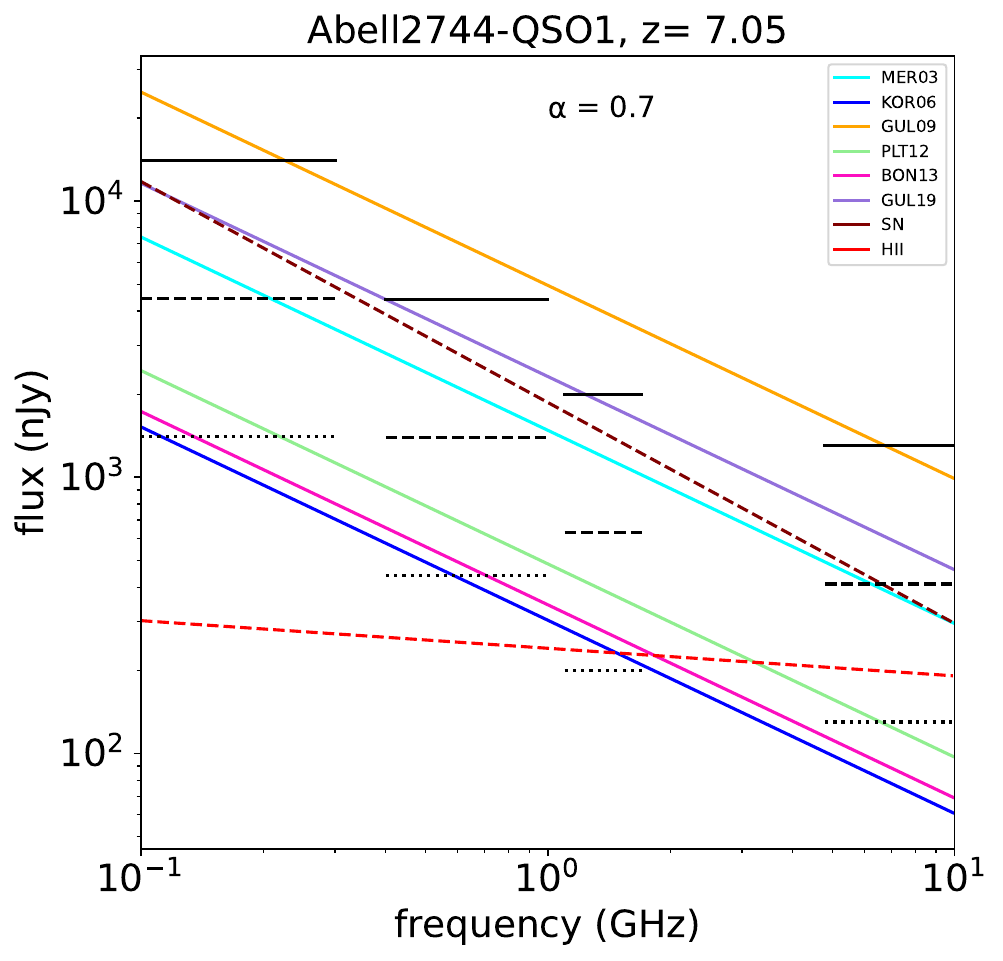}
\includegraphics[scale=0.45]{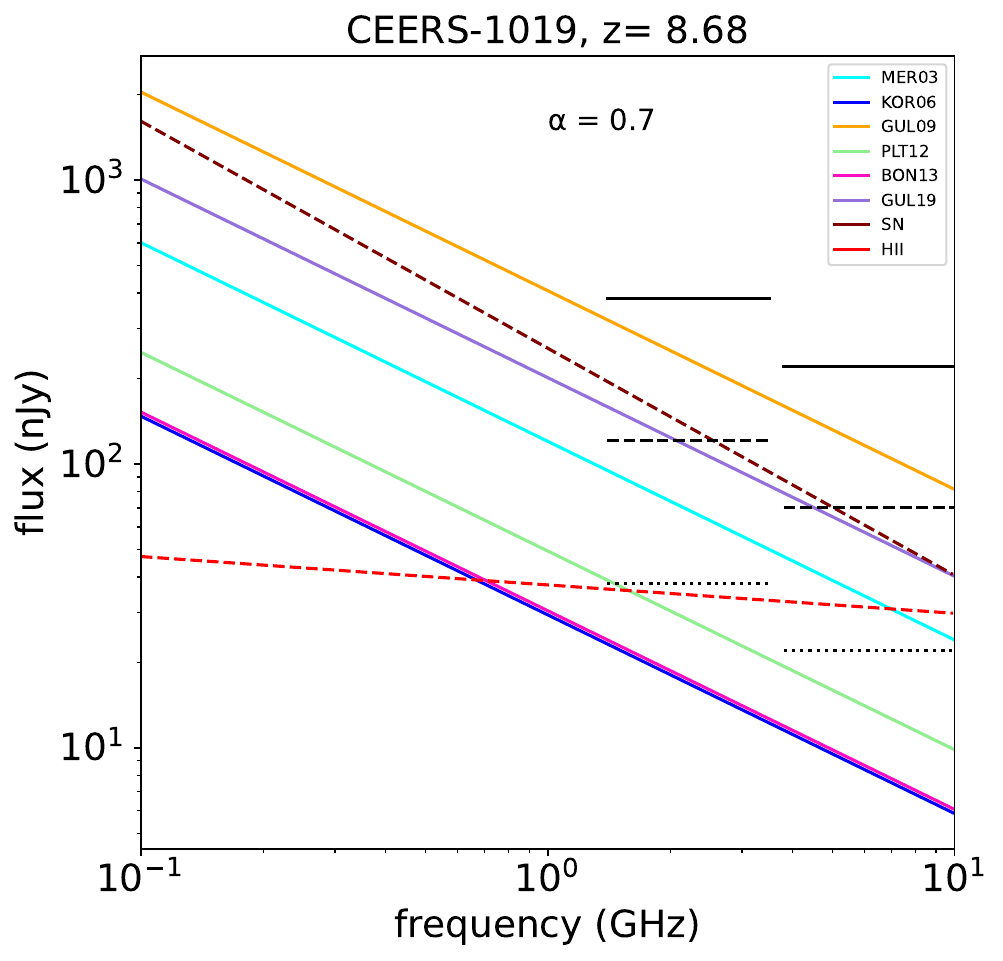}
\includegraphics[scale=0.45]{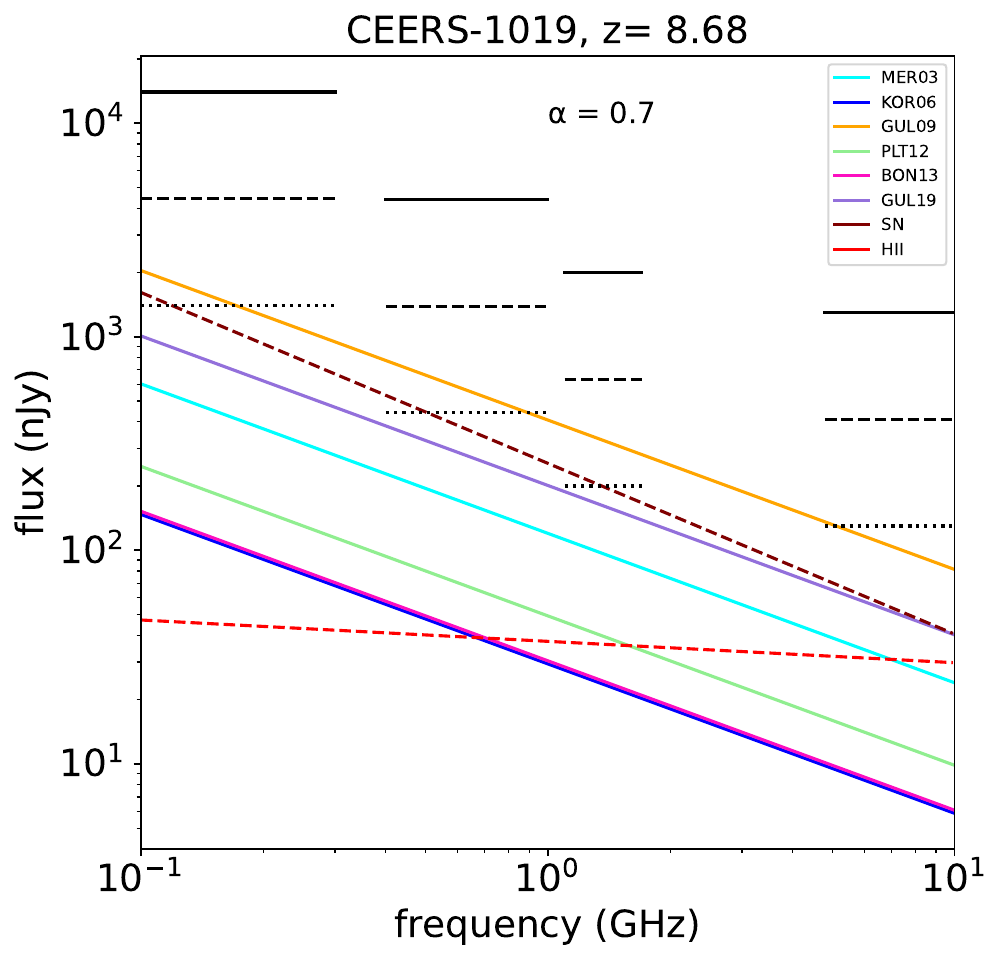}
\includegraphics[scale=0.45]{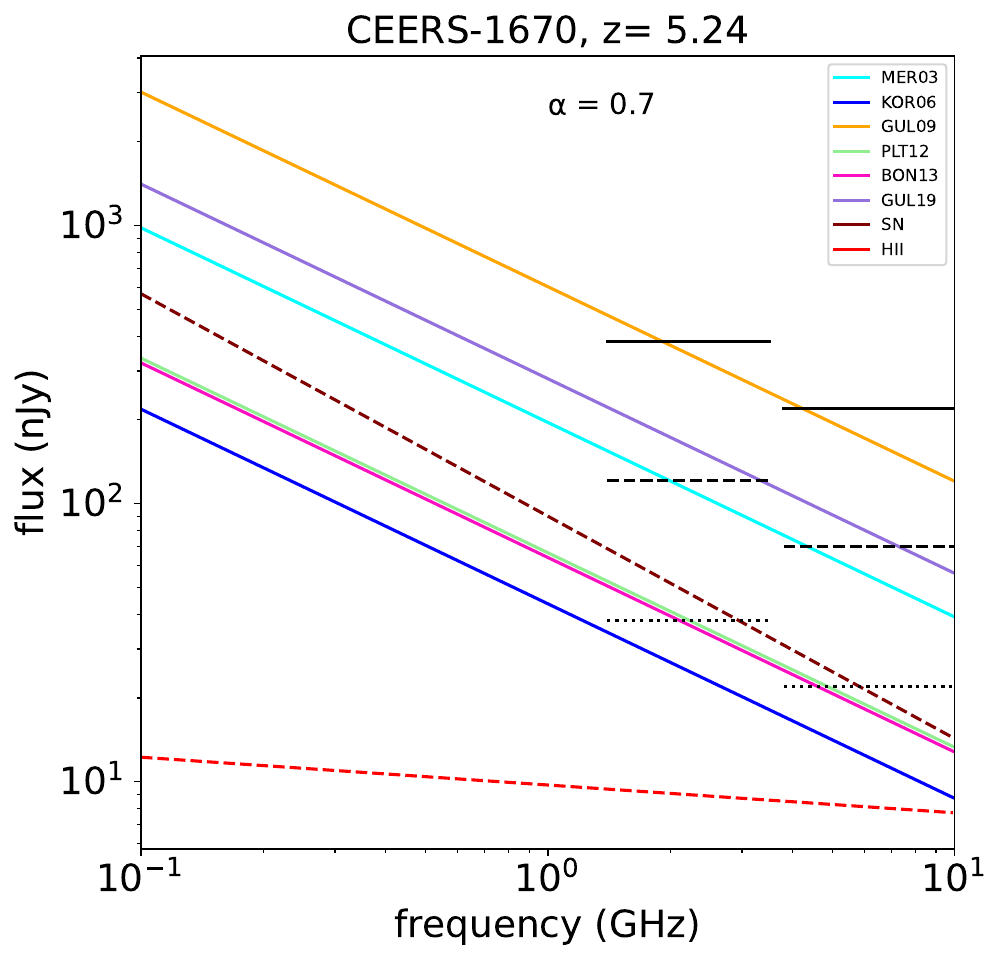}
\includegraphics[scale=0.45]{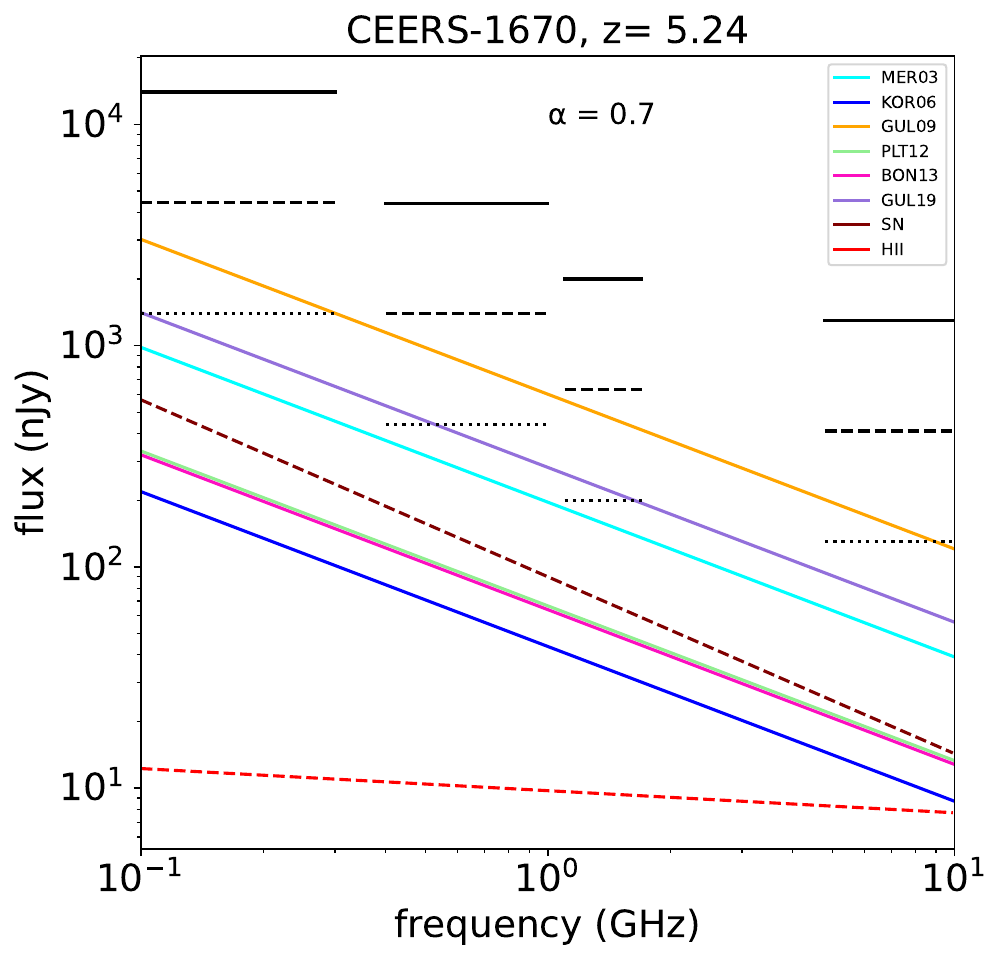}
\end{center}
\vspace{-0.1cm}
\caption{Radio flux densities for AGNs observed in CEERS, JADES and UNCOVER for $\alpha$ = 0.7 with detection limits for ngVLA (left column) and SKA (right column). The dotted, dashed and dot-dashed red and brown lines are H II region and SN flux densities for SFRs of 1, 3 and 10 \Ms\ yr$^{-1}$, respectively. The black solid, dashed and dotted horizontal bars show ngVLA limits for integration times of 1, 10 and 100 hr, respectively.}
\label{fig:f11}
\end{figure*}

\begin{figure*} 
\begin{center}
\includegraphics[scale=0.45]{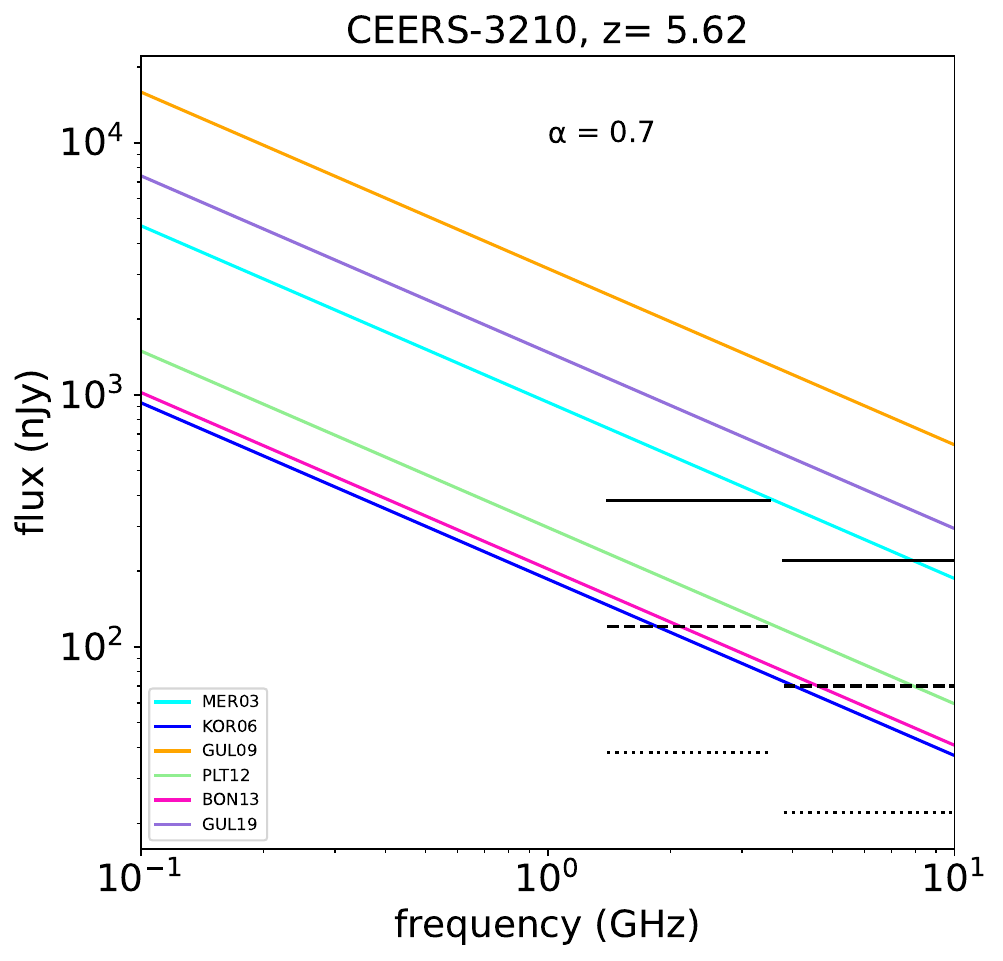} 
\includegraphics[scale=0.45]{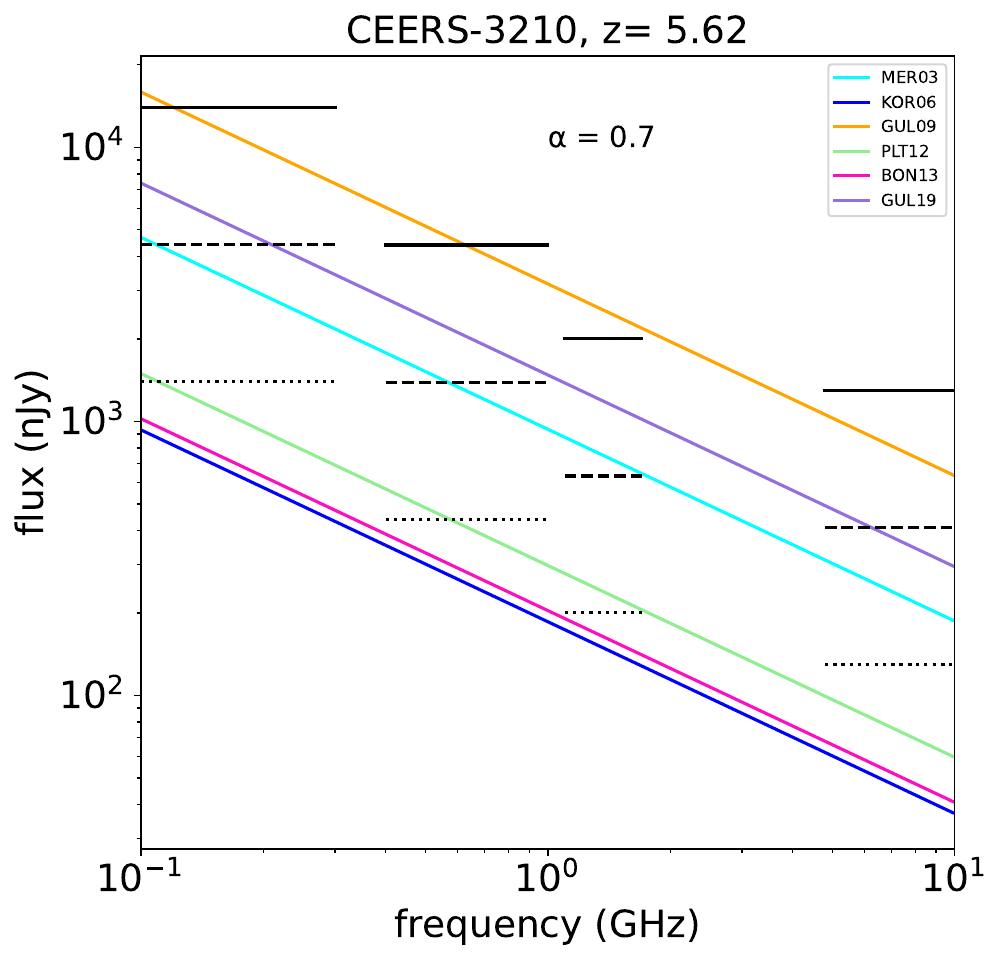} 
\includegraphics[scale=0.45]{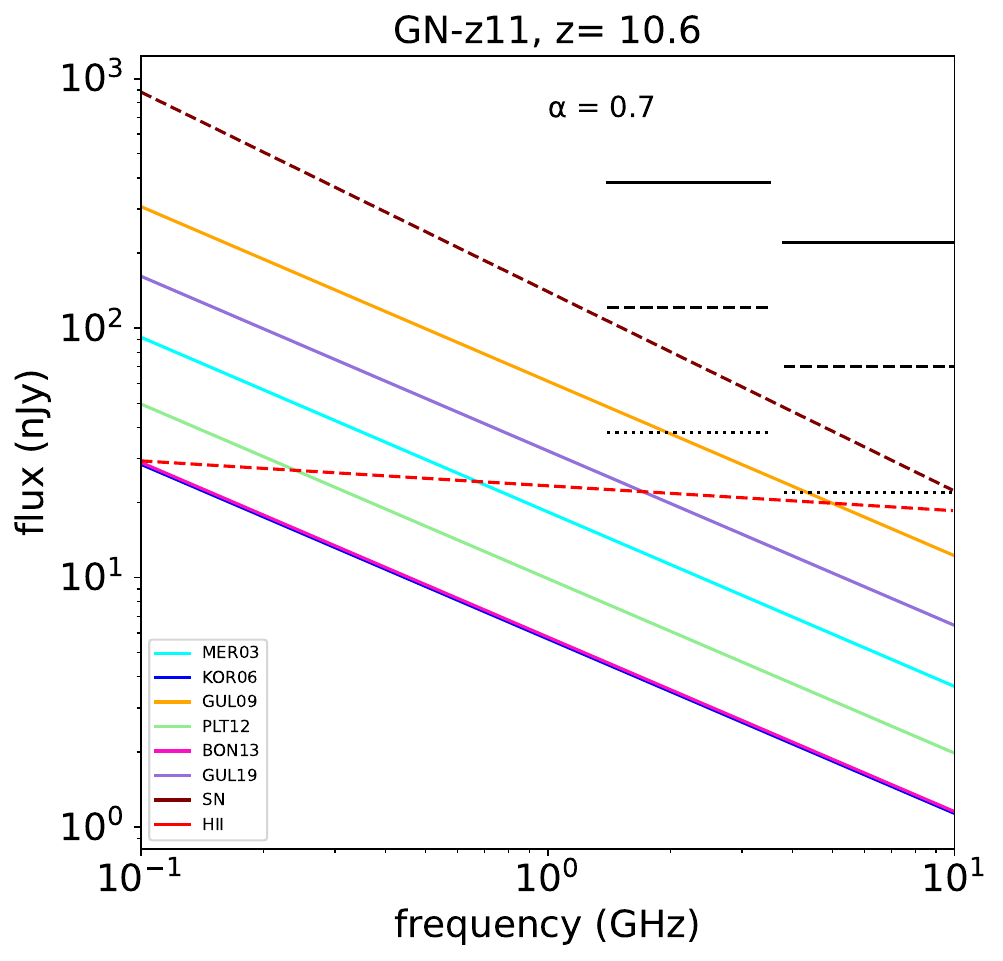}
\includegraphics[scale=0.45]{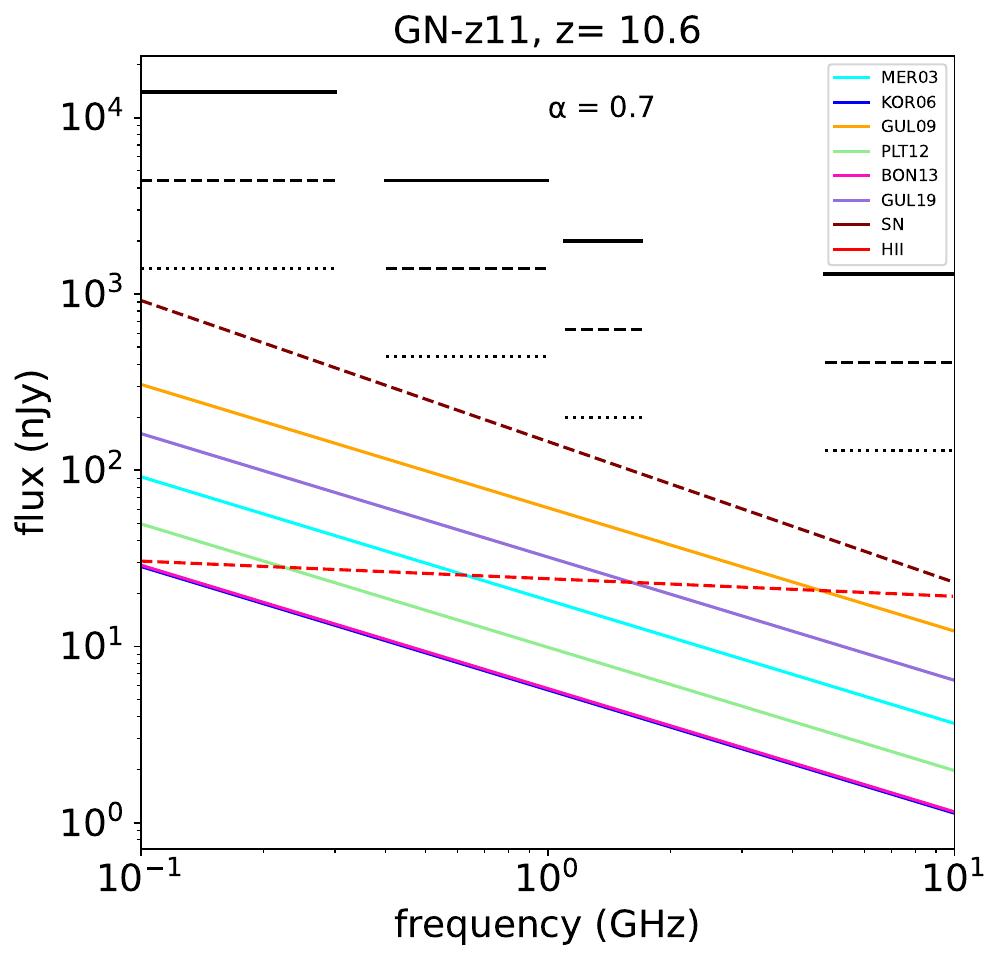}
\includegraphics[scale=0.45]{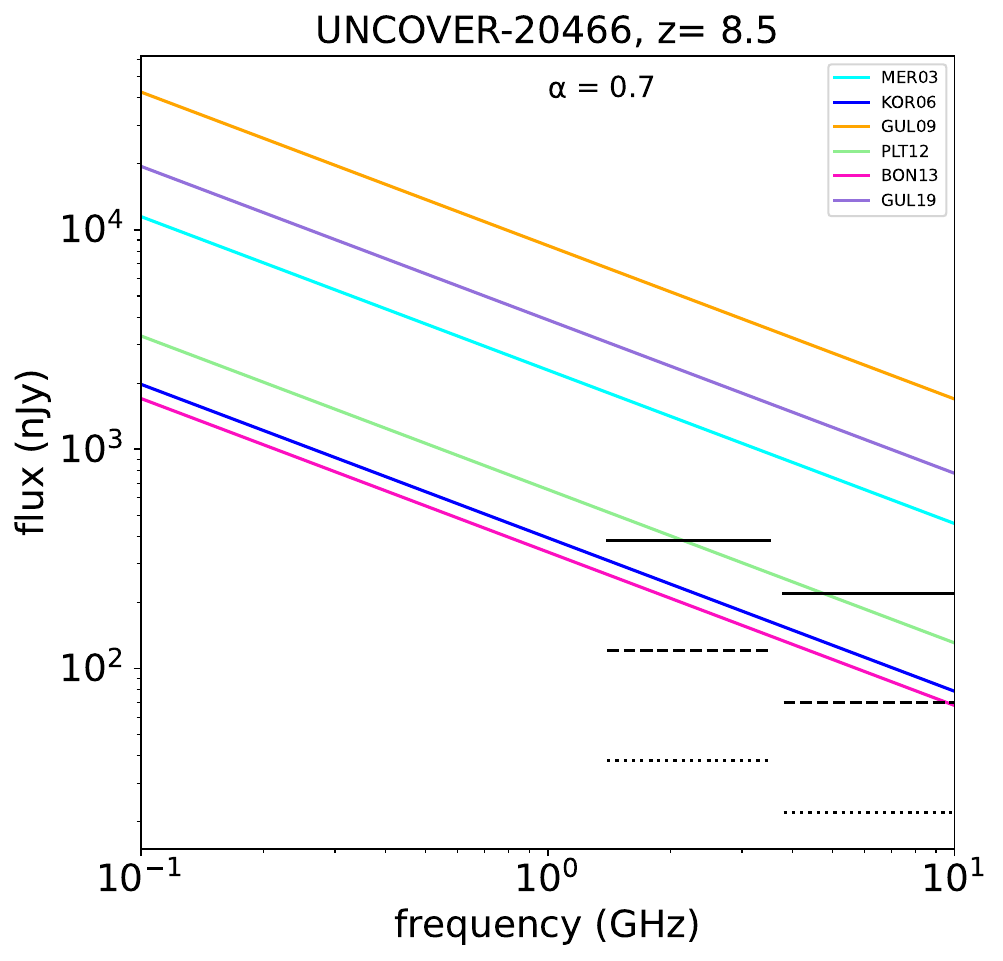}
\includegraphics[scale=0.45]{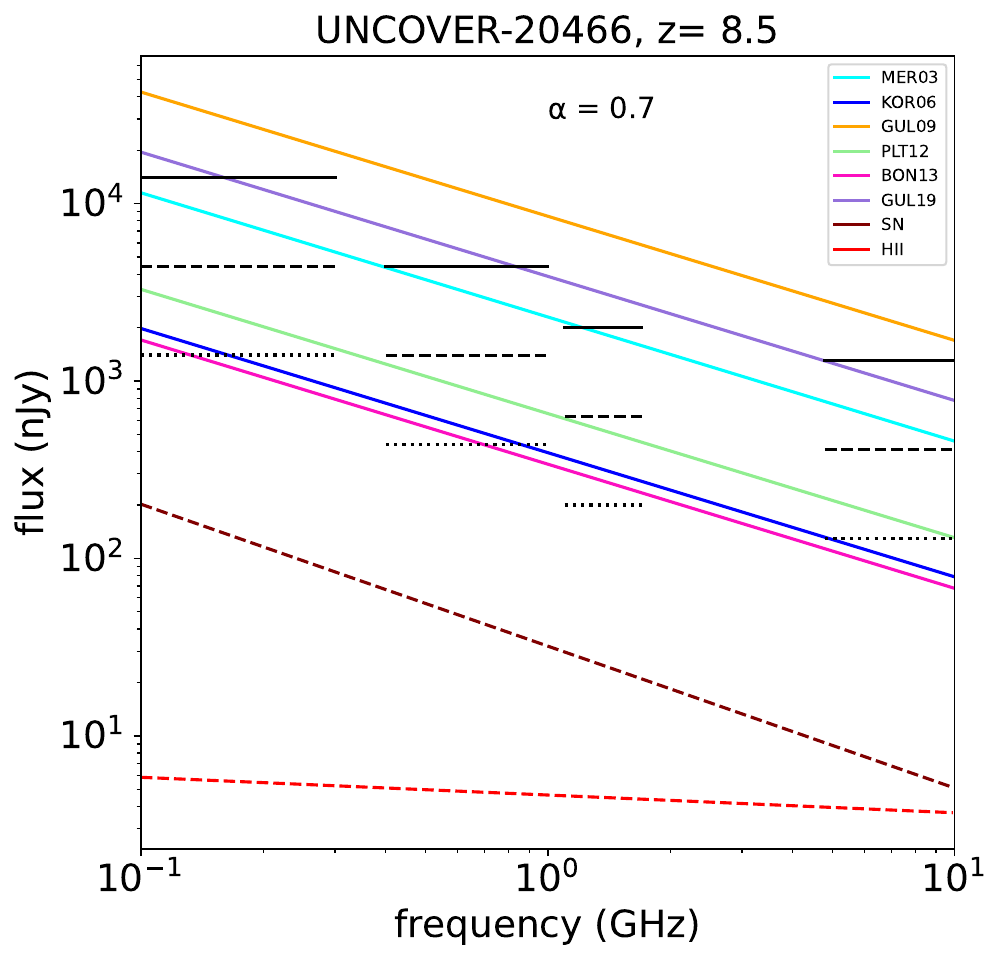}
\end{center}
\vspace{-0.1cm}
\caption{Radio flux densities for AGNs observed in CEERS, JADES and UNCOVER for $\alpha$ = 0.7 with detection limits for ngVLA (left column) and SKA (right column). The dotted, dashed and dot-dashed red and brown lines are H II region and SN flux densities for SFRs of 1, 3 and 10 \Ms\ yr$^{-1}$, respectively. The black solid, dashed and dotted horizontal bars show ngVLA limits for integration times of 1, 10 and 100 hr, respectively.}
\label{fig:f12}
\end{figure*}

We show radio flux densities for UHZ1 for both values of $\alpha$ in Figure \ref{fig:f13}, Jades sources with $\rm alpha =0.3$ in Figures \ref{fig:f4}-\ref{fig:f6} and all sources with $\rm alpha =0.7$ in Figures~\ref{fig:f7}-\ref{fig:f12}.

%\section{Figure for $\rm alpha =0.7$}
%\ch{We show the plots for all sources with $\rm alpha =0.7$ in the appendix figures 8-13.}

\end{document}